\documentclass[prb,onecolumn,10pt,superscriptaddress,showpacs]{revtex4}
\usepackage{graphicx}

\begin{document}

\title{
Effects of lattice geometry and interaction range on polaron dynamics
}

\author{J.P. Hague}
\affiliation{Department
 of Physics, Loughborough University, Loughborough, LE11 3TU, United Kingdom}

\author{P.E. Kornilovitch} 
\affiliation{Hewlett-Packard Company, 1000 NE Circle Blvd, 
Corvallis, Oregon 97330, USA}

\author{A.S. Alexandrov}
\affiliation{Department
 of Physics, Loughborough University, Loughborough, LE11 3TU, United Kingdom}

\author{J.H. Samson}
\affiliation{Department
 of Physics, Loughborough University, Loughborough, LE11 3TU, United Kingdom}

\begin{abstract}

We study the effects of lattice type on polaron dynamics using a
continuous-time quantum Monte-Carlo approach. Holstein and screened
Fr\"{o}hlich polarons are simulated on a number of different Bravais
lattices. The effective mass, isotope coefficients, ground state
energy and energy spectra, phonon numbers, and density of states are
calculated. In addition, the results are compared with weak and strong
coupling perturbation theory. For the Holstein polaron, it is found
that the crossover between weak and strong coupling results becomes
sharper as the coordination number is increased. In higher dimensions,
polarons are much less mobile at strong coupling, with more phonons
contributing to the polaron. The total energy decreases monotonically
with coupling. Spectral properties of the polaron depend on the
lattice type considered, with the dimensionality contributing to the
shape and the coordination number to the bandwidth. As the range of
the electron-phonon interaction is increased, the coordination number
becomes less important, with the dimensionality taking the leading
role.


\end{abstract}

\maketitle

\section{Introduction}

Interest in the role of electron-phonon (e-ph) interactions and
polaron dynamics in contemporary materials has recently gone through a
large revival.  Electron-phonon interactions have been shown to be
relevant in the cuprate superconductors through isotope substitution
experiments \cite{zhao1997a}, recent high resolution angle resolved
photoemission \cite{lanzara2001a}, and a number of earlier optical
\cite{mic,ita} and neutron-scattering \cite{ega} spectroscopies.  In
the colossal magnetoresistance manganites, isotope substitution also
shows a significant effect on the physical properties
\cite{zhao1996a}.  It has been suggested that the long-range
Fr\"ohlich e-ph interaction is relevant in cuprates and manganites at
any doping because of poor screening \cite{alebra}, although the
spatial extent of the interactions is still part of an ongoing
discussion \cite{edwards2002}.

The polaron problem has been actively researched for a long time (for
a review see Refs. \onlinecite{alemot,dev}).  For weak electron-phonon
coupling $\lambda\ll 1$ in the adiabatic limit $\hbar\omega /E_F \ll
1$, Migdal theory describes electron dynamics \cite{migdal1958a}.
With increasing strength of interaction and increasing phonon
frequency $\omega$ finite bandwidth \cite{alemaz,mar} and vertex
corrections \cite{hag} become important.  The neglect of vertex
corrections in the Migdal approximation breaks down entirely at
$\lambda \sim 1$ for any value of the adiabatic ratio $\hbar
\omega/E_F$ because the bandwidth is narrowed and the Fermi energy is
renormalized down exponentially.  As a result, the effective parameter
$\lambda\hbar\omega/E_F$ becomes large \cite{alebook}.  In the strong
coupling limit $\lambda \gg 1$, a canonical Lang--Firsov (LF)
transformation can be used to determine the properties of the small
polaron \cite{lang1962a}.  The effective mass does not diverge at a
critical coupling but instead increases exponentially.  No general
solution to the polaron problem exists for all values of the coupling
constant, except in the infinite dimensional limit where dynamical
mean-field theory (DMFT) can be applied \cite{ciuchi1997a}.  It is the
enormous differences between weak and strong coupling limits and
adiabatic and antiadiabatic limits which make the polaron problem in
the intermediate $\lambda$ regime difficult to study analytically and
numerically.

Several methods exist for the numerical simulation of the polaron
problem at intermediate $\lambda$.  They include exact diagonalization
\cite{alekab}, advanced variational methods \cite{tru} (effects of
dimensionality are also discussed), conventional discrete-time path
integral quantum Monte-Carlo (QMC) algorithms \cite{qmc,DeRaedt} as
well as the newer density matrix renormalization group \cite{White}
(DMRG), continuous-time QMC \cite{kornilovitch1998a,spencer2005a}, LF
QMC \cite{hohen}, and diagrammatic QMC \cite{Prokofev,Macridin}.  The
methods vary in accuracy and versatility but none can provide all the
polaron properties of interest in the entire space of model
parameters.  For instance, exact diagonalizations suffer from the
necessary truncation of the phonon Hilbert space, especially at strong
couplings and low phonon frequencies (even then, the total Hilbert
space is huge, reducing the number of sites and leading to poor
momentum resolution), DMRG cannot easily handle long-range
interactions, diagrammatic QMC and exact diagonalization are
inconvenient in calculating the density of states, path-integral QMC
slows down at small frequencies, and so on.  In numerical analysis of
polaron models, a complex approach is needed where each method is
employed to calculate what it does best.

In this paper we use one of these methods, the continuous-time
path-integral quantum Monte-Carlo algorithm (CTQMC) to investigate the
effects of lattice geometry on the properties of the polaron from weak
to strong coupling.  This algorithm was developed previously by one of
us \cite{kornilovitch1998a} and was based on the analytical
integration of phonons introduced by Feynman \cite{Feynman} and on an
earlier numerical implementation in discrete time by De Raedt and
Lagendijk \cite{DeRaedt}.  CTQMC introduced two critical improvements.
Firstly, formulation in continuous imaginary time eliminated errors
caused by the Trotter slicing and made the method numerically exact
for any strength of electron-phonon interaction.  Secondly,
introduction of twisted boundary conditions in imaginary time
\cite{KornilovitchPike,kornilovitch1998a} enabled calculation of
polaron effective masses, spectra and even densities of states.
Although the method is not ideal (it slows down at low phonon
frequencies because of the condition on the inverse temperature $\beta
\gg (\hbar \omega)^{-1}$ and suffers from a sign problem at small
coupling and large polaron momenta) it is quite versatile.  In
particular, it has enabled accurate analysis of models with long-range
electron-phonon interactions \cite{alexandrov1999a,spencer2005a} and a
model with anisotropic electron hopping \cite{Kornilovitch1999b}.  In
addition, it is currently the only method that can provide numerically
{\em exact} polaron densities of states \cite{kornilovitch1998a} and
isotope exponents \cite{spencer2005a,KornilovitchAlexandrov2004}. We note that the variational approach of Bon\v{c}a \emph{et al.} has been used to compute the mass isotope effect of the bipolaron, and presumably such a method could be used to compute the isotope exponent for the polaron \cite{tru}.
  
In this paper, we develop the algorithm further.  Because it is
formulated in real space it is ideally suited for analyzing more
complex lattices than the conventional linear, square and cubic
Bravais lattices.  All that is needed is a redefined set of
nearest-neighbor hops (or kinks) by which the QMC process moves the
polaron path through the space.  We will analyze and compare several
Bravais lattices: linear, square and triangular in $d=2$, simple
cubic, face-center-cubic (FCC), hexagonal and body-center-cubic (BCC)
in $d=3$, and simple hypercubic in $d=4$.  To our knowledge, this is
the first investigation of this kind in the polaron literature. In
section \ref{sec:method}, we introduce a model of electron-phonon
interactions and briefly describe the method applied in this work. In
section \ref{sec:limits} we describe the limiting behavior of the
polaron problem on different lattices.  In section \ref{sec:results},
we apply CTQMC to the lattice Holstein polaron and calculate dynamical
quantities such as the effective mass, polaron dispersion and density
of states. The effects of changing the length scale of the interaction
are discussed in section \ref{sec:screening}. Finally we discuss the
relevance of these results in section \ref{sec:summary}.

\section{Model and Method}
\label{sec:method}

We restrict the model to have only one electron Wannier state $\vert
{\bf n} \rangle$ and one phonon degree of freedom $\xi_{\bf m}$ per
lattice unit cell.  The unit cells are numbered by the indices ${\bf
  n}$ or ${\bf m}$.  The electron hopping is assumed to be isotropic
and between the nearest neighbors only.  The phonon subsystem is a set
of independent oscillators with frequency $\omega$ and mass $M$.  In
the real space representation the Hamiltonian reads
\begin{equation}
H = H_{\mathrm{e}} + H_{\mathrm{ph}} + H_{\mathrm{e-ph}} \: ,
\label{eq:one}
\end{equation}
with
\begin{equation}
H_{\mathrm{e}} = - t \sum_{\langle \mathbf{nn'} \rangle} c^{\dagger}_{\mathbf{n'}}
c_{\mathbf{n}} \: ,
\label{eq:two}
\end{equation}
\begin{equation}
H_{\mathrm{ph}} = \frac{1}{2M} \sum_{\mathbf{m}} \hat{P}^{2}_\mathbf{m} +
\frac{M\omega^2}{2} \sum_{\mathbf{m}} \xi^{2}_{\mathbf{m}} \: ,
\label{eq:three}
\end{equation}
\begin{equation}
H_{\mathrm{e-ph}} = - \sum_{\mathbf{n}\mathbf{m}} f_{\mathbf{m}}(\mathbf{n})
c^{\dagger}_{\mathbf{n}} c_{\mathbf{n}} \xi_{\mathbf{m}} \: .
\label{eq:four}
\end{equation}
Here $t$ is the hopping amplitude, $\langle \bf{nn'} \rangle$ denote pairs of
nearest neighbors, and $\hat{P}_{\bf m} = -i\hbar\partial/\partial \xi_{\bf m}$ is 
the ion momentum operator.  The spin indices are absent because the system contains
only one electron. 

The form of electron-phonon interaction is specified via the {\em force} function
$f_{\bf m}({\bf n})$.  The latter is defined as the force with which an electron in 
state $\vert {\bf n} \rangle$ interacts with the ion degree of freedom $\xi_{\bf m}$.
In this paper, we consider two types of the force function: (i) the short-range 
Holstein interaction $f_{\bf m}({\bf n}) = \kappa \, \delta_{\bf nm}$ \cite{Holstein},
and (ii) a screened discrete Fr\"ohlich interaction 
\begin{equation}
f_{\mathbf{m}}(\mathbf{n}) = \frac{\kappa}
{[(\mathbf{m}-\mathbf{n})^2 + 1]^{3/2}} \exp \left(-\frac{|\mathbf{m}-\mathbf{n}|}
{R_{\mathrm{sc}}} \right) .
\label{eq:five}
\end{equation}
In $d=2$ (or $d=1$), it describes the screened {\em isotropic} Coulomb
interaction of an electron with {\em linearly} polarized lattice
distortions.  The distortions are polarized perpendicular to the plane
(chain) of the electron motion, and their equilibrium positions are
shifted perpendicular to the plane (chain) by one lattice period.  In
Eqn. (\ref{eq:five}), $\kappa$ is a constant and $R_{sc}$ is the
screening radius in units of the lattice vector.  Note that the
Holstein model is recovered for $R_{\mathrm{sc}} = 0$ and the lattice
Fr\"{o}hlich model for $R_{\mathrm{sc}} \rightarrow \infty$.  The
interaction (\ref{eq:five}) was introduced in
Ref. \onlinecite{spencer2005a} to model hole-phonon interaction
in doped cuprates.  Its various limiting cases were studied in
Refs. \onlinecite{alexandrov1999a,Kornilovitch1999b,Fehske2000,Bonca2001}.

The focus of this paper is the effects of the lattice geometry on the polaron 
properties.  Each lattice is characterized by its coordination number $z$ and the
bare electron dispersion 
$E_{\bf k} = -tN^{-1} \sum_{<\bf n,n'>} \exp{[i{\bf k}\cdot({\bf n}-{\bf n'})]}$.
(For example, for the triangular lattice, $E_{\bf k} = 
-2t [ \cos{k_xa} + \cos{(\frac{1}{2}k_x + \frac{\sqrt{3}}{2} k_y)a} 
+ \cos{(\frac{1}{2}k_x - \frac{\sqrt{3}}{2} k_y)a}]$.)  In particular, we will be
interested in comparing Bravais lattices with equal $z$ but different dimensionality, $d$.
We will analyze the cases $z=2$ (linear chain), $z=4$ (square in $d=2$), $z=6$ (triangular in $d=2$ and
simple cubic in $d=3$), $z=8$ (body-centered cubic in $d=3$, simple hexagonal in 
$d=3$ ($t_\parallel=t_\perp$), and hyper cubic in $d=4$) and $z=12$ (face-center-cubic in $d=3$).
Note that the minimum bare electron energy is $-zt$ for all these lattices.


%
%
%
%
%
%

The dimensionless coupling constant of the electron-phonon interaction $\lambda$ 
is defined as follows:
\begin{equation}
\lambda = \frac{1}{2M \omega^2 zt} \sum_{\bf m} f^2_{\bf m}(0) \: .
\label{eq:six}
\end{equation}
This coupling constant is the ratio of the polaron energy in the atomic limit
(i.e. at $t = 0$) to the kinetic energy of the free electron $zt$.  It is also
convenient to introduce the dimensionless phonon frequency 
$\bar \omega \equiv \hbar\omega/t$, the dimensionless inverse temperature
$\bar \beta \equiv \beta t$, and the dimensionless force 
${\bar f}_{\bf m}({\bf n}) \equiv \kappa^{-1} f_{\bf m}({\bf n})$.    

The CTQMC method employed here has been described in detail elsewhere
\cite{kornilovitch1998a,spencer2005a} so here we give a quick overview of the 
algorithm.  The initial step is to determine the effective electron (polaron) 
action that results when the phonon degrees of freedom have been integrated
out analytically.  The action is a functional of the polaron path in imaginary
time ${\bf r}(\tau)$ and is given by the following double integral
\begin{eqnarray}
A[{\bf r}(\tau)] & = & \frac{z\lambda\bar{\omega}}{2\Phi_0(0,0)}
\int_0^{\bar\beta} \int_0^{\bar\beta} d \tau d \tau'
e^{-\bar{\omega} \bar\beta/2}\left( e^{\bar{\omega}(\bar\beta/2-|\tau-\tau'|)} + 
                               e^{-\bar{\omega}(\bar\beta/2-|\tau-\tau'|)} \right) 
\Phi_0[\mathbf{r}(\tau),\mathbf{r}(\tau')] \nonumber \\
 & + & \frac{z\lambda\bar{\omega}}{\Phi_0(0,0)}
 \int_0^{\bar\beta} \int_0^{\bar\beta} d \tau d \tau' e^{- \bar{\omega} \tau} 
 e^{-\bar{\omega}( \bar\beta - \tau')} 
 \left( \Phi_{\Delta\mathbf{r}}[\mathbf{r}(\tau),\mathbf{r}(\tau')] - 
 \Phi_0[\mathbf{r}(\tau),\mathbf{r}(\tau')]\right) \: ,
\label{eq:seven} 
\end{eqnarray}
\begin{equation}
\Phi_{\Delta {\bf r}}[{\bf r}(\tau), {\bf r}(\tau')] = 
\sum_{\bf m} \bar{f}_{\bf m}[{\bf r}(\tau)] 
\bar{f}_{{\bf m} + \Delta {\bf r}} [{\bf r}(\tau')] \: ,
\label{eq:eight}
\end{equation}
where the vector $\Delta {\bf r} = {\bf r}(\beta) - {\bf r}(0)$ is the difference 
between the end points of the polaron path.  These expressions are valid when the
condition $\exp{(\bar\beta \bar\omega)} \gg 1$ is satisfied.  From this starting 
point, the polaron is simulated using the Metropolis Monte-Carlo method. The 
electron path is continuous in time with hopping events (or kinks) introduced or 
removed from the path with each Monte-Carlo step.  From this ensemble, various 
physical properties may be computed.  The ground state polaron energy is
\begin{equation}
\epsilon_0 = -\lim_{\beta\rightarrow\infty} \left[\left\langle \frac{\partial A}{\partial\beta} \right\rangle
- \frac{1}{\beta}\left\langle \sum_s N_s \right\rangle\right] \: ,
\label{eq:nine}
\end{equation}
where $N_s$ is the number of kinks of type $s$, and angular brackets denote ensemble
averaging.  The number of phonons is given by:
\begin{equation}
N_{\mathrm{ph}} = - \lim_{\beta\rightarrow\infty}\frac{1}{\bar{\beta}}
\left\langle \left. 
\frac{\partial A}{\partial \bar{\omega}}\right|_{\lambda\bar{\omega}}\right\rangle \: ,
\label{eq:ten}
\end{equation}
where the derivative is taken keeping $\lambda \bar\omega$ constant.  
The polaron band energy spectrum can be computed from:
\begin{equation}
\epsilon_{\bf k} - \epsilon_0= - \lim_{\beta\rightarrow\infty} \frac{1}{\beta} \ln
\langle \cos ({\bf k} \cdot \Delta{\bf r}) \rangle  \: ,
\label{eq:eleven}
\end{equation}
where ${\bf k}$ is the quasi momentum.  By expanding this expression in small
${\bf k}$, the $i$-th component of the inverse effective mass is obtained as
\begin{equation}
\frac{1}{m^*_i} = \lim_{\beta\rightarrow\infty}\frac{1}{\beta \hbar^2} \langle (\Delta {\bf r}_i)^2 \rangle \: .
\label{eq:twelve}
\end{equation}
Thus the inverse effective mass is the diffusion coefficient of the polaron
path in the limit of the infinitely long ``diffusion time'' $\beta$. 
Finally, the mass isotope coefficient, $\alpha_{m^*_i} = d \ln m^*_i / d \ln M$,
is calculated as follows
\begin{equation}
\alpha_{m^* _i}= \lim_{\beta\rightarrow\infty}\frac{\bar{\omega}}{2} 
\frac{1}{\langle(\Delta {\bf r}_i)^2 \rangle}
\left[\left\langle (\Delta {\bf r}_i)^2 
\left. \frac{\partial A}{\partial \bar{\omega}} \right\vert_{\lambda}
\right\rangle - \langle (\Delta {\bf r}_i)^2 \rangle
\left\langle \left. \frac{\partial A}{\partial \bar{\omega}} \right\vert_{\lambda}
\right\rangle \right] \: .
\label{eq:thirteen}
\end{equation}
Here the action derivatives must be taken at constant $\lambda$.
$\alpha_{m^*_i}$ and effective mass averaged over dimensions are related to the critical temperature 
isotope coefficient, $\alpha=-d \ln T_c/d\ln M$, of a (bi)polaronic superconductor as
\begin{equation}
\alpha = \alpha_{m^*} \left( 1 - \frac{m_0/m^*}{\lambda - \mu_c} \right) \: ,
\label{eq:fourteen}
\end{equation}
where $\mu_c$ is the Coulomb pseudo-potential \cite{aleiso,alebook}.

Several of the lattices studied here are not bipartite, and
observables will quantitatively change with sign of $t$. In this paper
$t$ is always taken to be positive (corresponding to $s$ orbitals
etc.) Negative $t$ would lead to a sign problem, since the probability
of the update is proportional to t. One could probably proceed by
mapping the negative $t$ problem onto a modified positive $t$
problem with long range hopping. We leave such a study for a future publication.

\section{Limiting behavior}
\label{sec:limits}

In this section, we discuss limiting behaviors for the polaron problem to 
demonstrate the physical differences between the two limits.  The results presented 
here can also be used to check the accuracy of the QMC algorithm, and are compared 
with the numerical results in the next section.

\subsection{Weak coupling}
\label{sec:weakcoupling}

The weak coupling behavior can be computed using a simple second-order perturbation 
theory. In such a theory, the polaron dispersion is given by \cite{fro}:
\begin{equation}
\epsilon^{(2)}_{\mathbf{k}} = E_{\mathbf{k}} -
\frac{z\lambda\bar{\omega}t^2}{\sum_{\mathbf{m}}f^2_{\mathbf{m}}(0)}
\frac{1}{N} \sum_\mathbf{q}\frac{|f_{\mathbf{q}}|^2} {W(\mathbf{k},\mathbf{q})} \: ,
\label{eqn:wcenergy}
\end{equation}
\begin{equation}
W(\mathbf{k},\mathbf{q}) = 
E_{\mathbf{k}-\mathbf{q}} + \hbar\omega - E_{\mathbf{k}} \: ,
\label{eq:sixteen}
\end{equation}
\begin{equation}
f_{\bf q} = \sum_{\bf{m}} f_{\bf m}(0) e^{-i {\bf q} \cdot {\bf m}} \: ,
\label{eq:seventeen}
\end{equation}
where $N$ is the number of unit cells.  Thus the ground state polaron energy 
(at ${\bf k} = 0$) is $\epsilon^{(2)}_0 = -t [ z + \lambda \Gamma_{\epsilon_0}(\bar{\omega})]$,
which defines a dimensionless coefficient $\Gamma_{E_0}$.  A second derivative of
Eqn. (\ref{eqn:wcenergy}) yields the effective mass  
\begin{equation}
\frac{1}{m^{*(2)}_i} = 
\left. \frac{1}{\hbar^2} \frac{\partial^2 \epsilon_{\bf k}}{\partial k^2_i} \right|_{{\bf k}= 0} 
= \frac{1}{m_{0i}} - \frac{z \lambda \bar{\omega} t^2}{\hbar^2 N} 
\sum_{\bf q} \frac{|f_{\bf q}|^2}{\sum_{\bf m} f^2_{\bf m}(0)} 
\frac{ 2W_{i}'^{2}-W''_{i} W }{ W^3({\bf k}, {\bf q}) }
\equiv \frac{ 1- \lambda\Gamma_{m^{*}_i}(\bar{\omega})}{m_{0i}} \: ,
\label{eqn:wcmass}
\end{equation}
with $W'_{i} = \partial W / \partial k_{i}$ etc.  In calculating the coefficients
$\Gamma_{m^{*}_{i}}$  
one should take into account that the expression for the bare effective mass changes 
with lattice type: $m^{-1}_{0i} = 2 a^2 t/\hbar^2$ for the hypercubic, face-center-cubic, and body-center-cubic 
lattices, but $m^{-1}_{0i} = 3 a^2 t/\hbar^2$ for the triangular lattice. The bare masses are not isotropic on the hexagonal lattice, with in plane
$m^{-1}_{0xy} = 3 a^2 t/\hbar^2$ and out of plane $m^{-1}_{0z}=2a^2t/\hbar^2$, so a little care has to be taken in the following.  
The mass isotope coefficient is defined as 
%
\begin{equation}
\alpha^{(2)}_{m^*_i} = \frac{\partial\ln(m^{*}_i)}{\partial\ln(M)}
= \frac{\bar\omega}{2 (m_{0i}/m^*_i)}
\frac{\partial}{\partial \bar\omega}\left(\frac{m_{0i}}{m^{*}_i}\right) \: .
\label{eqn:isoexp}
\end{equation}
Substituting here Eqn. (\ref{eqn:wcmass}) one obtains to leading order in $\lambda$
\begin{equation}
\alpha^{(2)}_{m^*_i} = - \lambda \frac{\bar\omega}{2} 
\frac{\partial \Gamma_{m^{*}_i}(\bar\omega)}{\partial \bar\omega}
\equiv \lambda \Gamma_{\alpha_{m^*_i}}(\bar{\omega}) \: .
\label{eqn:wcisotope}
\end{equation}
Finally, the number of phonons associated with the polaron in the ground state is
\begin{equation}
N^{(2)}_{\rm ph} = \frac{ z \lambda\bar{\omega}t^2}{\sum_{\bf m} f^2_{\bf m}(0)}
\frac{1}{N} \sum_{\bf q} \frac{\vert f_{\bf q} \vert^2} {W^2(0,\mathbf{q})}
\equiv \lambda \Gamma_{N_{ph}}(\bar{\omega}) \: .
\label{eqn:wcnp}
\end{equation}
The resulting weak-coupling coefficients $\Gamma$ can be computed by integrating 
over $\mathbf{q}$ in the Brillouin zone.  For the Holstein interaction at 
$\bar\omega = 1$, $\Gamma$ are given in Table \ref{tab:wc}. 
Note that for all the lattices considered, $\Gamma_{m^*_i}$ and 
$\Gamma_{\alpha_{m^*_i}}$ are identical in all directions, so only one value is presented.

\begin{table}
\caption{
Coefficients of the weak coupling behavior for $\bar{\omega} = 1$ and the Holstein 
interaction computed by numerical integration in Eqns.   
(\ref{eqn:wcenergy}-\ref{eqn:wcnp}).  Note that the mass isotope coefficient in $d>1$ 
has a small negative value.
}
\begin{ruledtabular}
\begin{tabular}{lllll}
Lattice & $\Gamma_{m^*}$ & $\Gamma_{\epsilon_0}$ & $\Gamma_{N_{ph}}$ & $\Gamma_{\alpha_{m^{*}}}$  \\
\hline
Linear             & 0.5366(6) & 0.8944(4) & 0.5355(9) &  0.12523(0) \\
Square             & 0.3610(7) & 1.0161(0) & 0.3610(4) & -0.00031(3) \\
Triangular         & 0.3216(8) & 1.1062(2) & 0.3217(2) & -0.00903(0) \\
Cubic              & 0.1533(6) & 1.0229(7) & 0.2300(1) & -0.04022(1) \\
BCC                & 0.2411(2) & 1.0378(8) & 0.1808(8) & -0.07974(4) \\
Hypercubic ($d=4$) & 0.1591(5) & 1.0168(6) & 0.1591(6) & -0.04533(4) \\
Hexagonal          & 0.2170(3) & 1.0536(0) & 0.1928(2) & -0.04461(3) \\
FCC                & 0.2856(0) & 1.1220(6) & 0.1564(8) & -0.11021(3) \\
\end{tabular}
\end{ruledtabular}
\label{tab:wc}
\end{table}

\subsection{Strong coupling}
\label{sec:strongcoupling}

For very strong coupling, the electron becomes very heavy and almost localized, since the act of hopping from one site to another by relaxing the local 
lattice and then distorting the lattice on a neighboring site is very unfavorable. 
The energy of the resulting small polaron is given as $E_p = - \lambda z t$. 
As the effective coupling is reduced, other terms due to hopping become relevant. 
The expansion about the atomic limit in the hopping parameter (which is small 
compared to the polaron energy) may be determined through a canonical 
LF transformation.  For nearest neighbor isotropic hopping the strong 
coupling polaron band dispersion is then given by \cite{alebook}:
\begin{equation}
\epsilon_{\bf k} = - zt \lambda + e^{-\gamma z \lambda/ \bar\omega} E_{\bf k}
+ {\mathcal{O}}(t/\lambda) \: ,
\label{eqn:smallpolaron}
\end{equation}
\begin{equation}
\gamma = 1 - \frac{\sum_{\bf m} f_{\bf m}(0) f_{\bf m}({\bf b})}
{\sum_{\bf m} f^2_{\bf m}(0)} \: ,
\end{equation}
where ${\bf b}$ is a nearest-neighbor lattice vector (it does not matter which one). 
Clearly $\gamma$ varies with interaction and lattice type.  For the Holstein interaction 
($R_{sc}=0$), which is purely site local, $\gamma=1$.  For the lattice Fr\"ohlich 
potential ($R_{sc} = \infty$) on a square lattice, $\gamma=0.292582$, and for the 
triangular lattice, $\gamma=0.197577$.  Such a polaron is {\em much} lighter than the 
Holstein polaron.  For the screened Fr\"ohlich interaction with $R_{sc} = 1$, the 
triangular lattice has $\gamma = 0.710852$ and the square lattice, $\gamma = 0.730176$.
%
%
The effective mass is computed as before, using the dispersion from
Eqn. (\ref{eqn:smallpolaron})
\begin{equation}
\frac{m_0}{m^*} \approx \exp\left(-\frac{z\gamma\lambda}{\bar{\omega}}\right) \: ,
\end{equation}
and the isotope exponent is
\begin{equation}
\alpha_{m^*}\approx\frac{\gamma\lambda z}{2\bar{\omega}} \: .
\end{equation}
The number of phonons is calculated directly from $\langle H_{\mathrm{ph}} \rangle$, which happens 
to have exactly half the magnitude of $\langle H_{\mathrm{el-ph}} \rangle$, so
\begin{equation}
N_{\mathrm{ph}} \approx - \frac{\epsilon_0}{\hbar \omega} = \frac{z\lambda}{\bar{\omega}} \: .
\end{equation}

The remainder of this paper will concern simulations of the polaron
problem using the CTQMC. The numerical results will be discussed with relation to the weak and strong coupling limits.

\section{The Holstein Polaron}
\label{sec:results}

\subsection{Ground state properties from weak to strong coupling}

The main difficulty in studying the polaron problem is the intermediate coupling 
regime.  For $\lambda\sim 1$, it is necessary to rely on numerical results. 
In this section, we present ground state properties of the polaron.  These may 
be determined from CTQMC with no sign problem.

\begin{figure}
\includegraphics[width=50mm,height=70mm,angle=270]{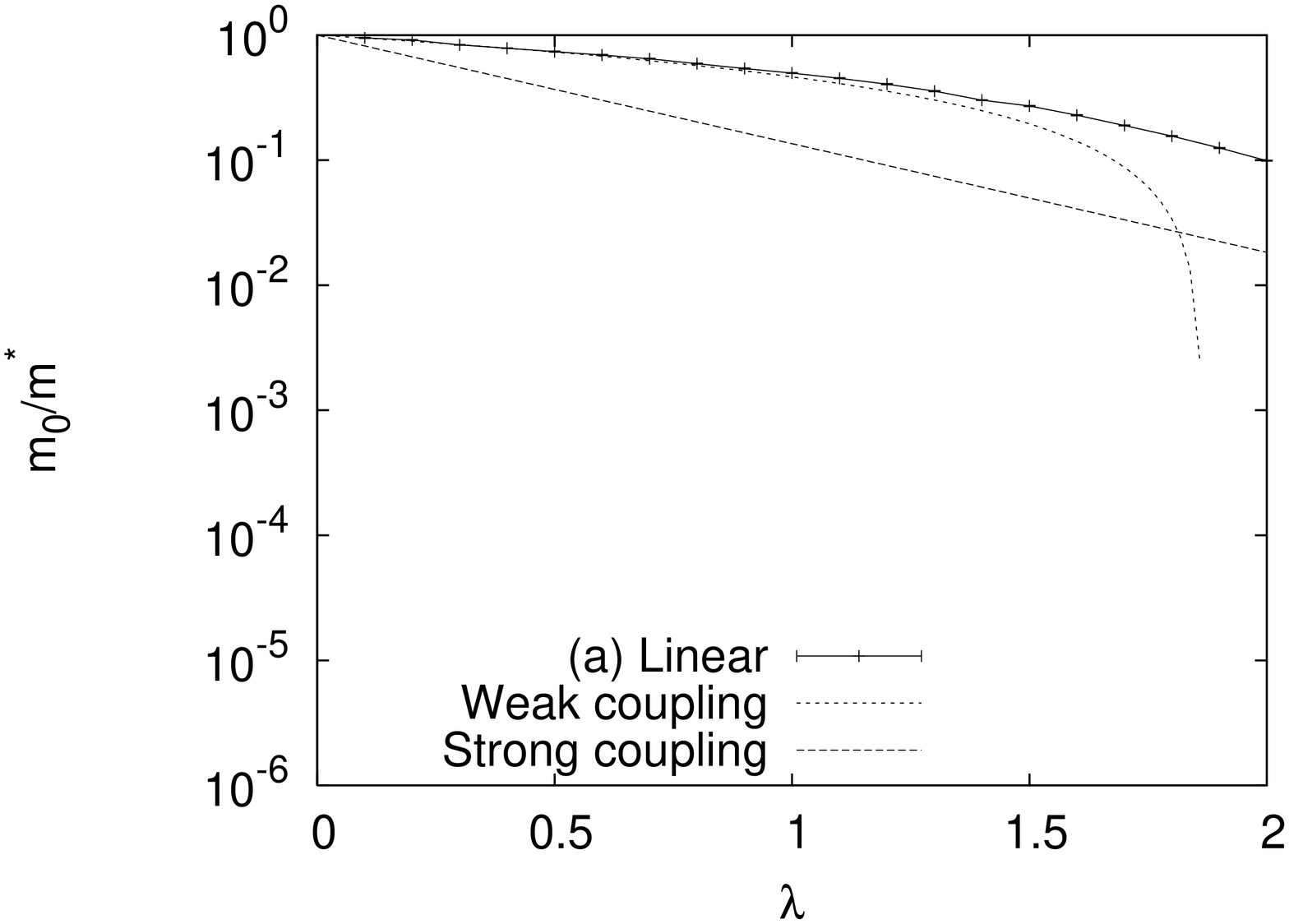}
\includegraphics[width=50mm,height=70mm,angle=270]{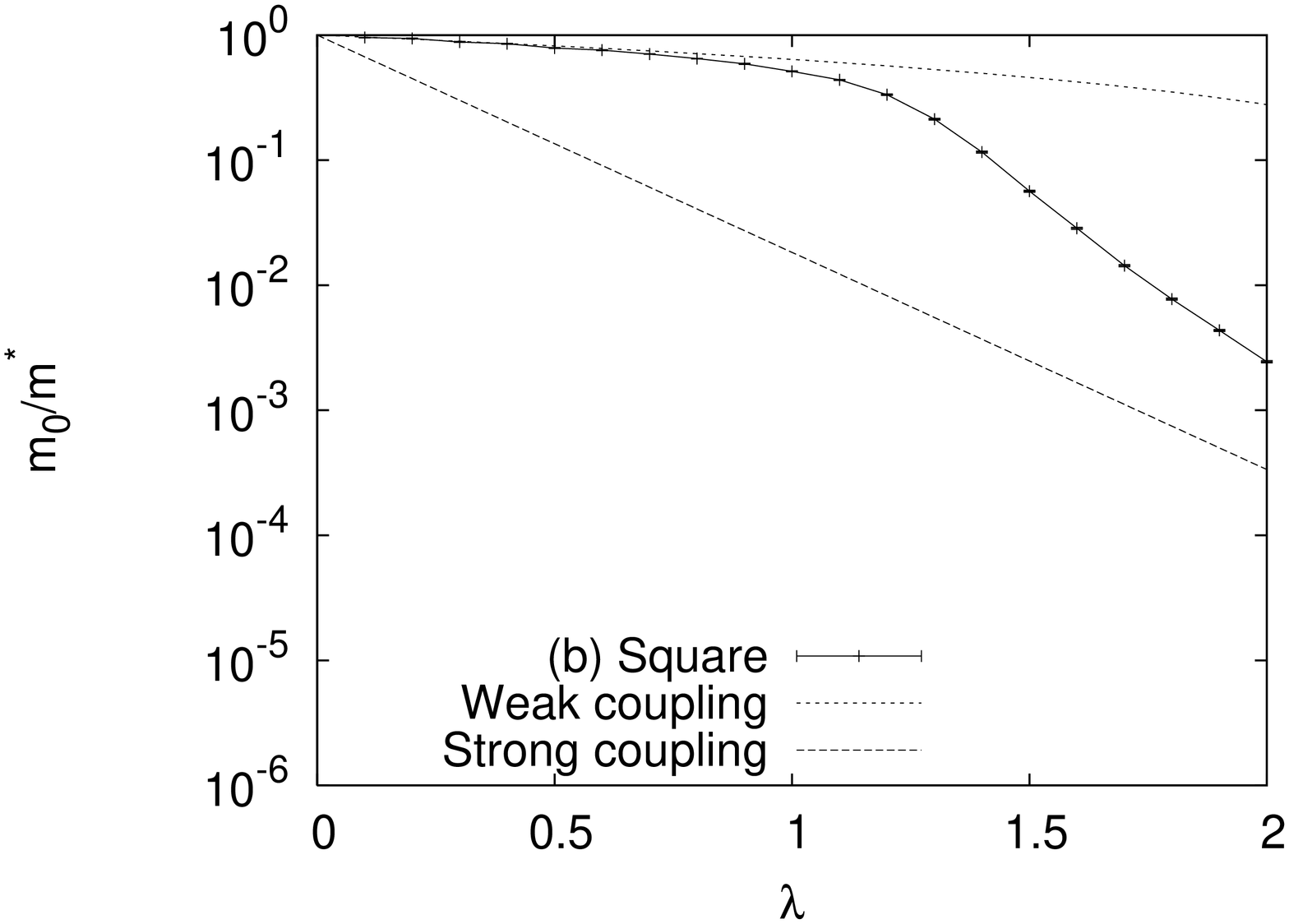}
\includegraphics[width=50mm,height=70mm,angle=270]{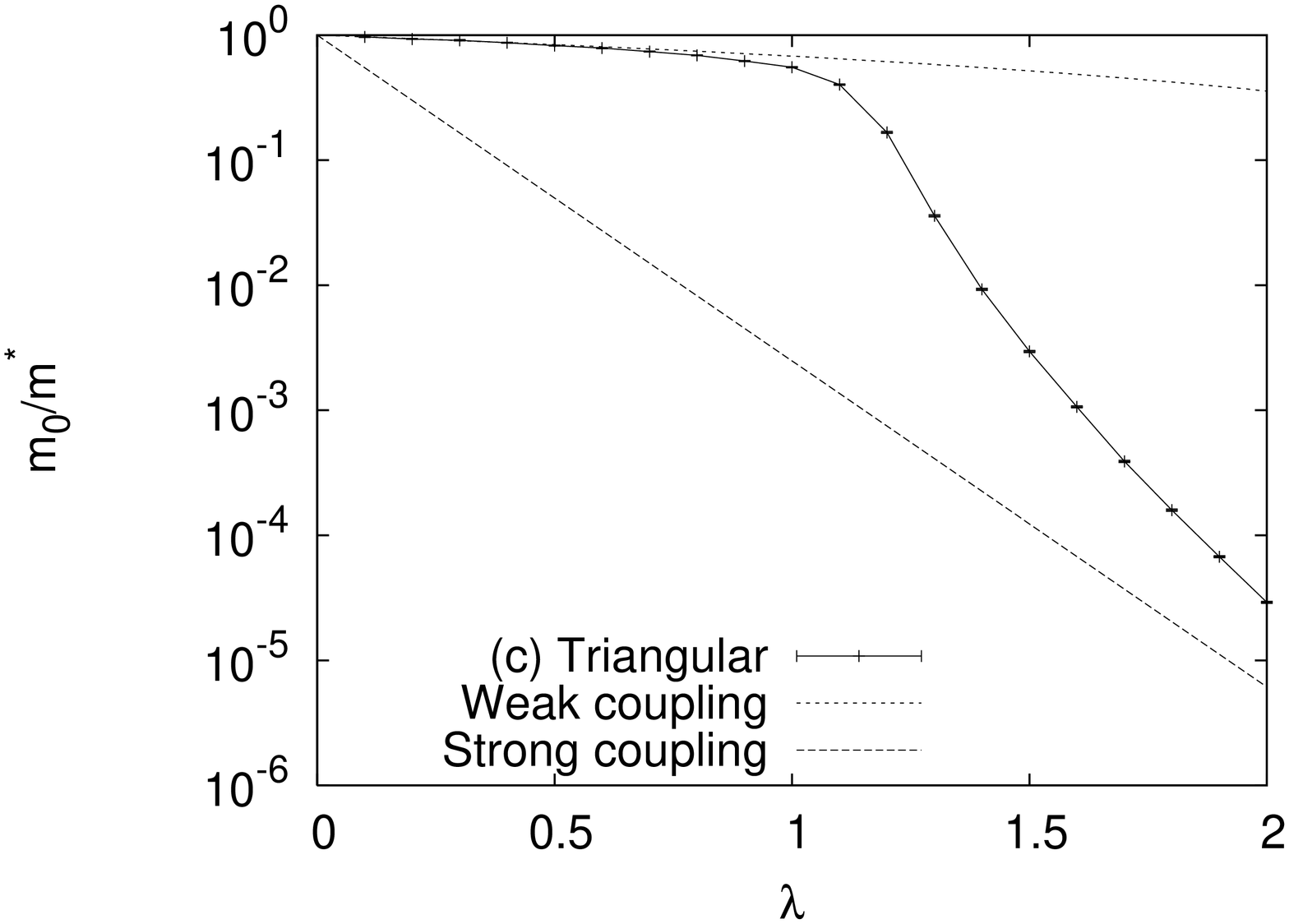}
\includegraphics[width=50mm,height=70mm,angle=270]{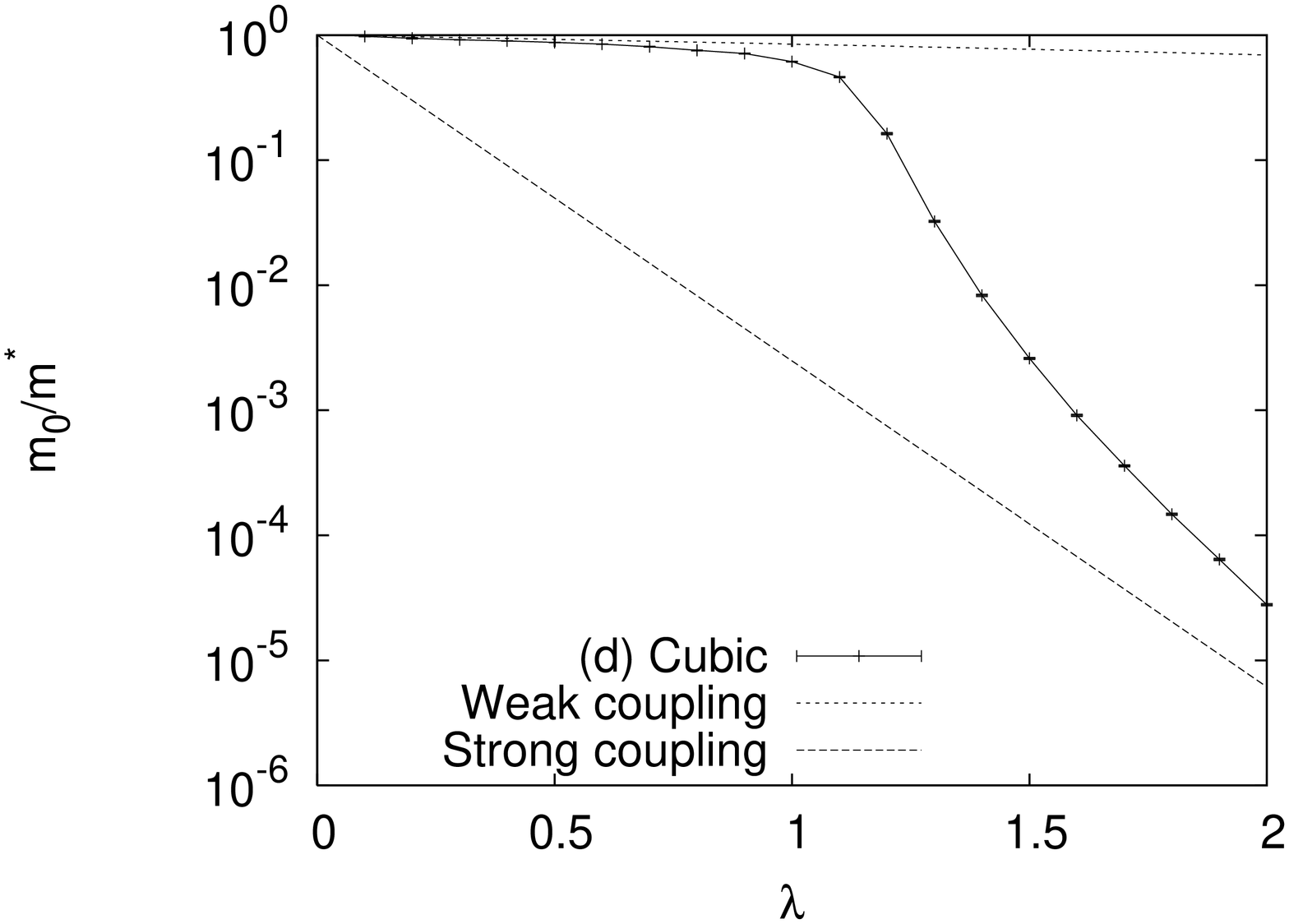}
\includegraphics[width=50mm,height=70mm,angle=270]{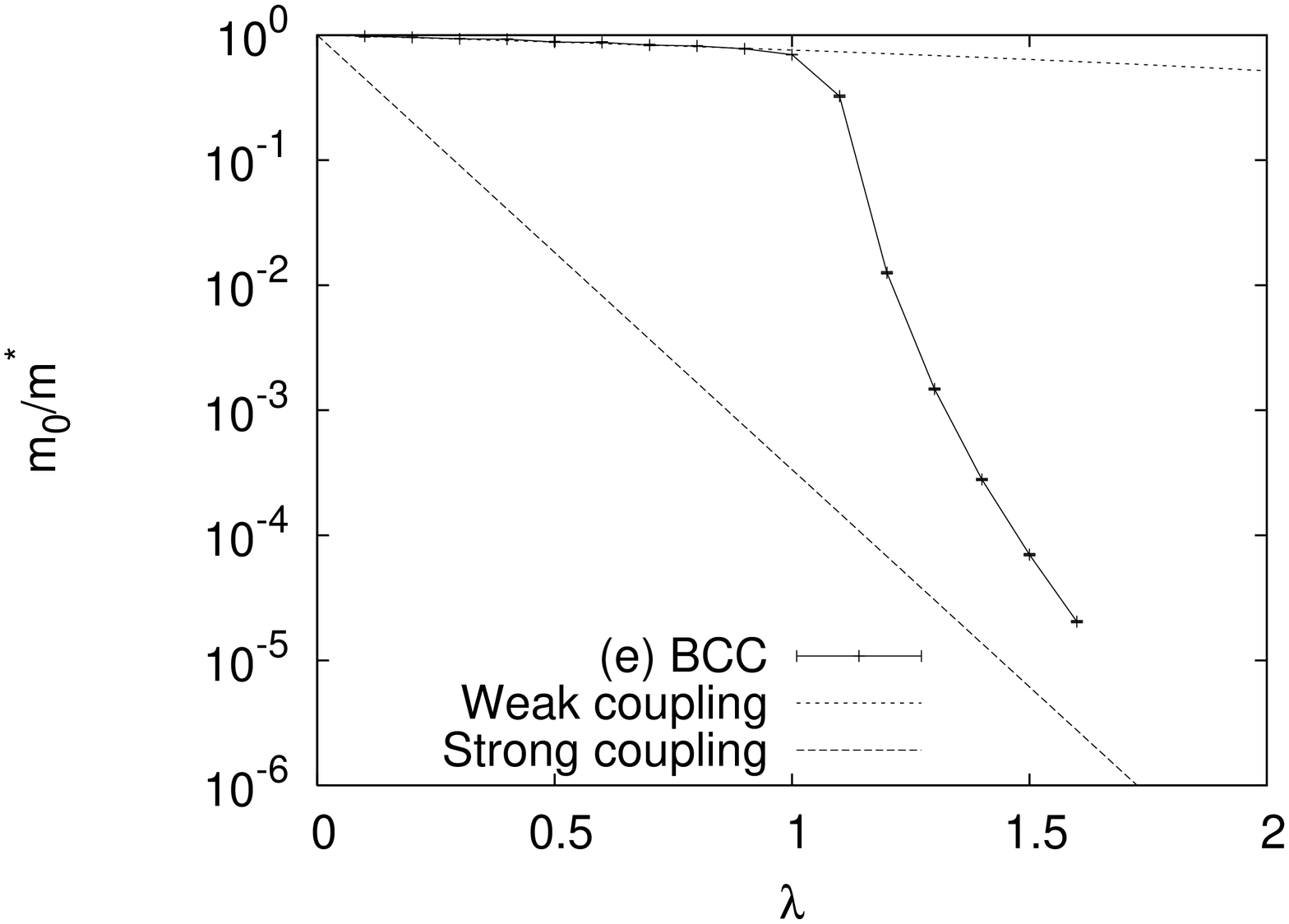}
\includegraphics[width=50mm,height=70mm,angle=270]{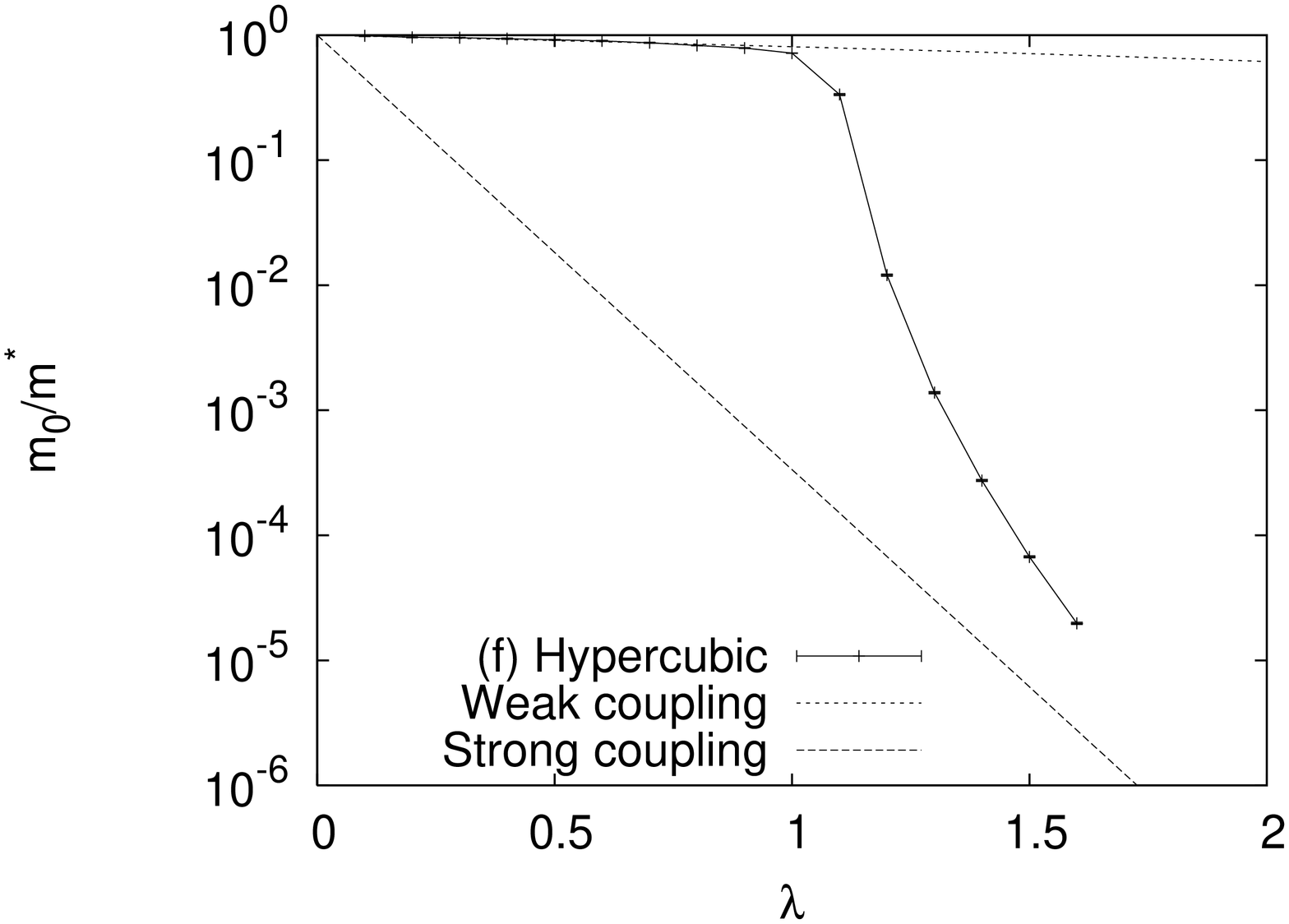}
\includegraphics[width=50mm,height=70mm,angle=270]{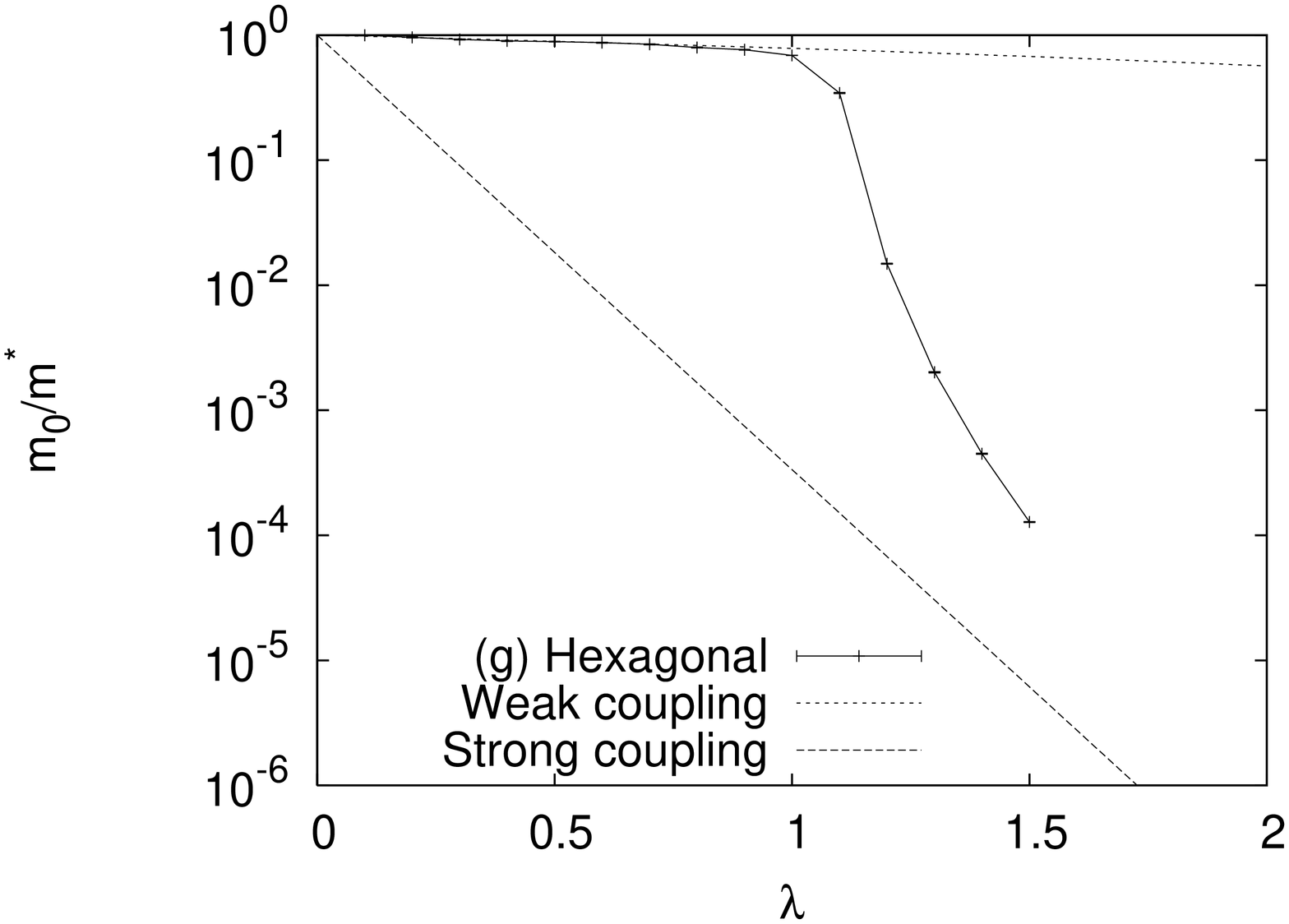}
\includegraphics[width=50mm,height=70mm,angle=270]{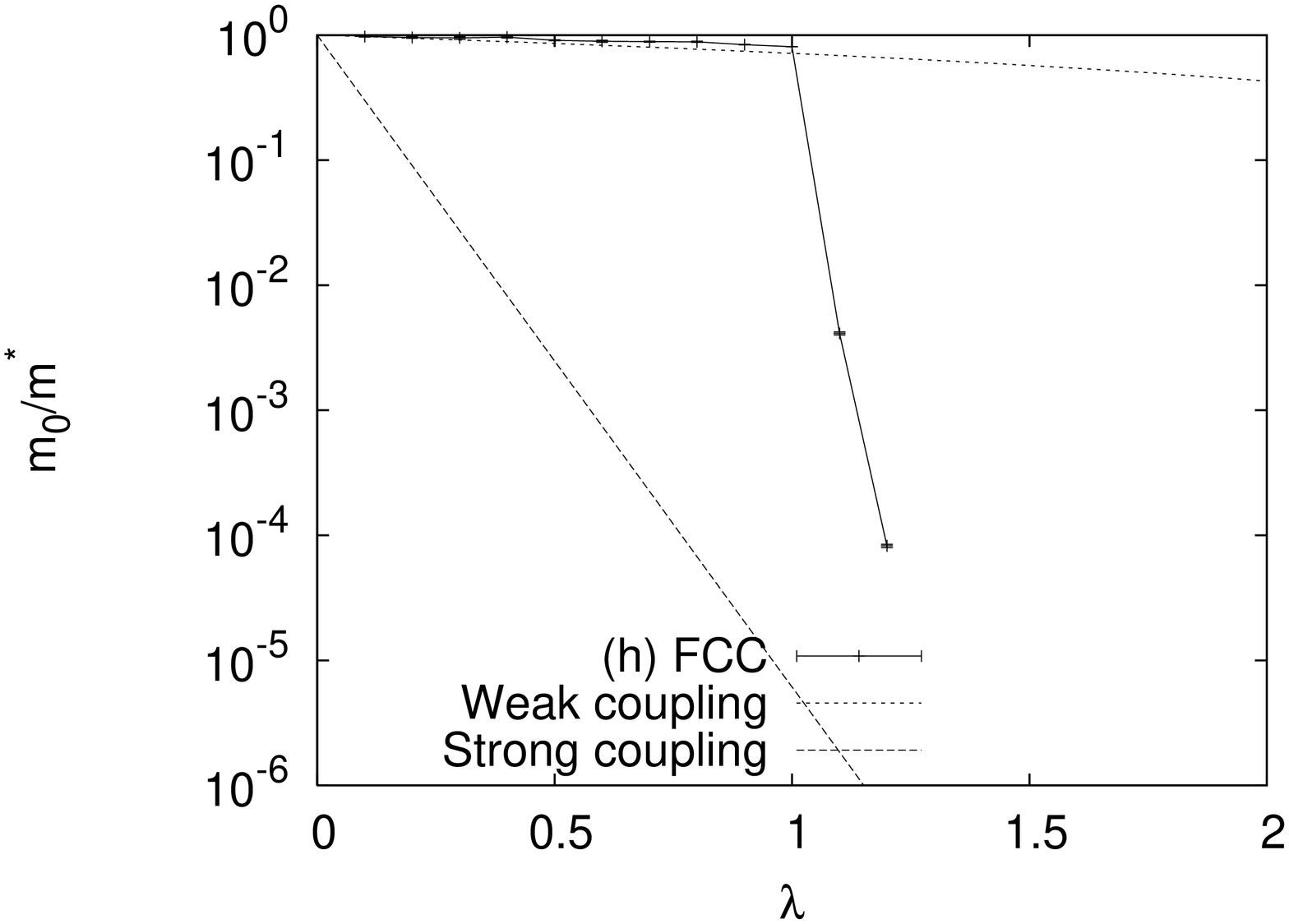}
\caption{ Inverse effective mass for the Holstein polaron as a
  function of coupling $\lambda$ on (a) linear (b) square (c)
  triangular (d) cubic (e) body-center-cubic (f) hypercubic ($d=4$)
  (g) hexagonal and (h) face-center-cubic lattices, and
  $\bar{\omega}=1$. Also shown are the weak and strong coupling
  asymptotes as calculated in section \ref{sec:limits}. Note that the
  effective masses of triangular and cubic lattices are almost
  equivalent for $\lambda>1$ (both lattices having $z=6$). Also the
  BCC, hexagonal and hypercubic lattices ($z=8$) have almost identical
  curves. It can also be seen that the coincidence occurs well before
  the approach to the strong coupling asymptote, indicating that the
  coordination number is very important to physical properties for
  most values of the coupling. Increased coordination number leads to
  more mobile polarons at weak coupling and more localized polarons
  for strong coupling with a very fast crossover between the two
  behaviors. Error bars show the standard error of the points computed
  using a blocking scheme.}
\label{fig:inversemass}
\end{figure}

\begin{figure}
\includegraphics[width=50mm,height=70mm,angle=270]{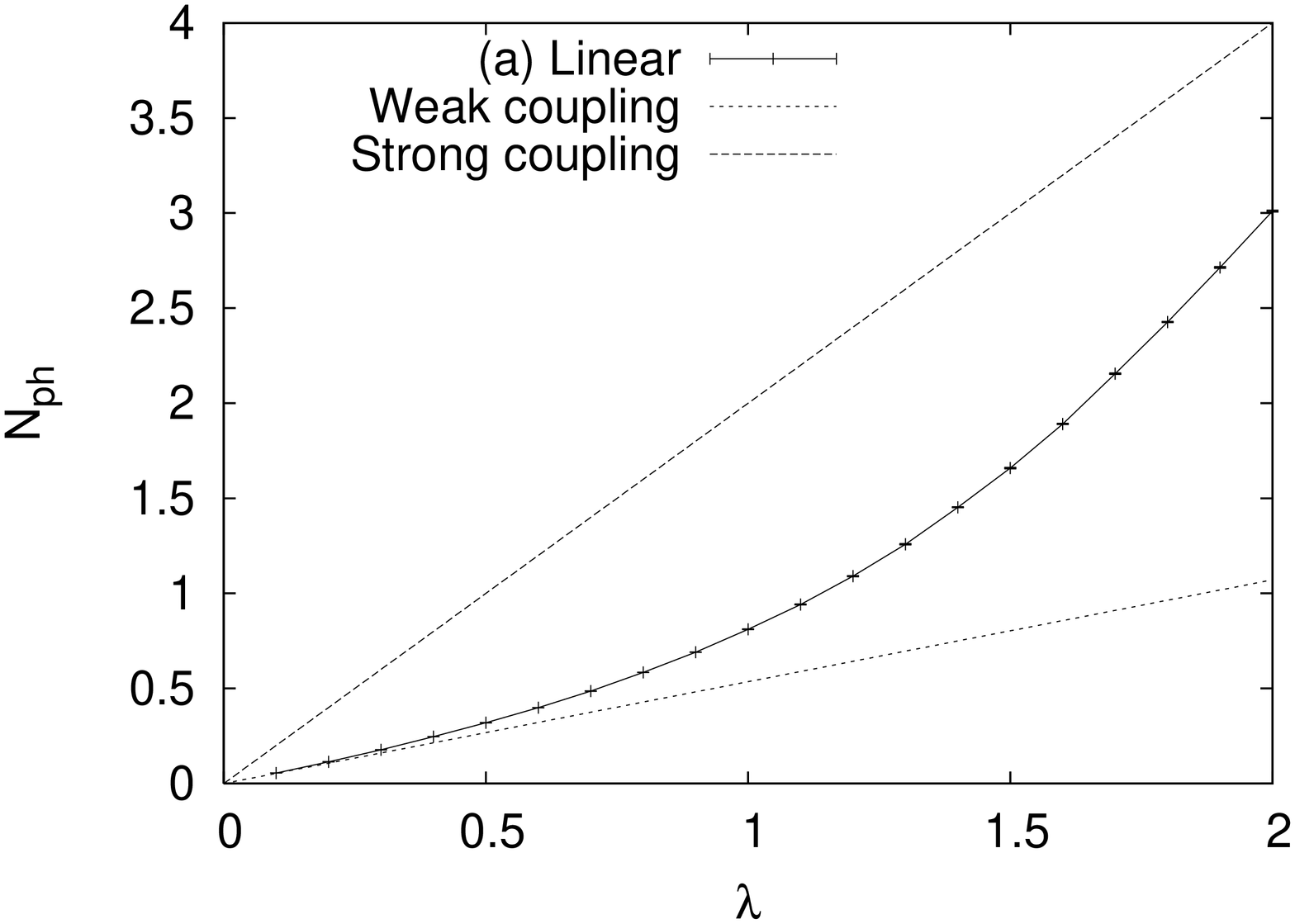}
\includegraphics[width=50mm,height=70mm,angle=270]{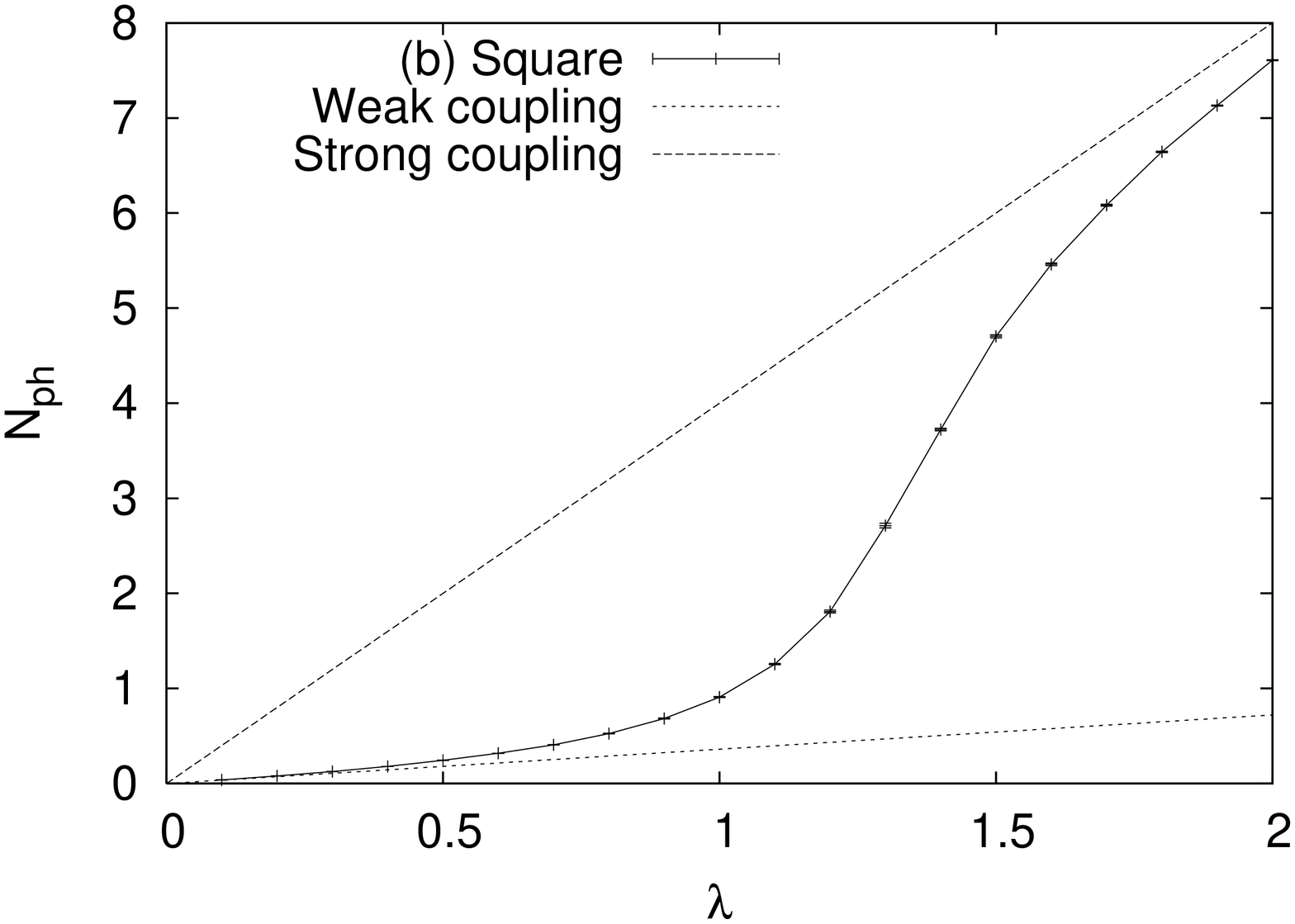}
\includegraphics[width=50mm,height=70mm,angle=270]{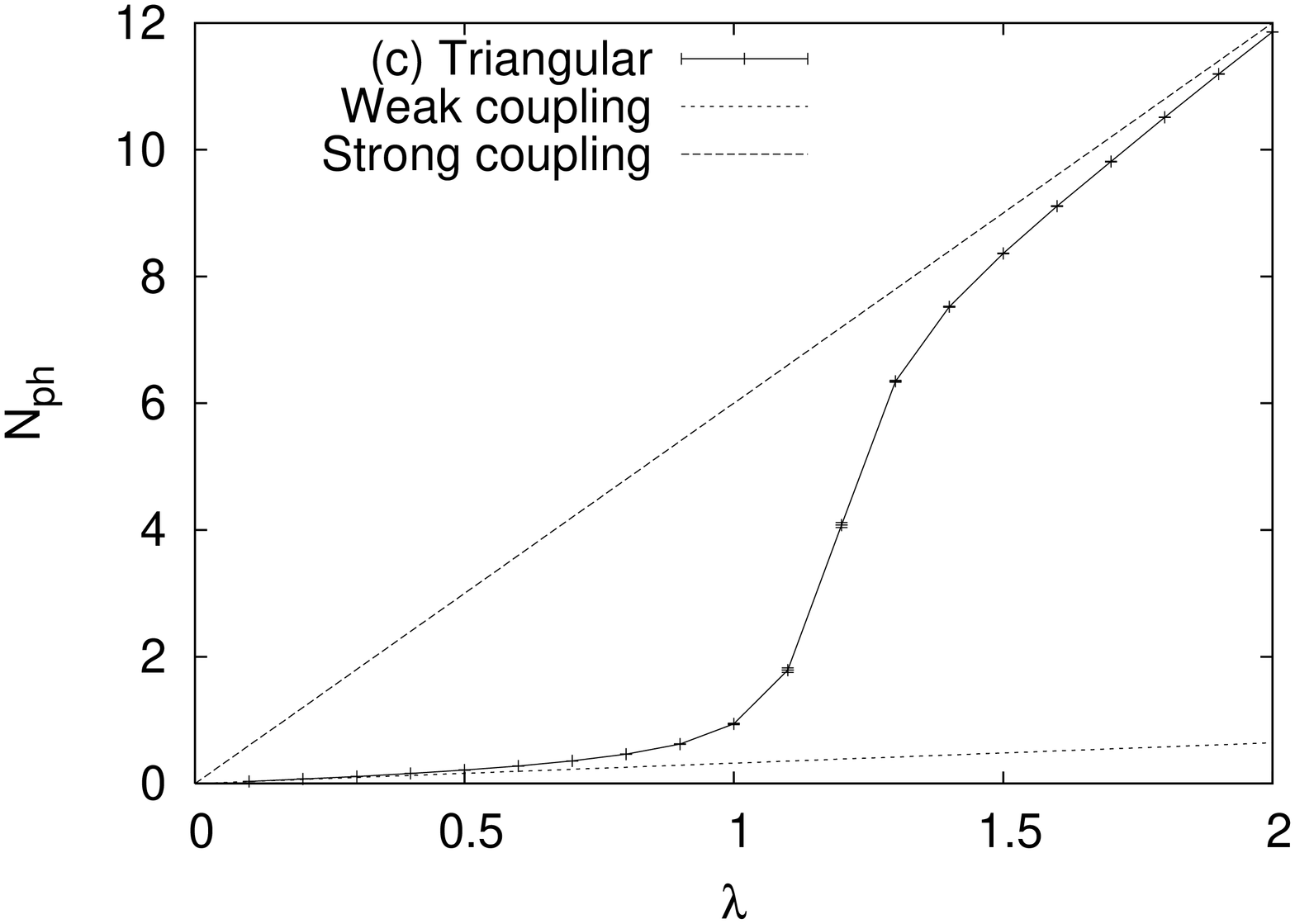}
\includegraphics[width=50mm,height=70mm,angle=270]{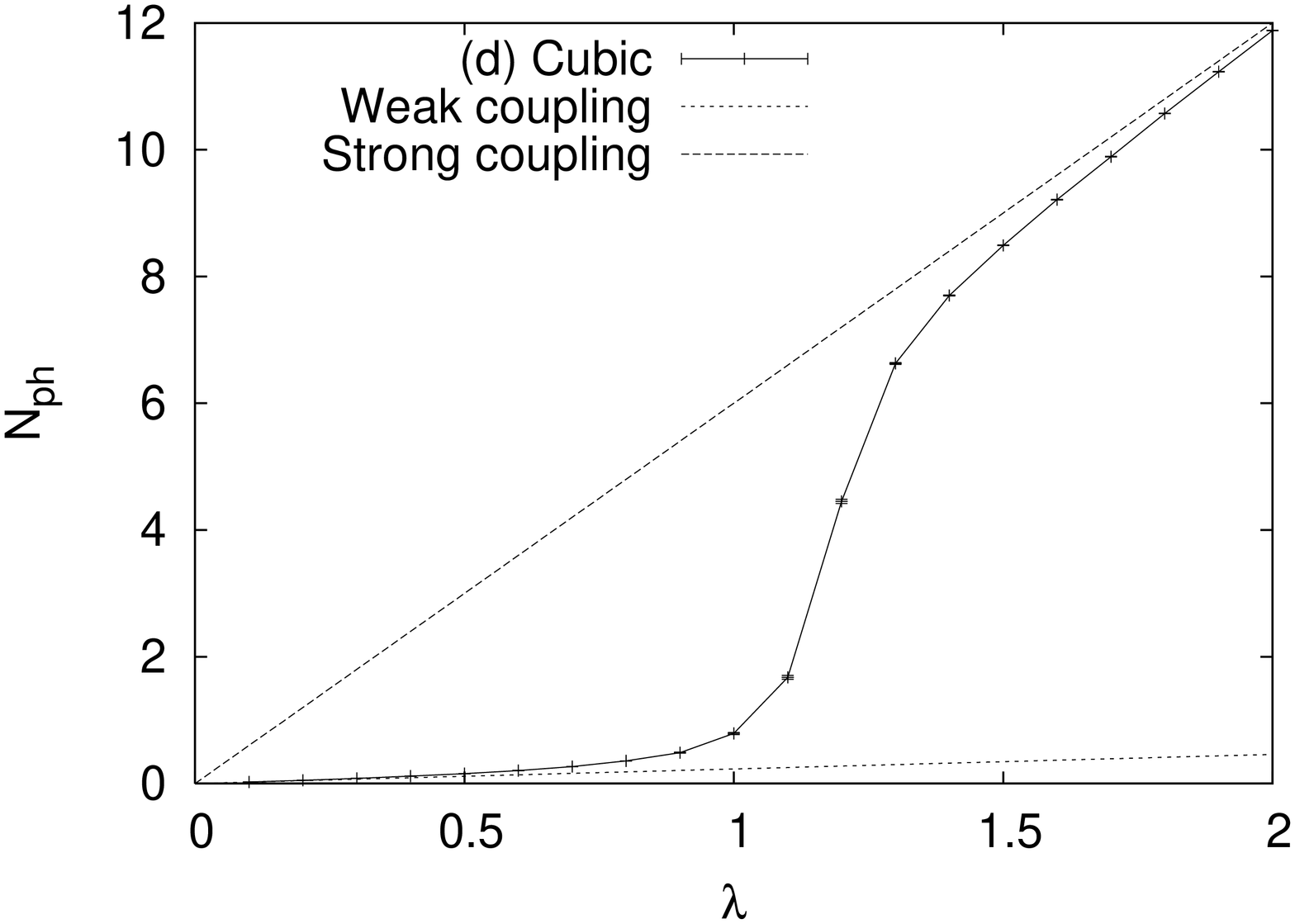}
\includegraphics[width=50mm,height=70mm,angle=270]{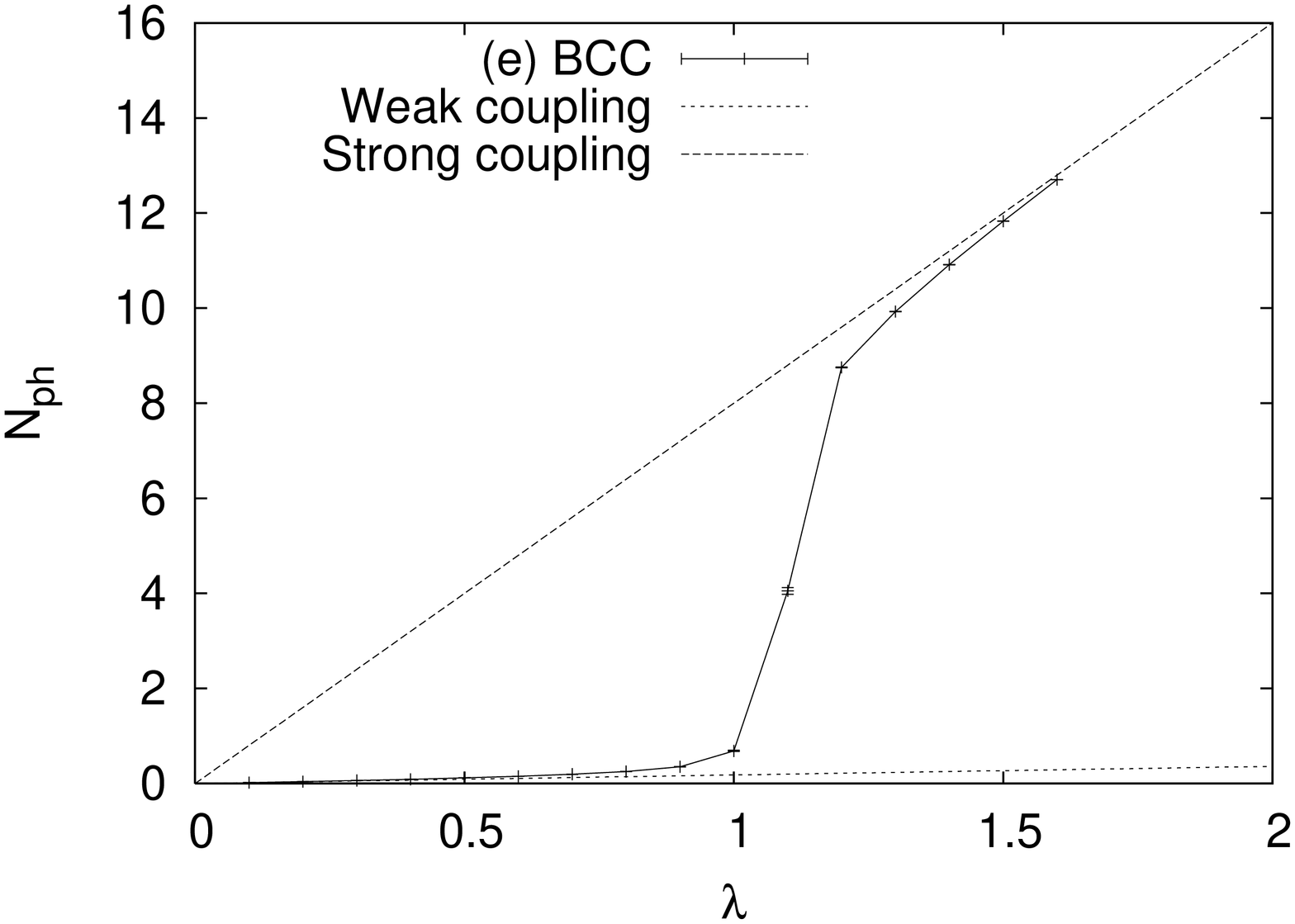}
\includegraphics[width=50mm,height=70mm,angle=270]{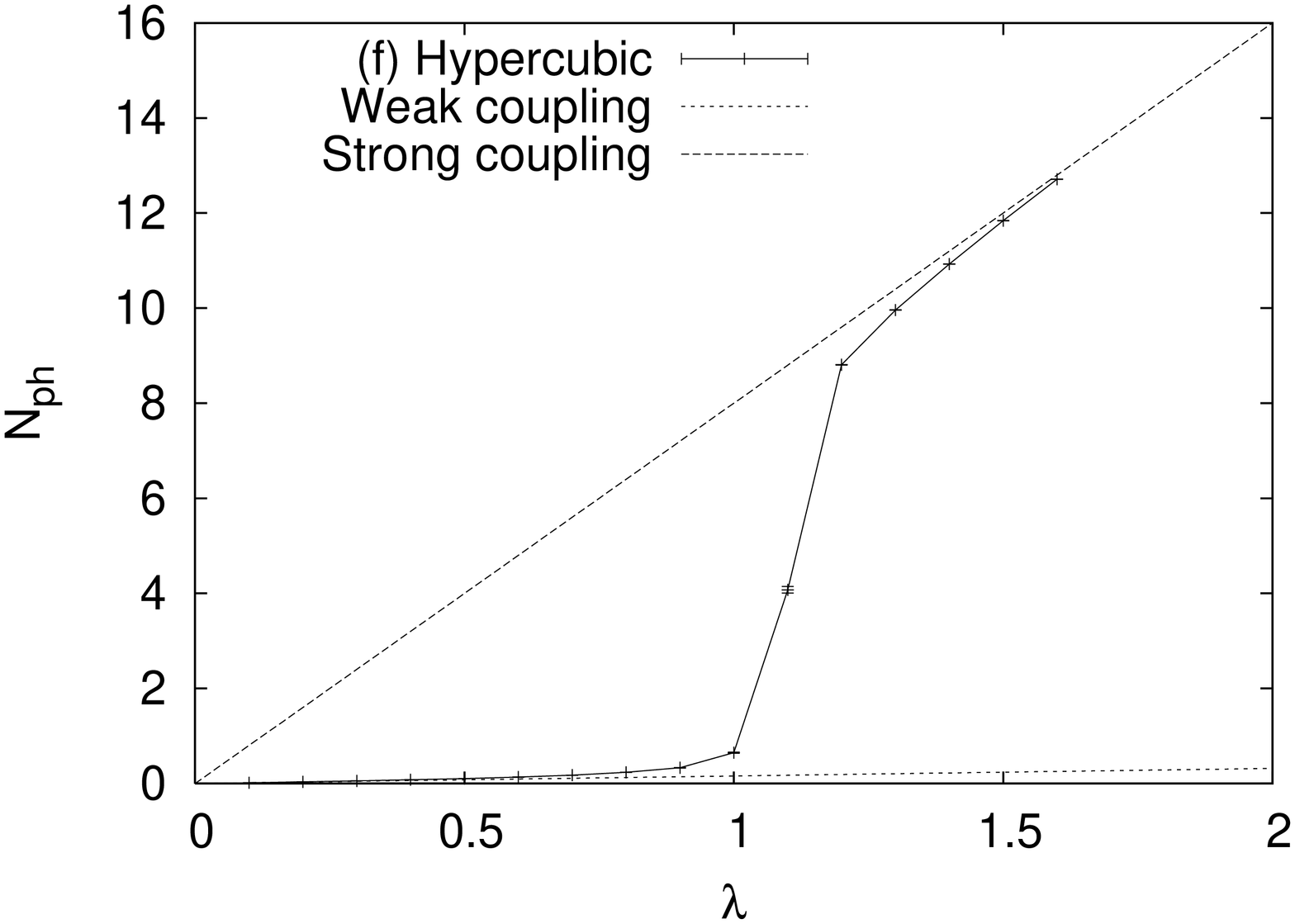}
\includegraphics[width=50mm,height=70mm,angle=270]{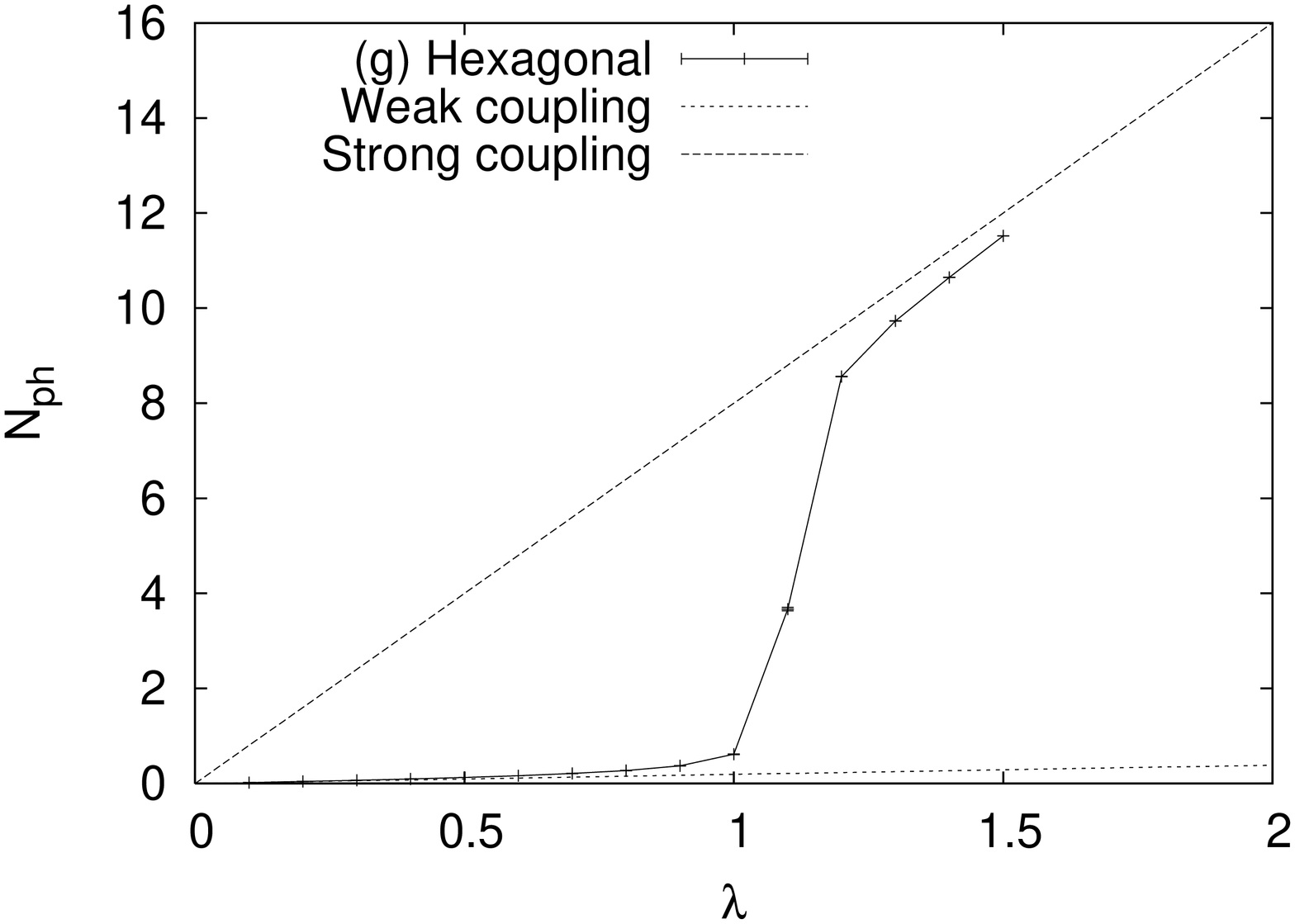}
\includegraphics[width=50mm,height=70mm,angle=270]{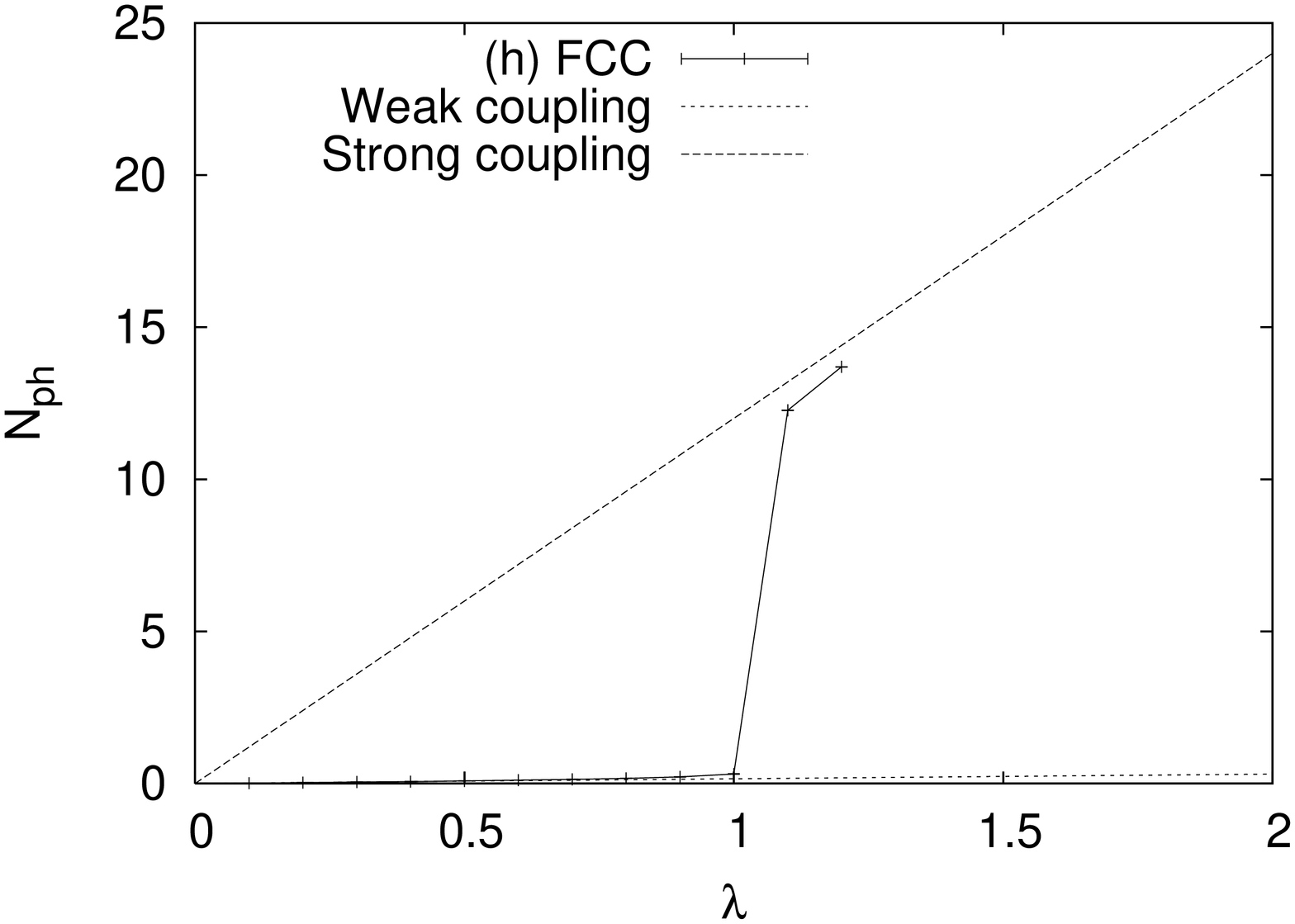}
\caption{Number of phonons associated with the Holstein polaron for
  increasing coupling with $\bar{\omega}=1$. For most couplings, the
  number of phonons is closely related to the exponent of the
  effective mass. This is expected in the very strong coupling limit
  where $m_0/m^*=\exp(-\gamma N_{\mathrm{ph}})$, and the continuation
  of this behavior to lower $\lambda$ values is of interest. Again,
  the speed of the crossover and the value of $\lambda$ at which the
  number of phonons reaches the strong coupling saturation value
  depends on the coordination number, not on the dimensionality.}
\label{fig:numberofphonons}
\end{figure}

\begin{figure}
\includegraphics[width=50mm,height=70mm,angle=270]{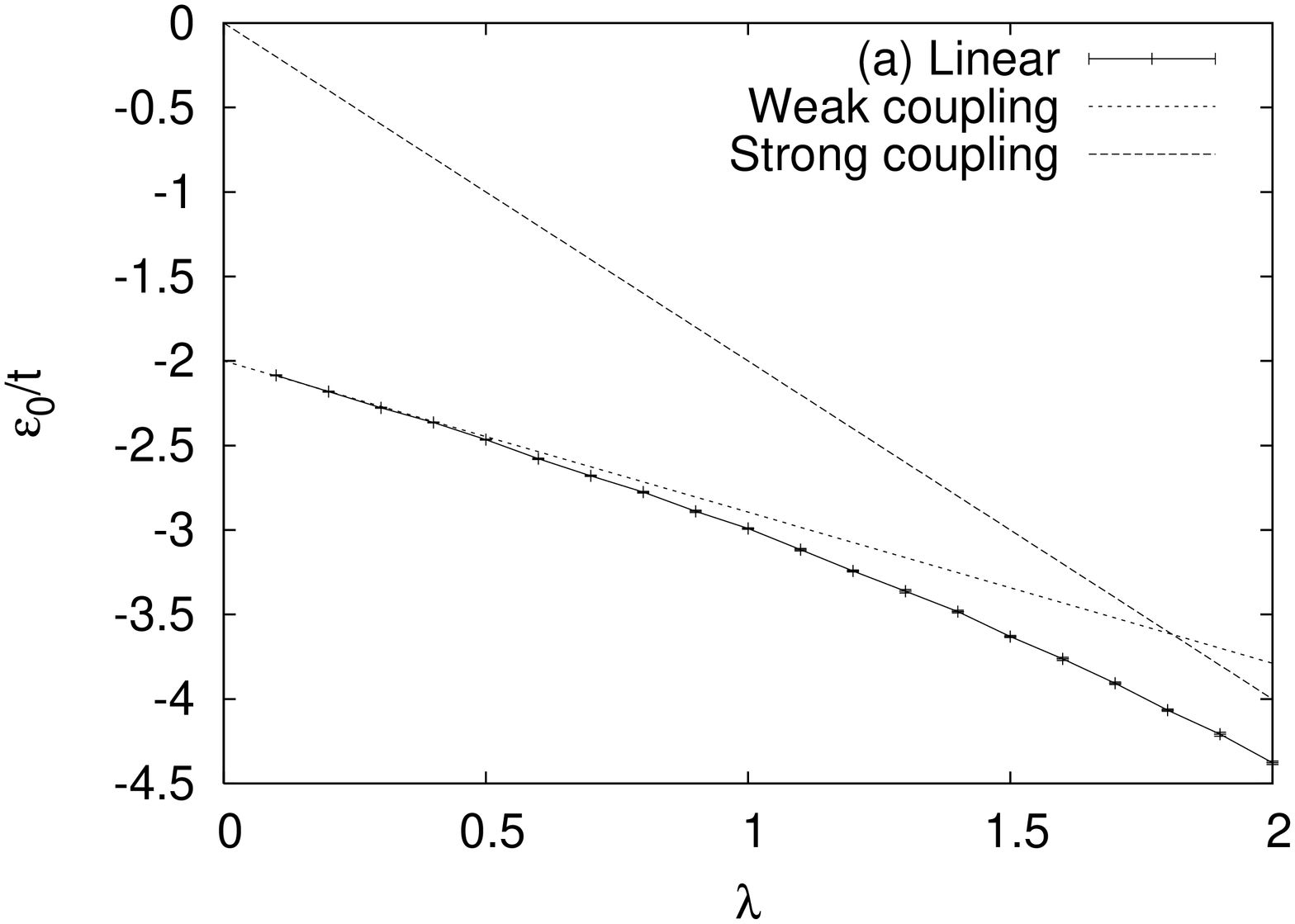}
\includegraphics[width=50mm,height=70mm,angle=270]{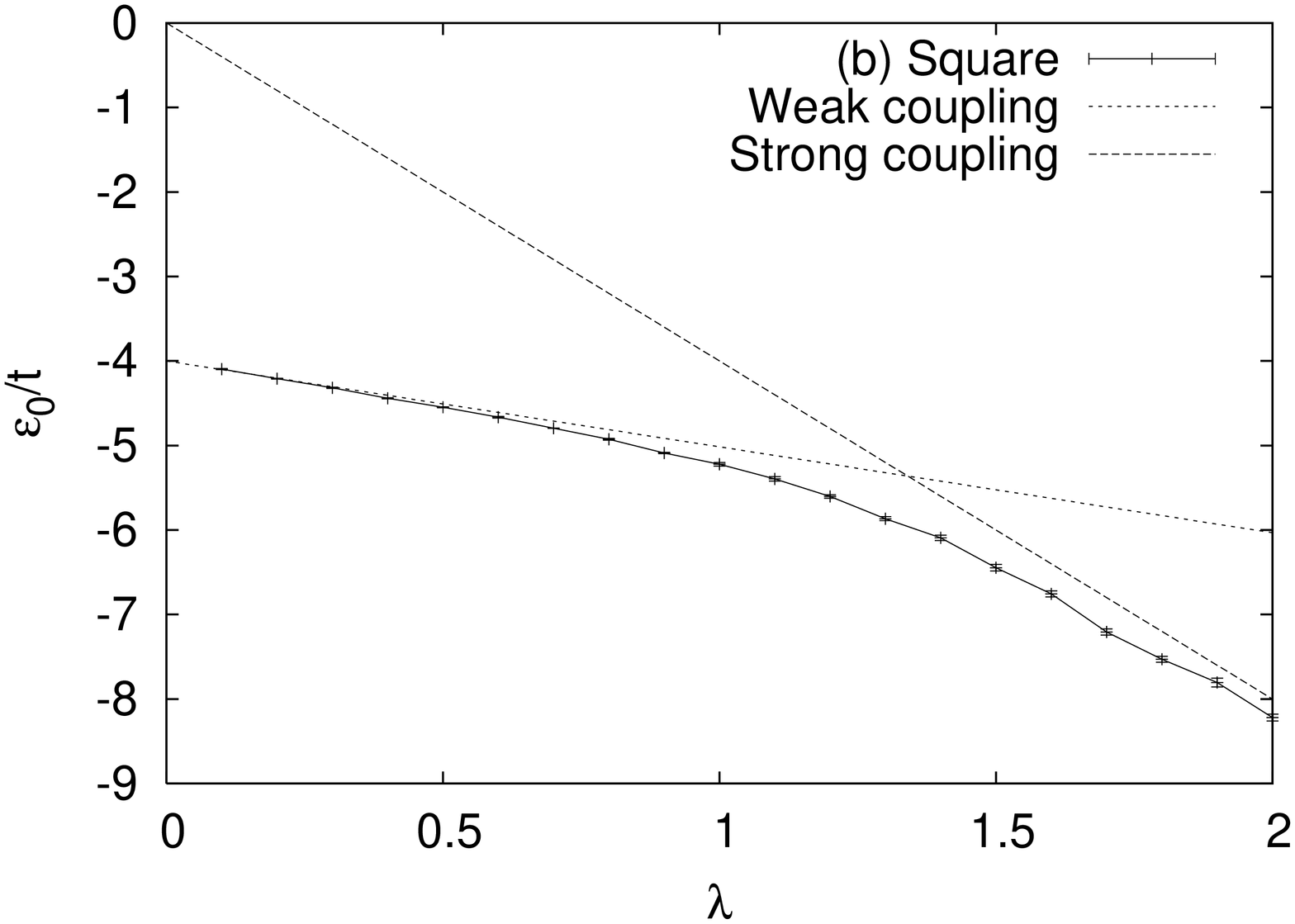}
\includegraphics[width=50mm,height=70mm,angle=270]{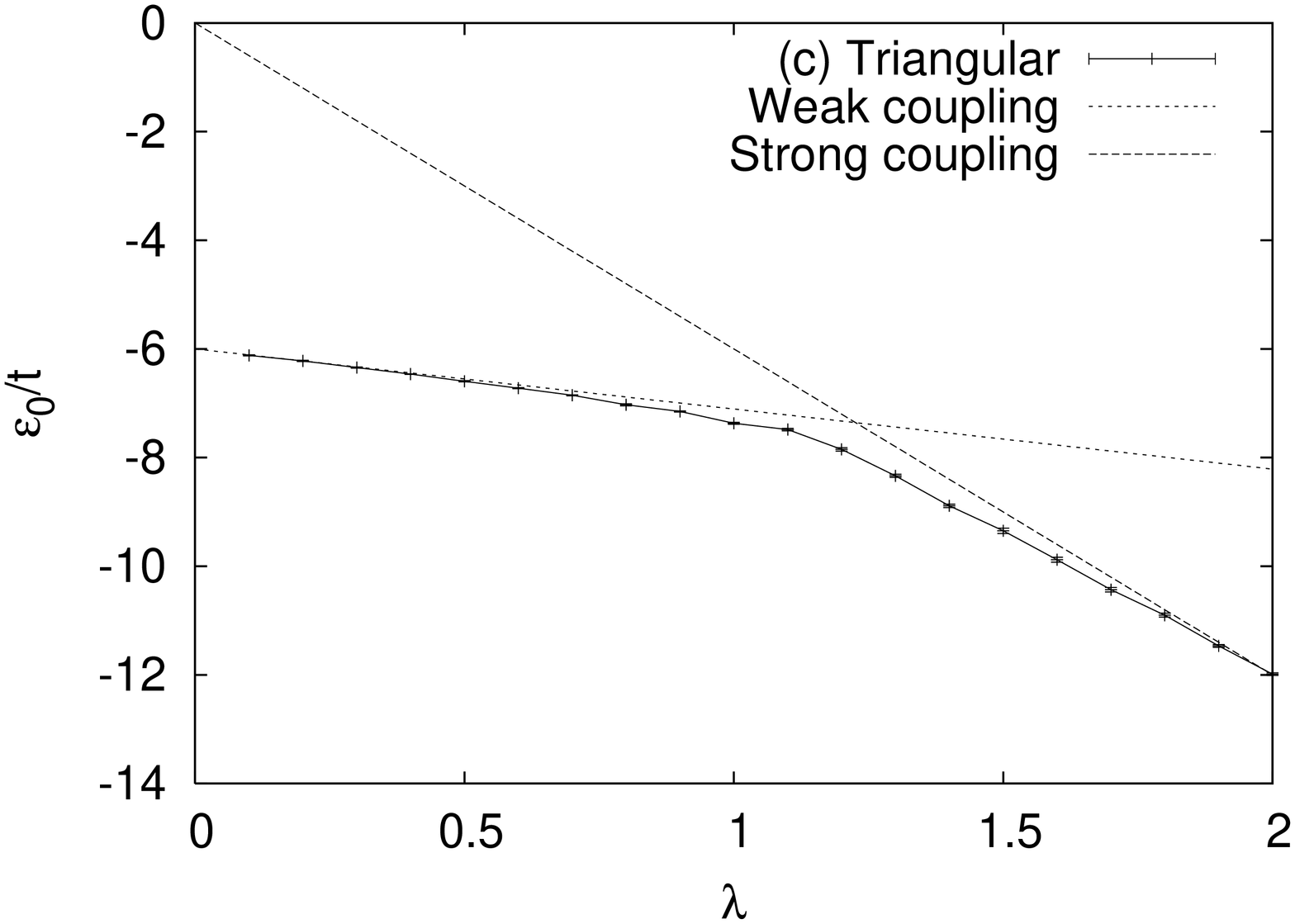}
\includegraphics[width=50mm,height=70mm,angle=270]{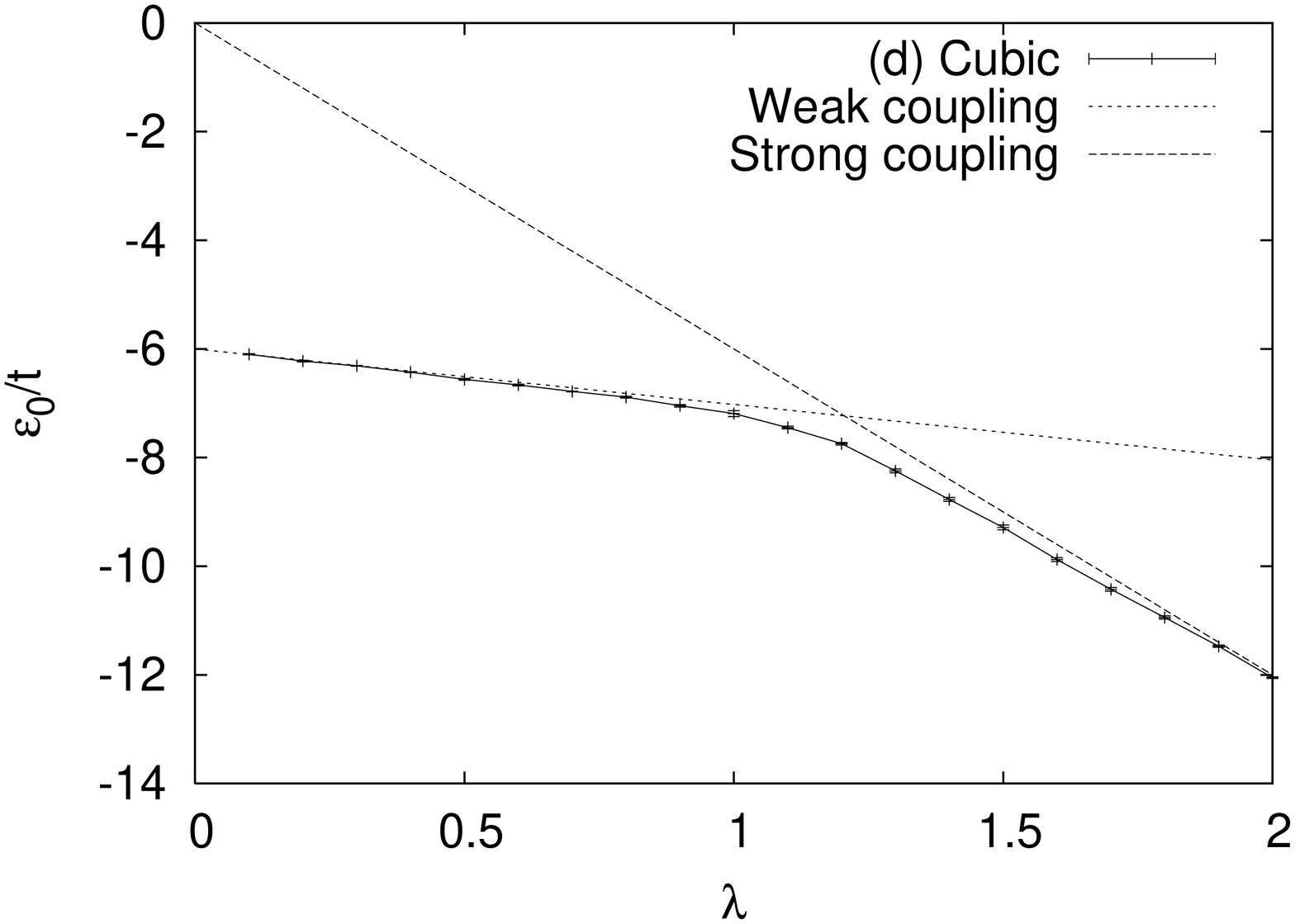}
\includegraphics[width=50mm,height=70mm,angle=270]{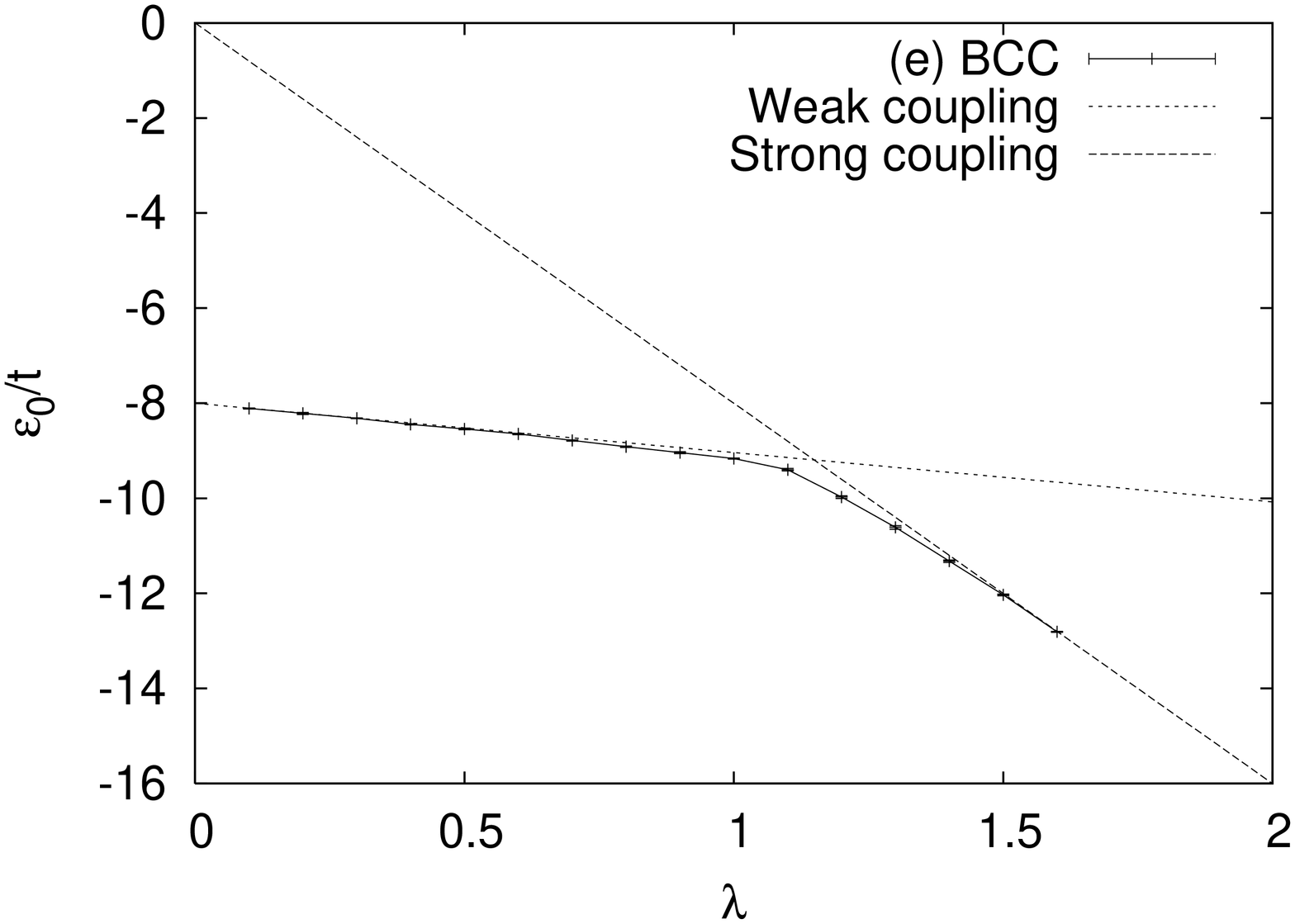}
\includegraphics[width=50mm,height=70mm,angle=270]{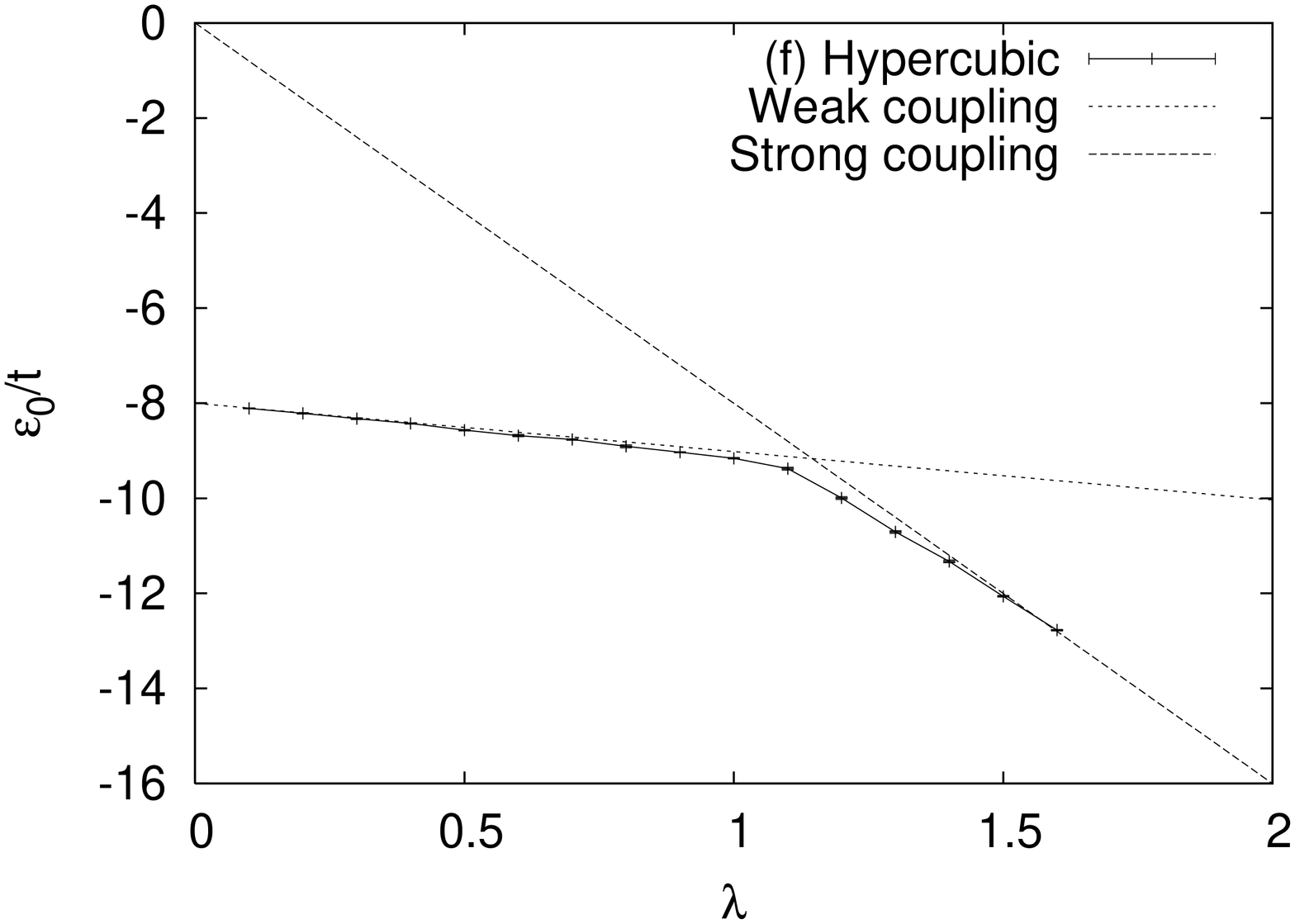}
\includegraphics[width=50mm,height=70mm,angle=270]{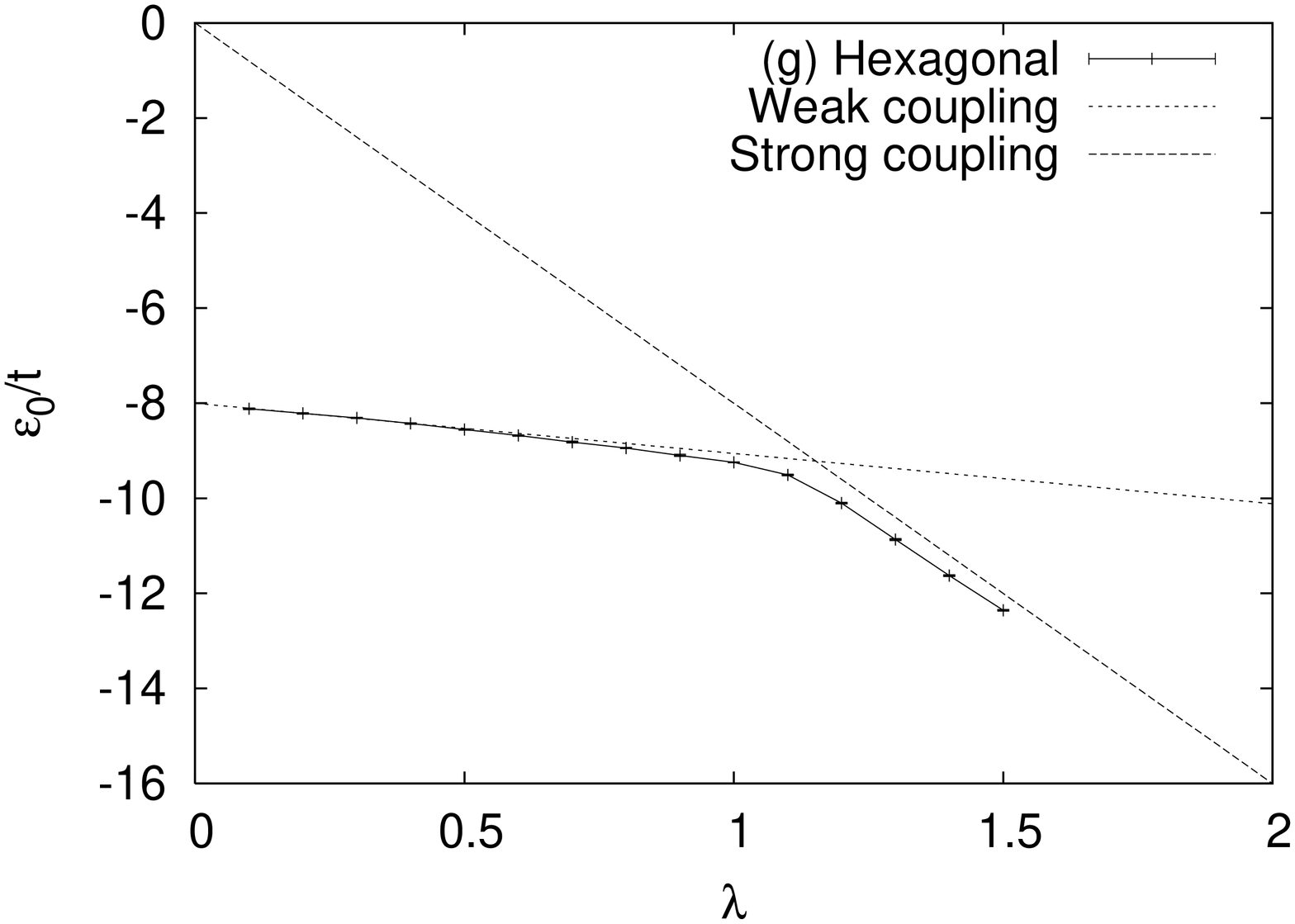}
\includegraphics[width=50mm,height=70mm,angle=270]{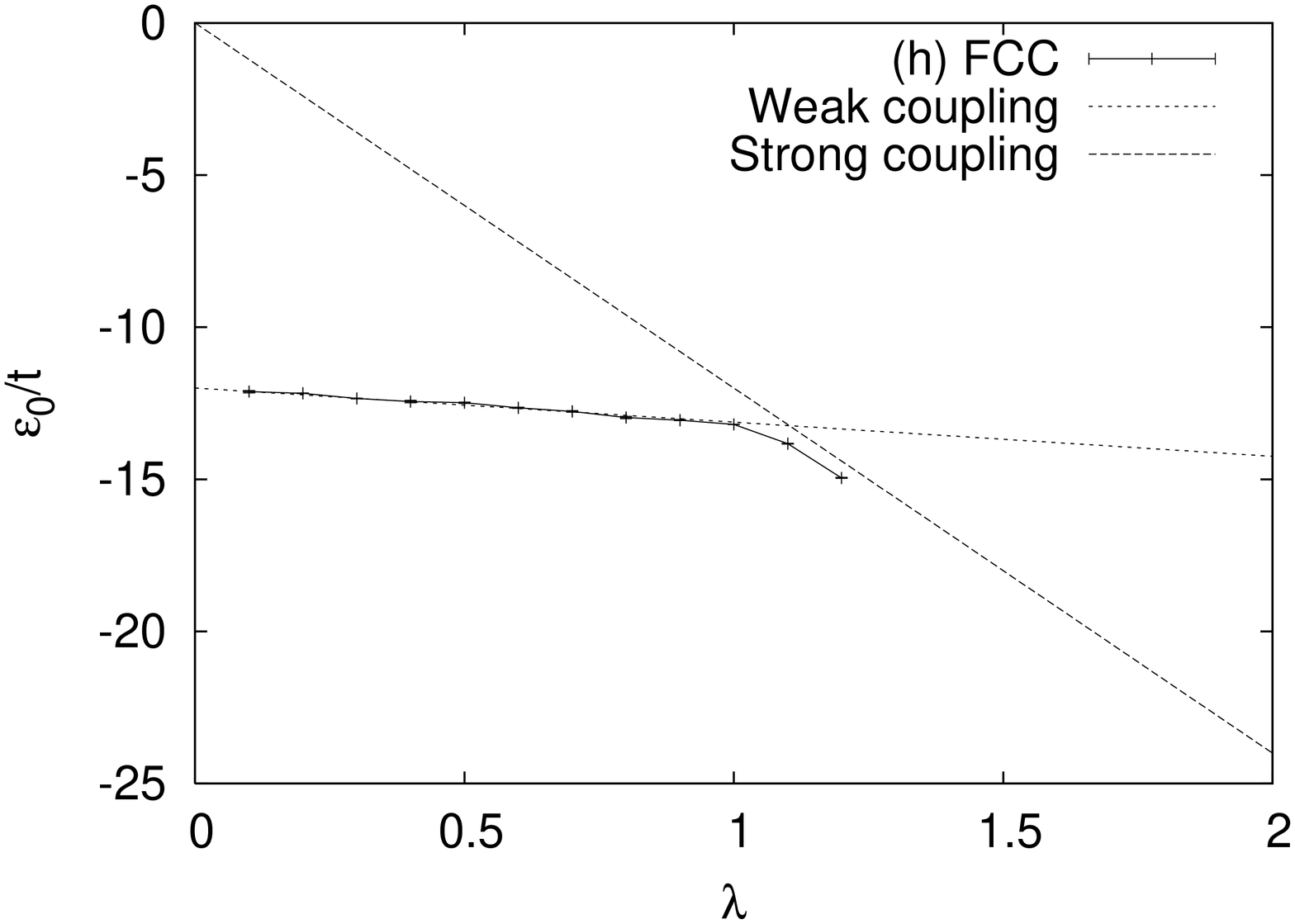}
\caption{Total energy of the Holstein polaron for increasing coupling $\lambda$
  with $\bar{\omega}=1$. Note that the crossover from weak to strong
  coupling behavior is very fast for this quantity on the triangular
  and cubic lattices, but is much slower for the square lattice. In
  general, the speed of the crossover is quicker for larger
  coordination number. The curve lies below the strong coupling asymptote for variational reasons and also lies below the weak coupling result.}
\label{fig:totalenergy}
\end{figure}

\begin{figure}
\includegraphics[width=50mm,height=70mm,angle=270]{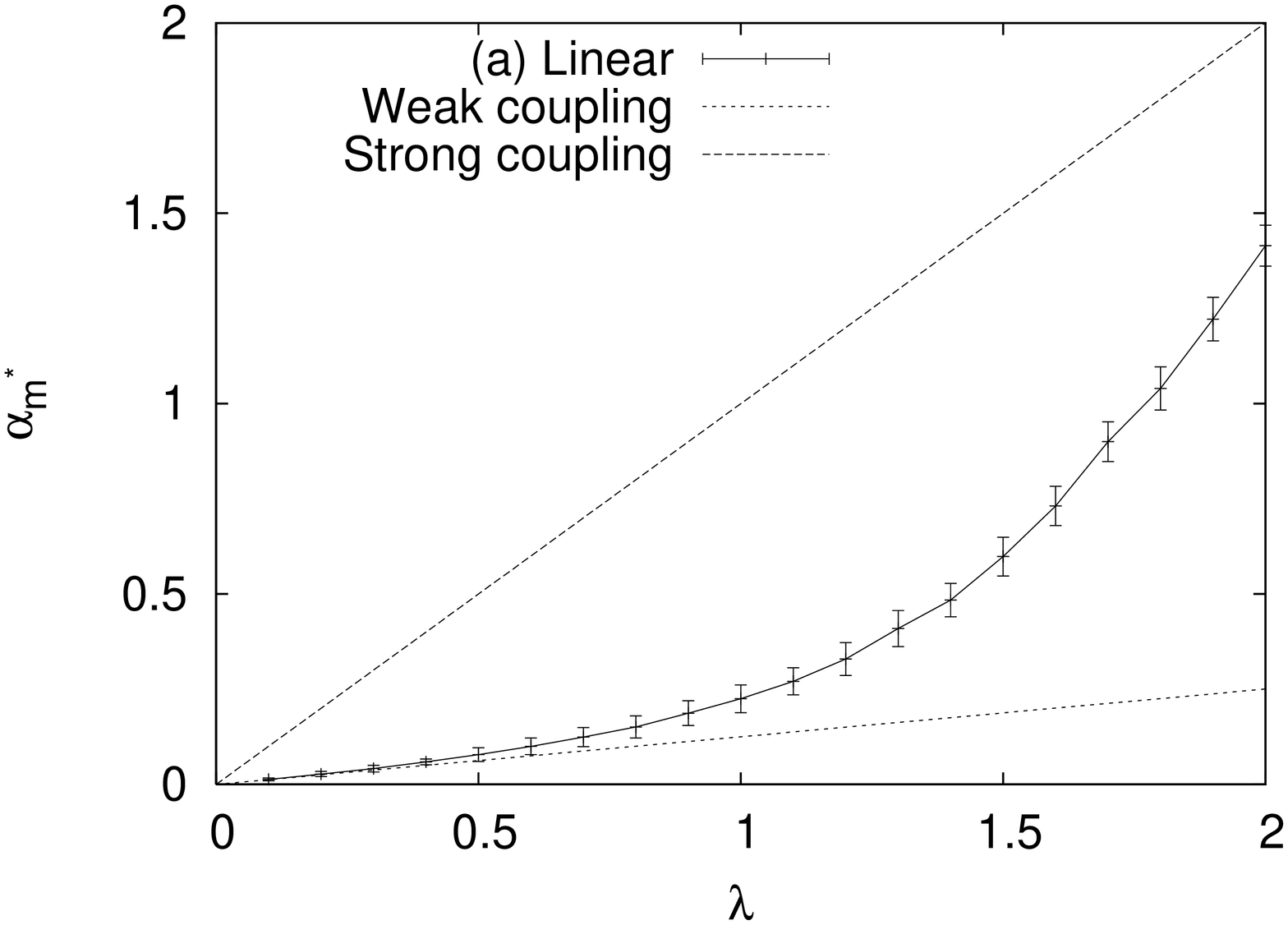}
\includegraphics[width=50mm,height=70mm,angle=270]{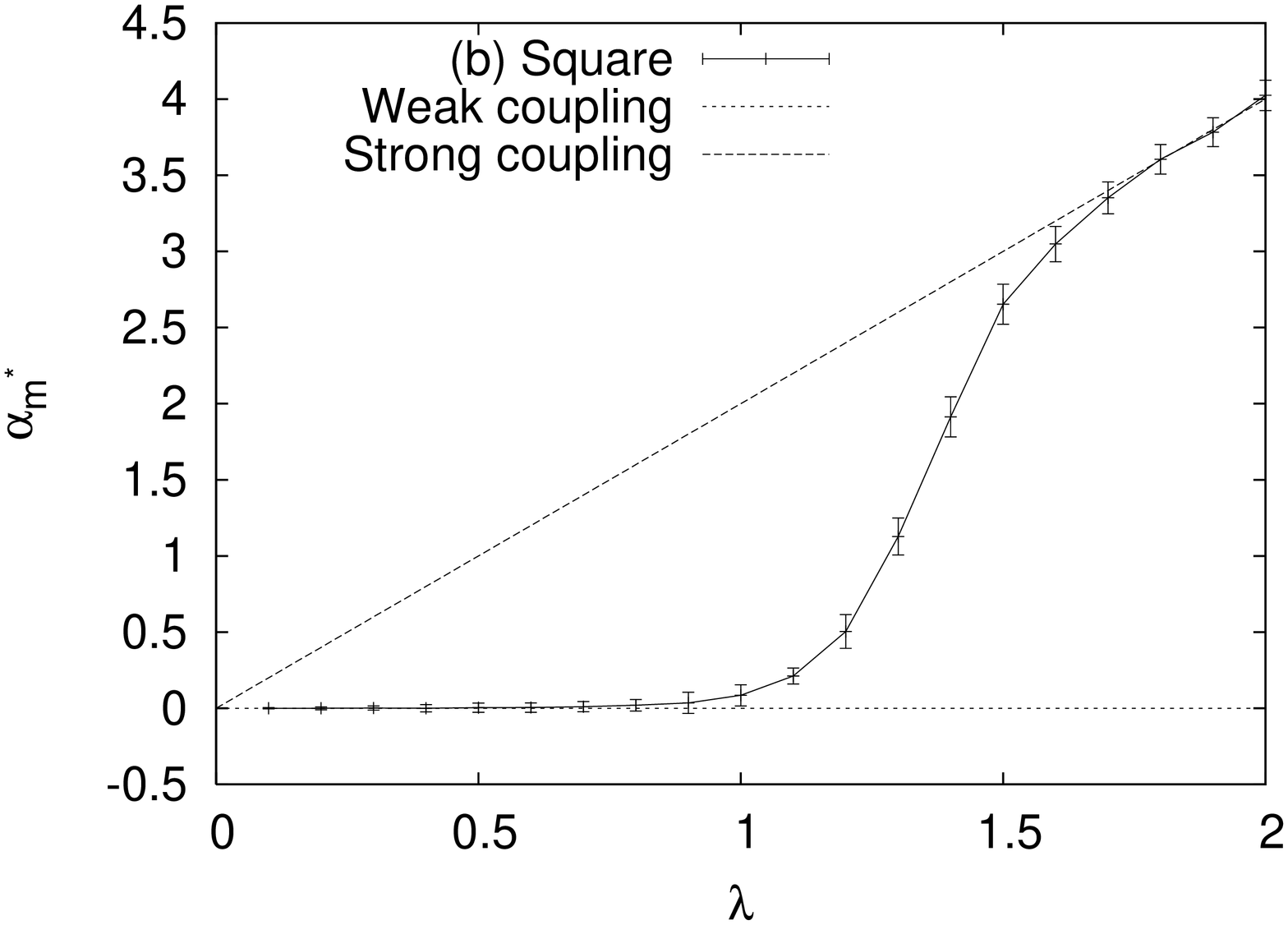}
\includegraphics[width=50mm,height=70mm,angle=270]{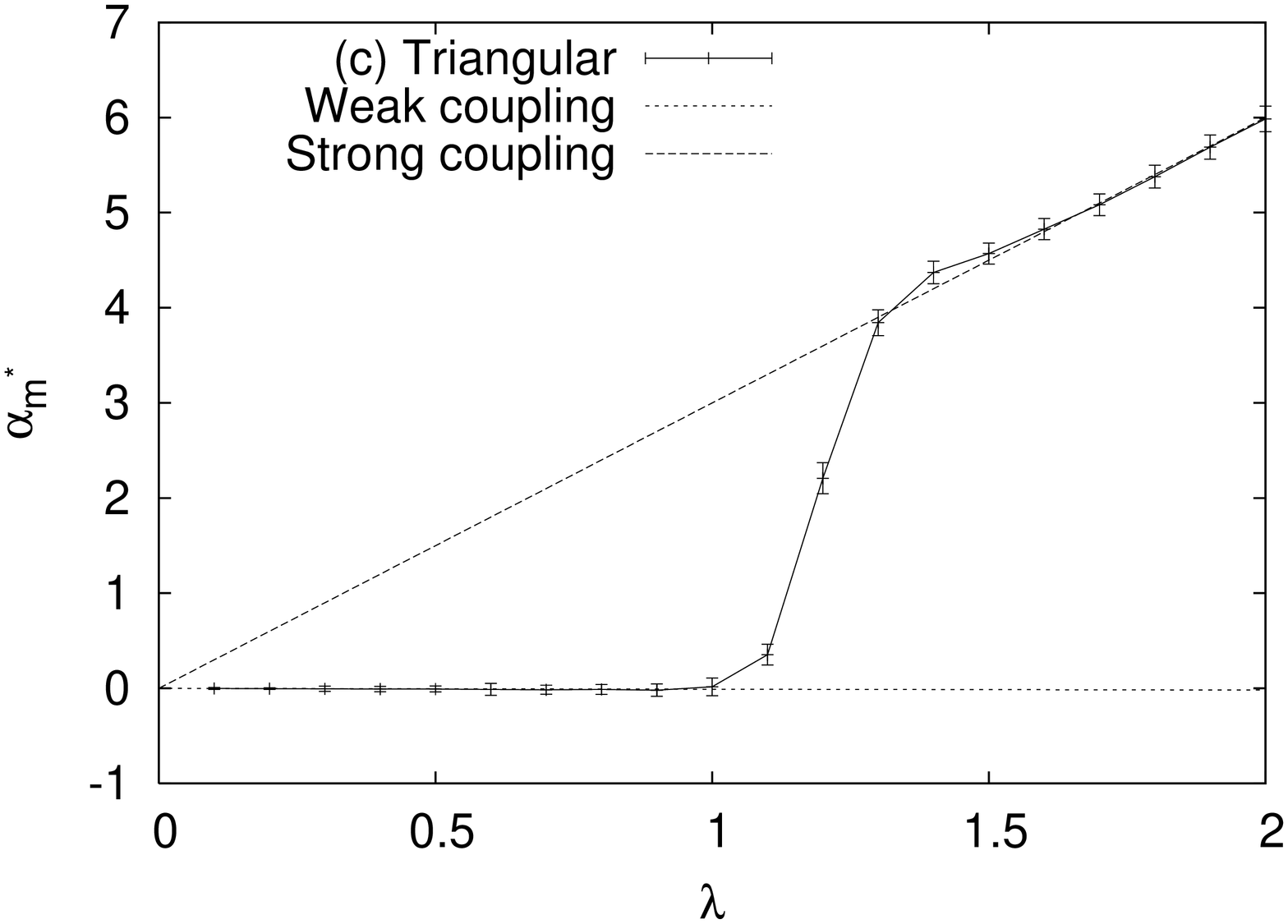}
\includegraphics[width=50mm,height=70mm,angle=270]{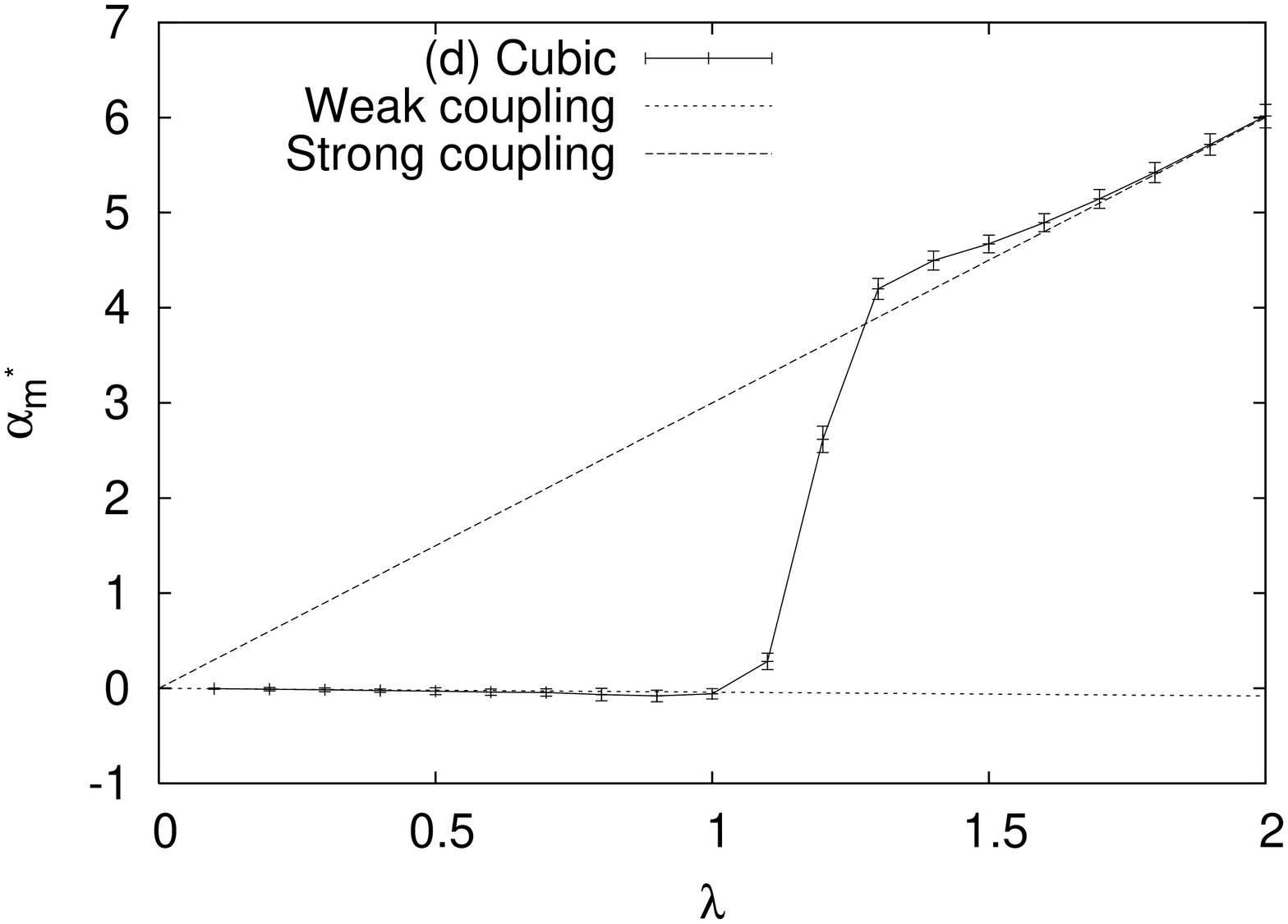}
\includegraphics[width=50mm,height=70mm,angle=270]{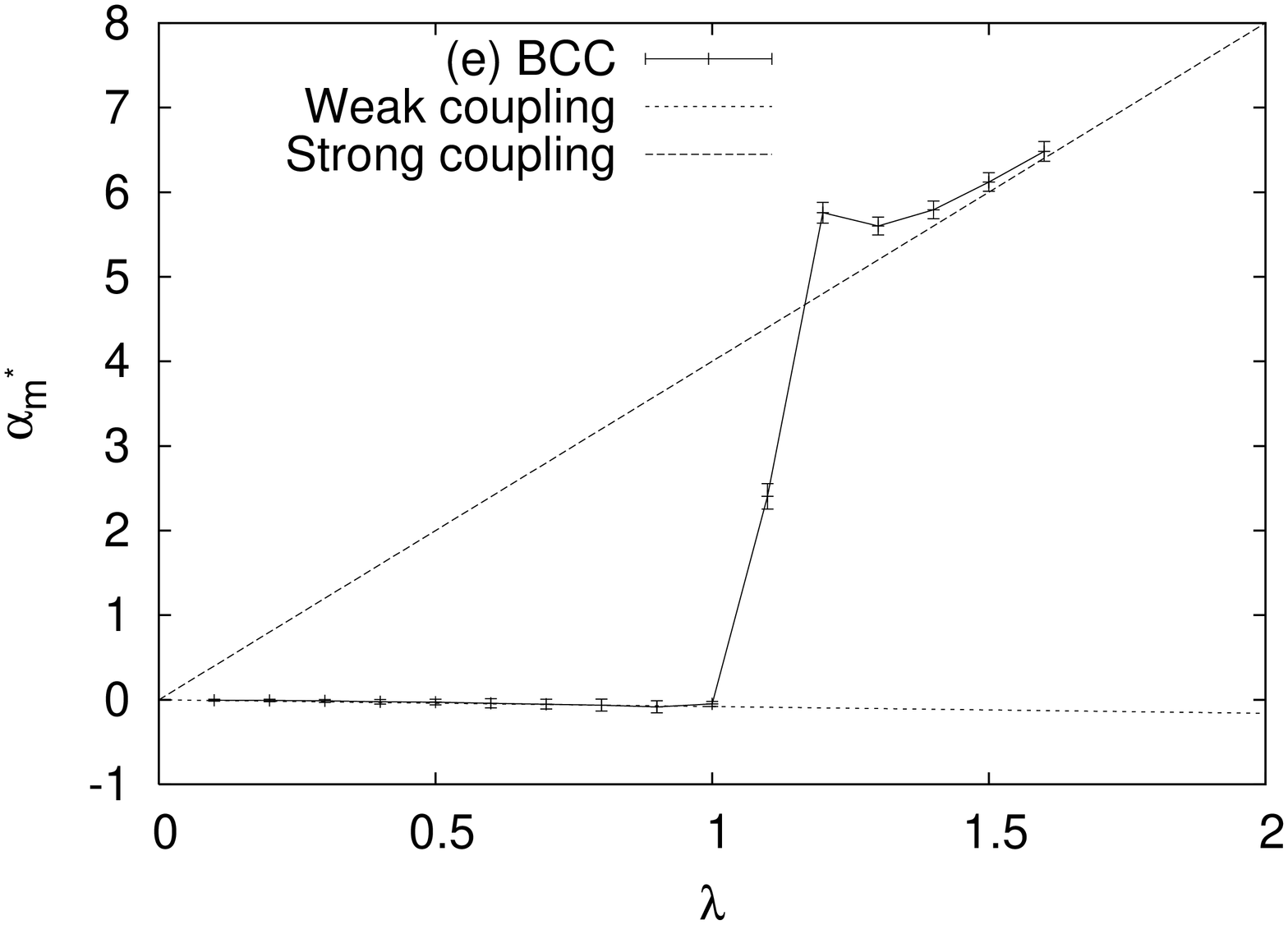}
\includegraphics[width=50mm,height=70mm,angle=270]{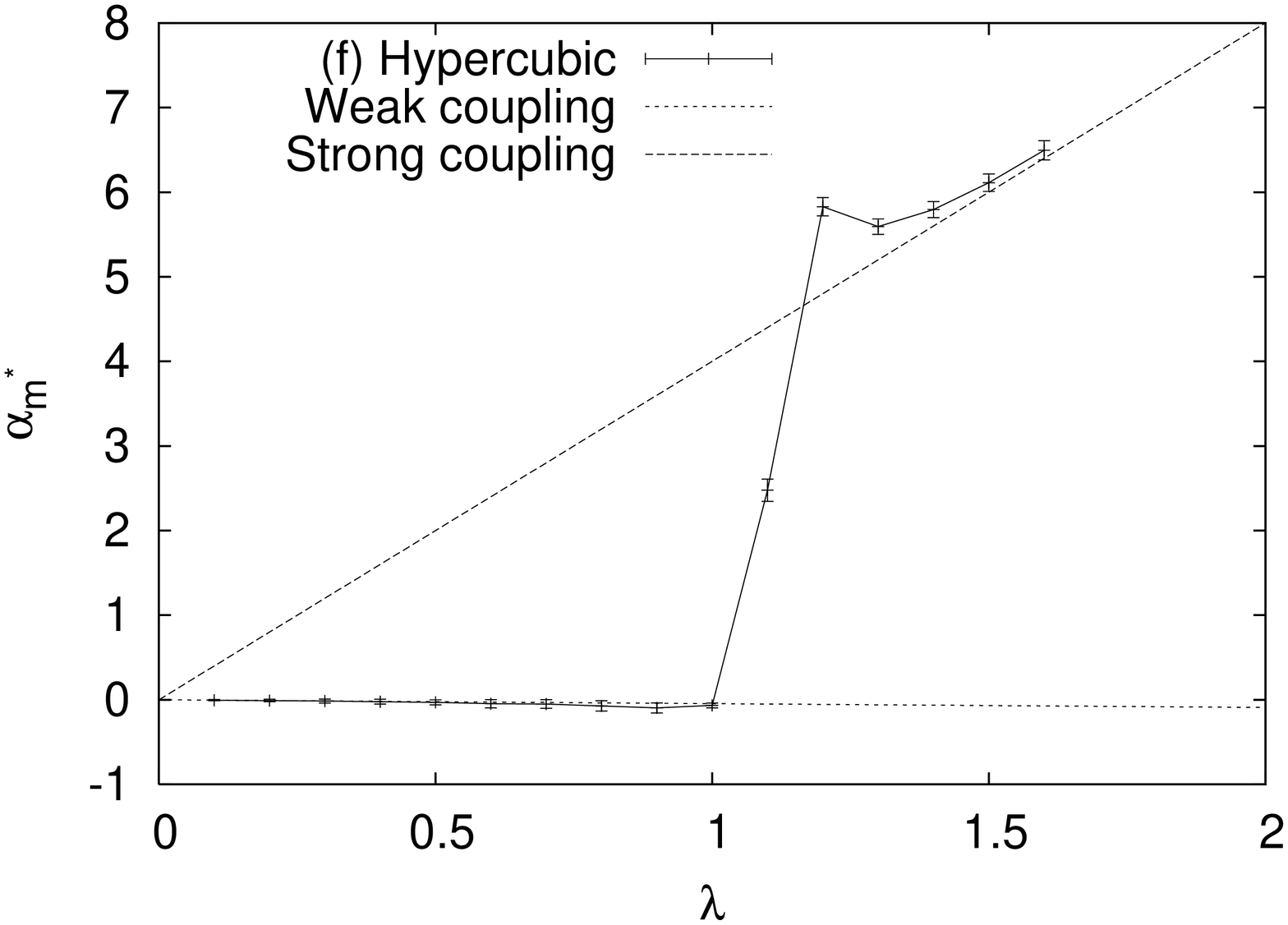}
\includegraphics[width=50mm,height=70mm,angle=270]{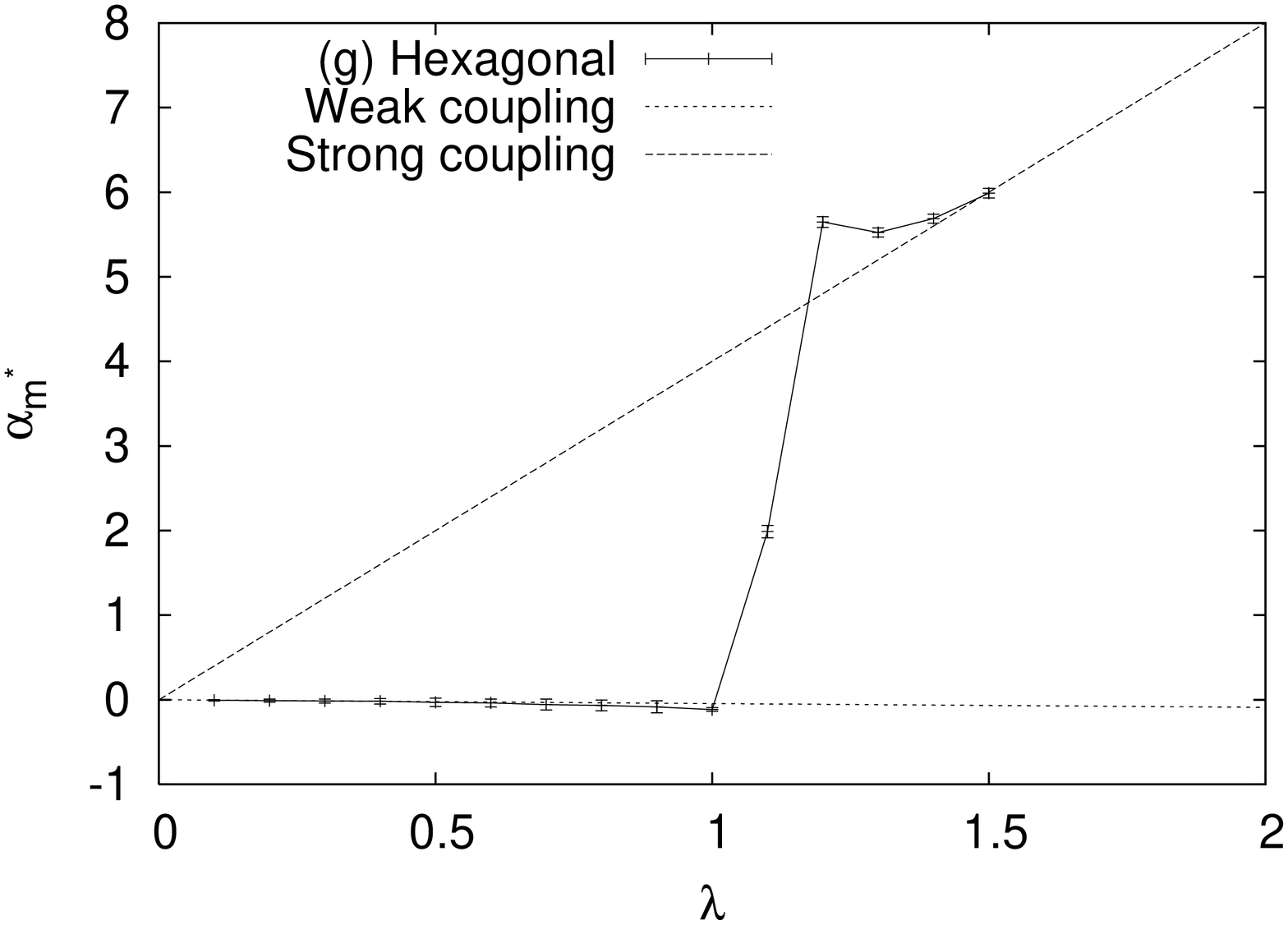}
\includegraphics[width=50mm,height=70mm,angle=270]{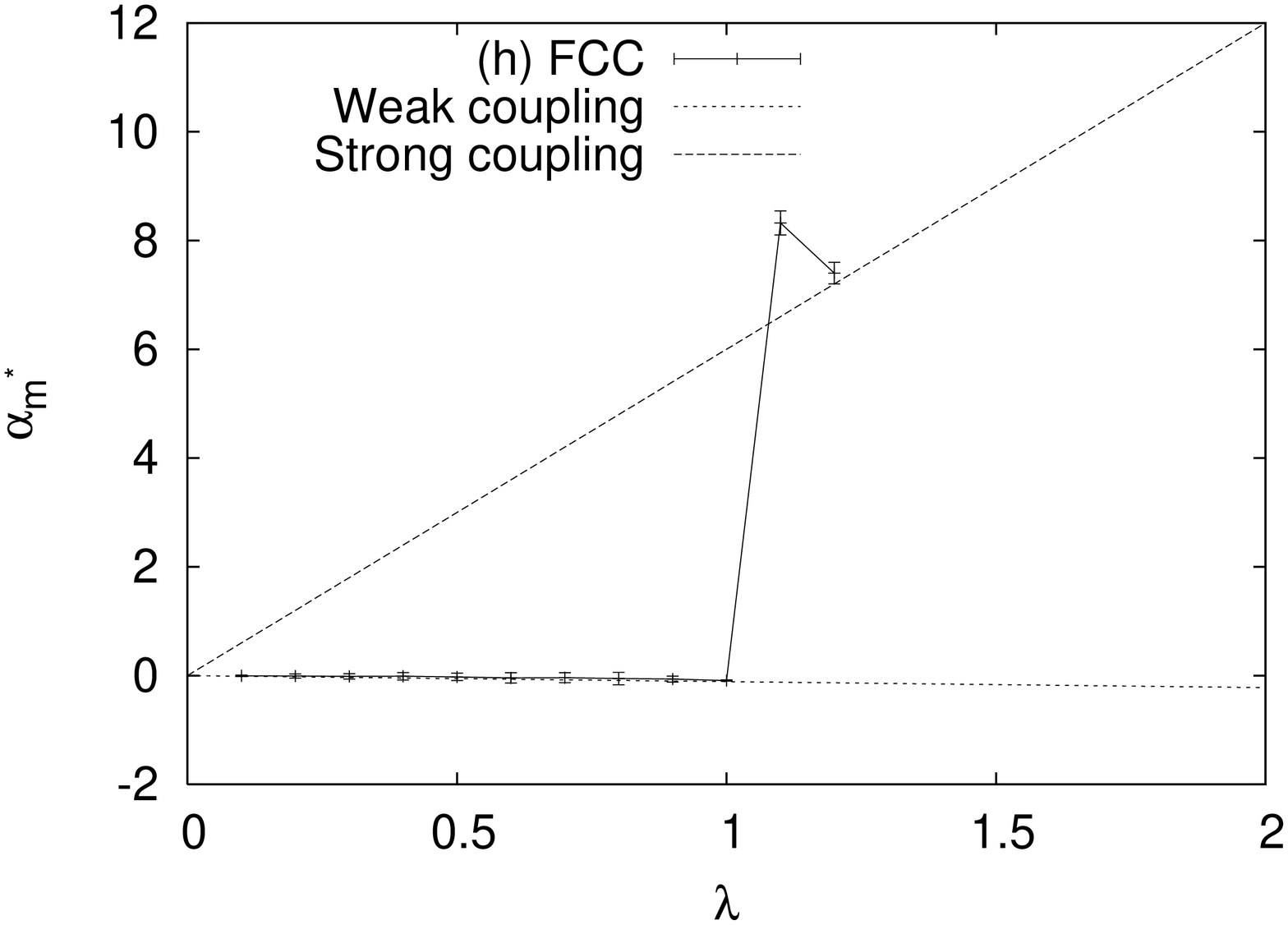}
\caption{Mass isotope coefficient of the Holstein polaron with
  increasing coupling and $\bar{\omega}=1$. At small $\lambda$, the
  $\alpha_{m^*}$ is negative and small, so the effective mass
  decreases with increased ion mass. Alternatively, the strong
  coupling behavior shows a clear increase of polaron mass with
  phonon mass. There is an anomalous bump in the isotope
  coefficient for lattices with $z>6$. }
\label{fig:isotopeeffect}
\end{figure}

\begin{figure}
\includegraphics[width=50mm,height=70mm,angle=270]{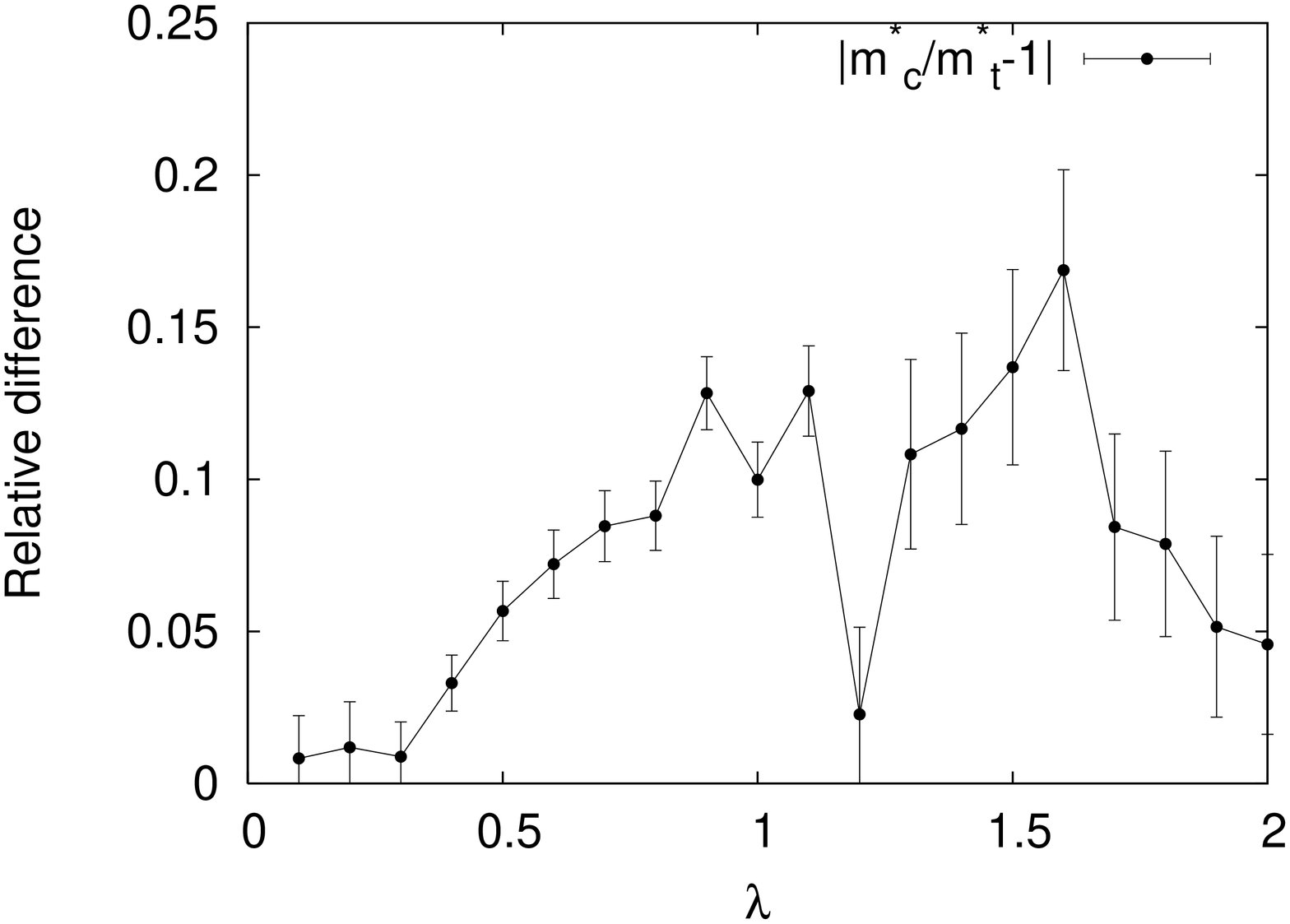}
\includegraphics[width=50mm,height=70mm,angle=270]{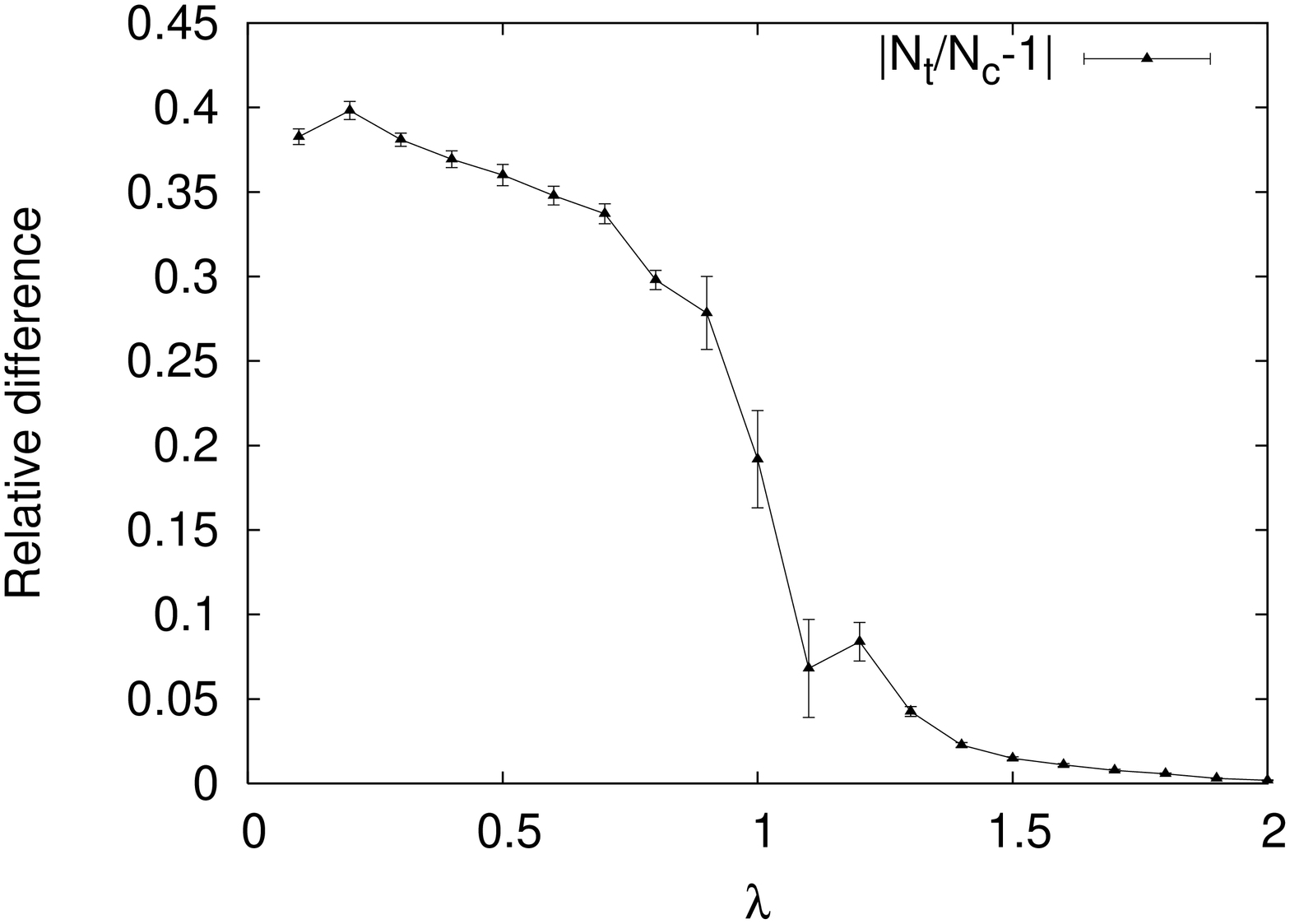}
\includegraphics[width=50mm,height=70mm,angle=270]{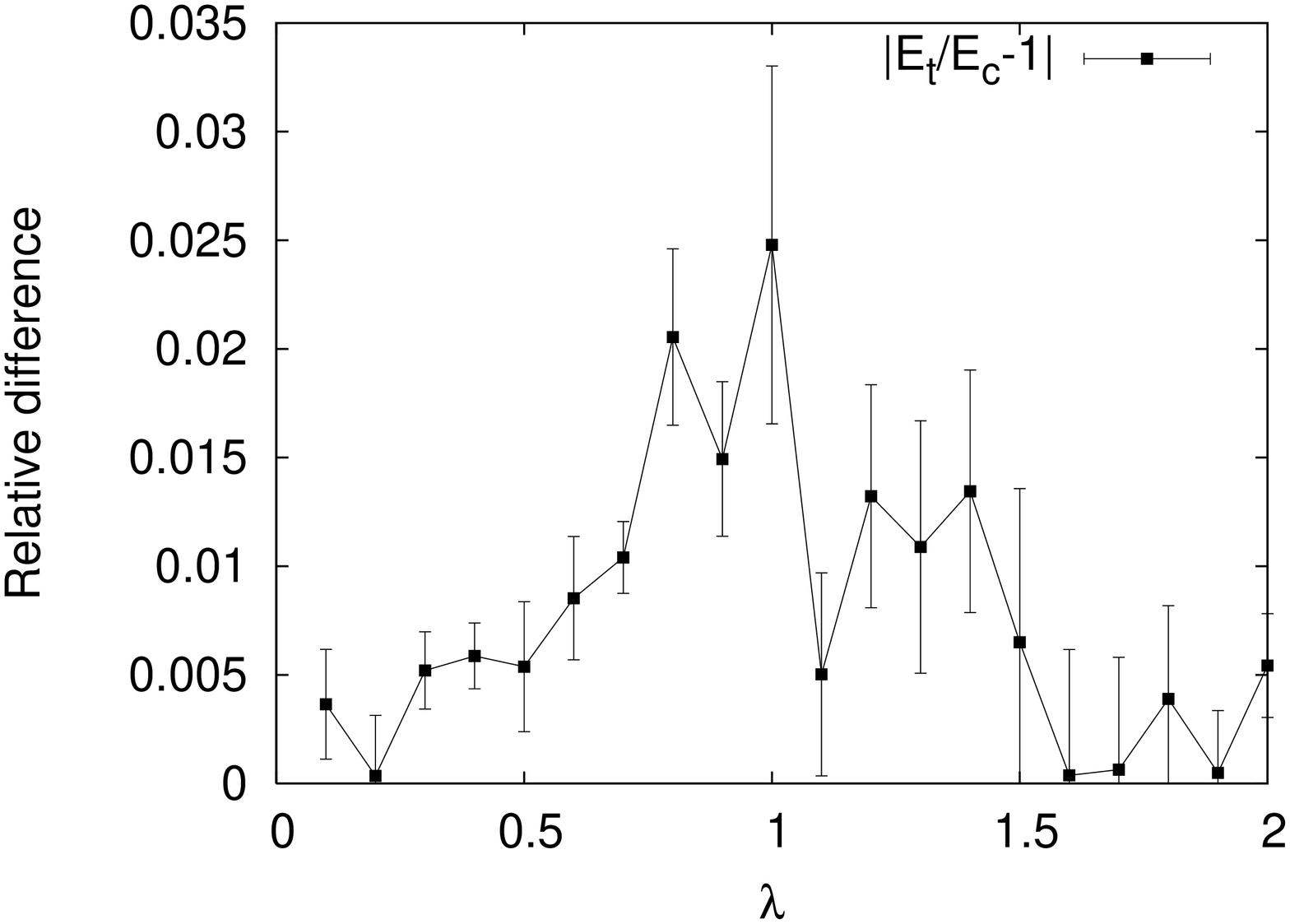}
\includegraphics[width=50mm,height=70mm,angle=270]{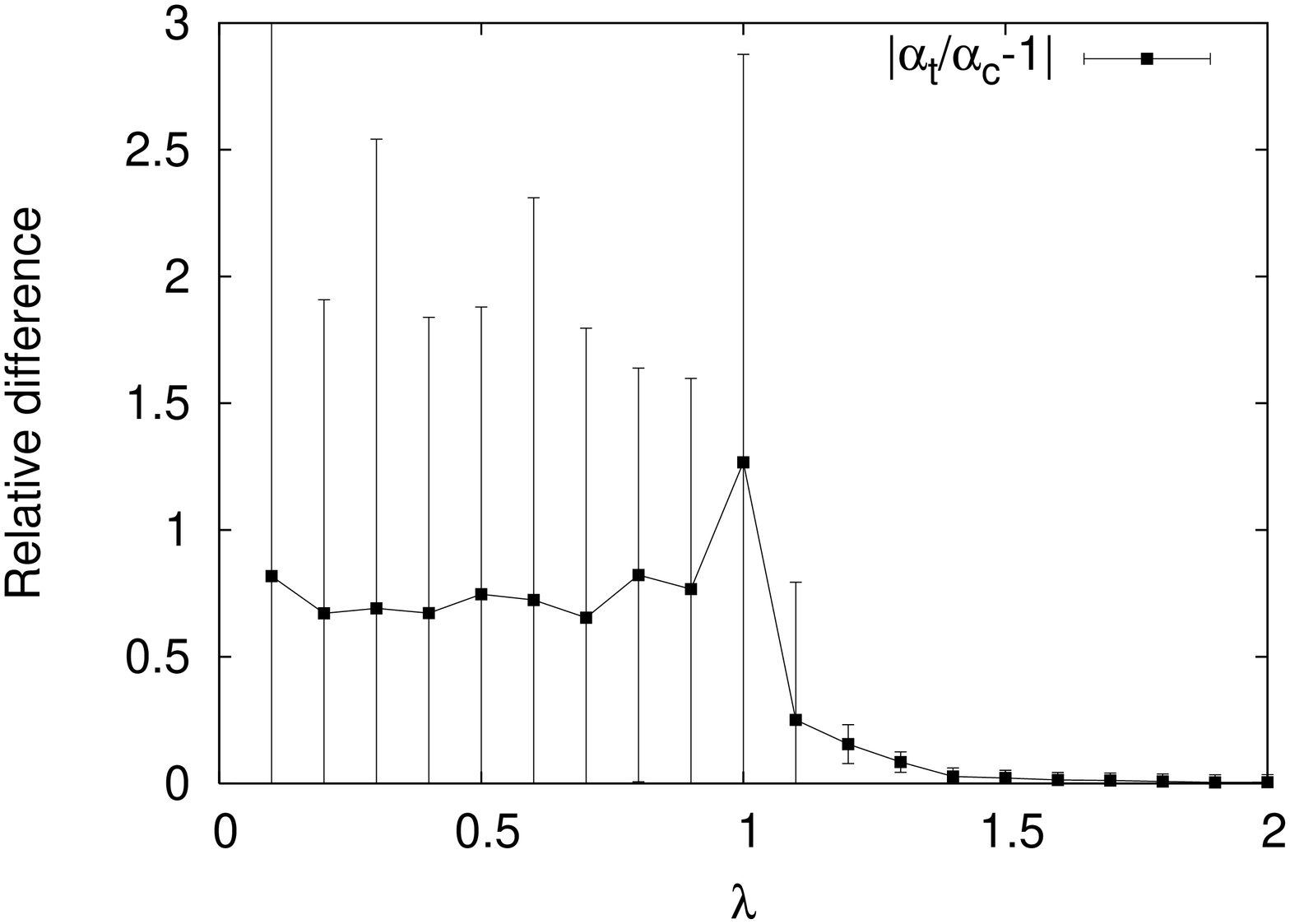}
\caption{Relative differences of physical observables on the cubic and
  triangular lattices. The total energy is identical to within 2\% on the triangular and cubic lattices, followed by the effective mass, which is within 20\% on both lattices. All quantities agree well for $\lambda>1$.}
\label{fig:reldiff}
\end{figure}

\begin{figure}
\includegraphics[width=50mm,height=70mm,angle=270]{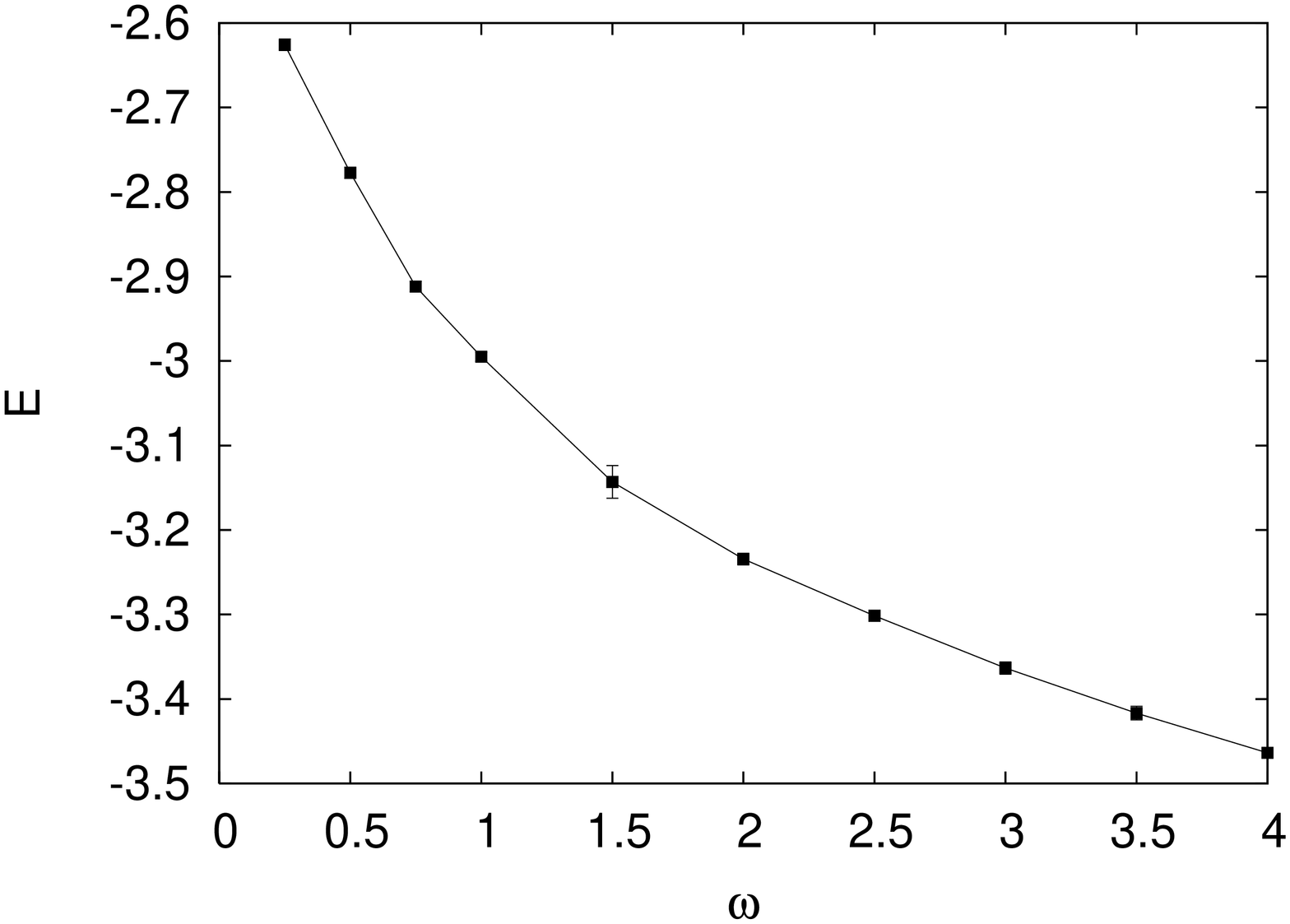}
\includegraphics[width=50mm,height=70mm,angle=270]{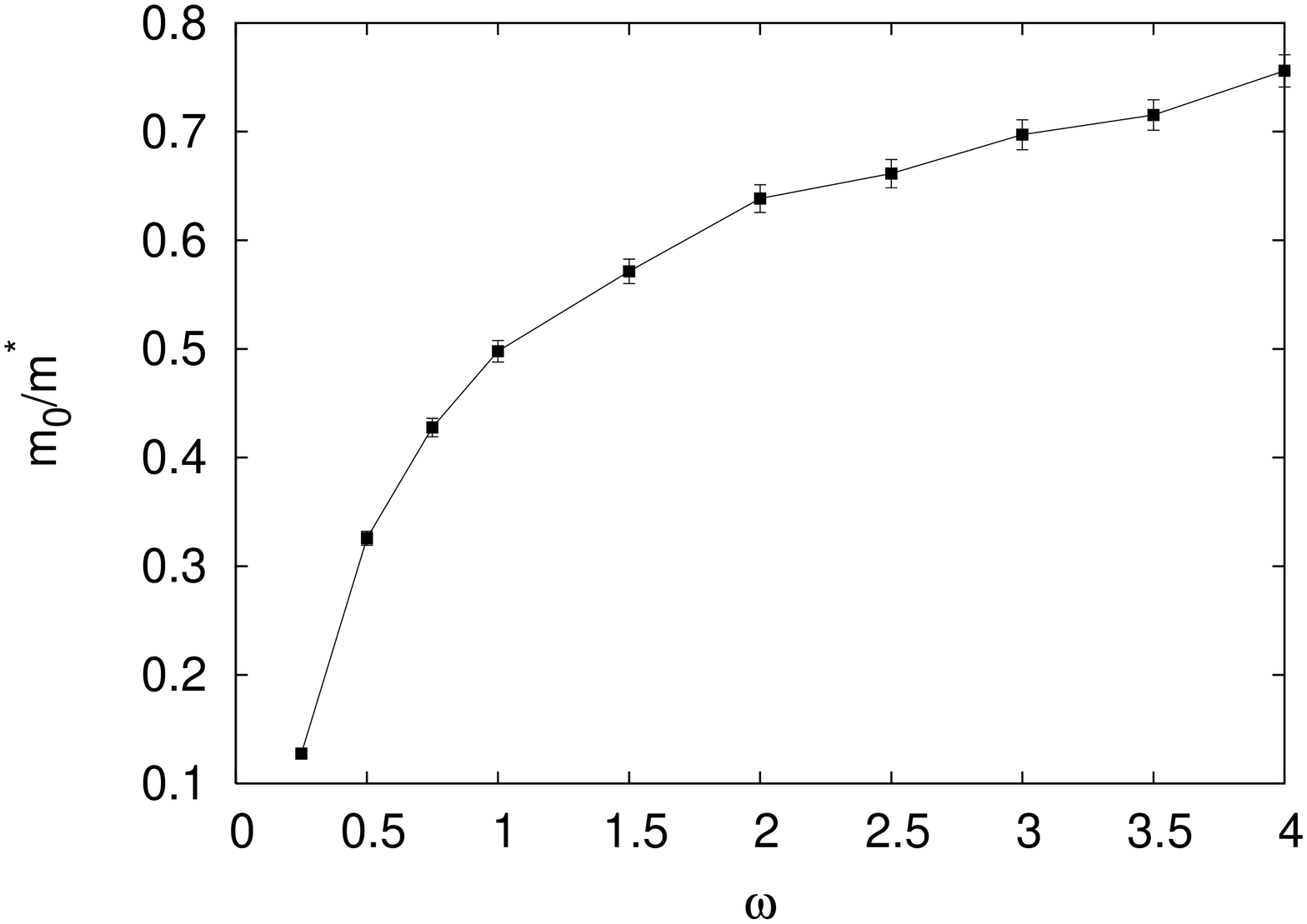}
\includegraphics[width=50mm,height=70mm,angle=270]{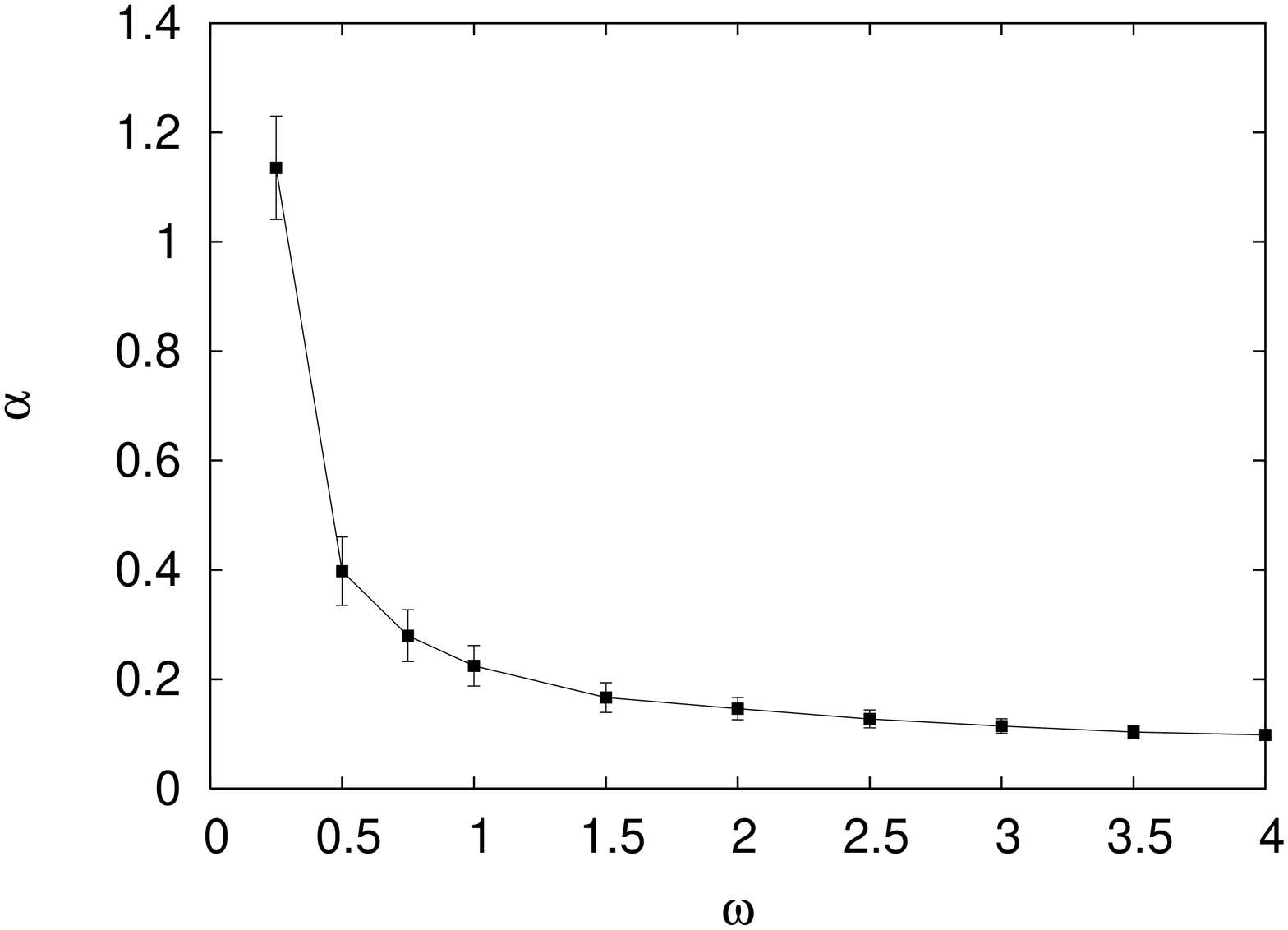}
\includegraphics[width=50mm,height=70mm,angle=270]{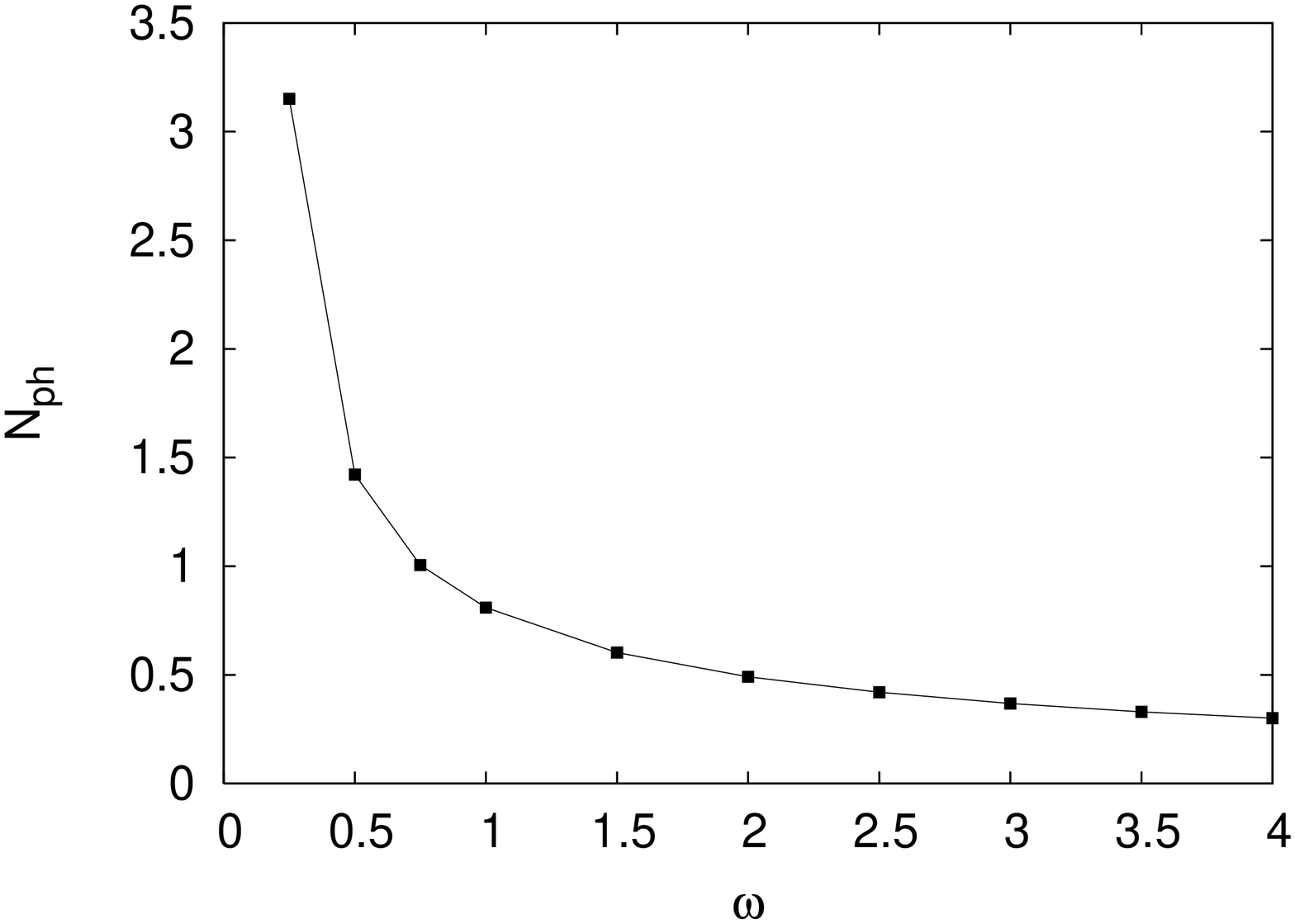}
\caption{Variation of physical observables with phonon frequency,
  $\bar{\omega}$ on a linear chain with $\lambda=1$. The physics
  remains qualitatively similar throughout, with no fast crossover or
  transition in any property.}
\label{fig:omegavar}
\end{figure}

In Fig. \ref{fig:inversemass}, we show the computed inverse effective
mass in the regime between weak and strong coupling.  Also shown on
the graphs are the weak and strong coupling results.  Panels are as
follows: (a) linear ($z=2$), (b) square ($z=4$), (c) triangular
($z=6$), (d) simple cubic ($z=6$), (e) body-center-cubic ($z=8$), (f)
hypercubic ($d=4, z=8$), (g) hexagonal ($z=8$), and (h)
face-center-cubic ($z=12$), with the same order of increasing
coordination number followed for all the graphs in this section.  The
first two panels are the results for the linear and square lattices,
which were presented first in Ref. \onlinecite{kornilovitch1998a}.
The linear lattice has the lowest coordination number of all lattices
computed here, and as such has the slowest crossover between weak and
strong coupling behavior.  The inverse mass diverges from the weak
coupling asymptote at couplings $\lambda\sim 1$, but the approach to
the strong coupling asymptote is very slow. The crossover is still
slow for the square lattice. In contrast, the change between the
limiting behaviors is much faster for the triangular lattice, with
much larger effective masses in the strong coupling limit.  The strong
coupling result depends only on the number of nearest neighbors $z$.
As such, strong coupling results for lattices with identical $z$ have
very similar profiles. For example, comparison with the cubic lattice,
which also has $z=6$, suggests that $z$ is the defining quantity for
the $\lambda>1$ curve, with only the weak coupling behavior dependent
on the lattice type (we have already seen how the strong coupling
limit of the effective mass is exponentially dependent on $z$ in
section \ref{sec:strongcoupling}).  For lattices with even larger
coordination number, this trend continues. For instance, the strong
coupling behavior of the polaron on the BCC lattice with $z=8$
corresponds well to that of the hypercubic lattice in $d=4$ and the
hexagonal lattice. Increasing $z$ to 12 (FCC lattice) shows an even
sharper transition between weak and strong coupling behavior. It
should be noted that in our calculation the strong coupling asymptote
of the effective mass depends exponentially on $z$. We note that in
the results from the DMFT with $z\rightarrow\infty$ presented in
Ref. \onlinecite{ciuchi1997a}, the ratio $\lambda/\bar{\omega}$ is
kept constant as $\lambda$ is changed, so the phonon frequency is
continuously increased and the $\lambda\rightarrow\infty$ results
correspond to the attractive Hubbard model and are not directly
comparable with our results. The computation of $m_0/m^*$ is limited
by its magnitude, since small $m_0/m^*$ means high effective mass and
rare hopping, so the ensemble has to be sampled for a long time to get
good statistics. The curve is truncated at weaker $\lambda$ for
lattices with large coordination number to take account of this
problem.

In Fig. \ref{fig:numberofphonons}, the number of phonons in the
polaron cloud is displayed. Again, a series of different lattices has
been considered. For most couplings, the number of phonons is closely
related to the exponent of the effective mass. For small
electron-phonon coupling, the polaron cloud is virtually empty. The
rate of increase, and the time taken to cross over from weak to strong
coupling behavior is highly dependent on the lattice type
considered. For example, on a square lattice, the phonon number is not
saturated to the strong coupling limit until $\lambda\sim 2$, in
contrast to the $z=6$ triangular and cubic lattices, which saturate at
$\lambda\sim 1.5$. For increasing $z$ the saturation value of
$\lambda$ decreases, with saturation at $\lambda\sim 1$ for the $z=12$
FCC lattice. It is instrumental to consider the ratio
$N_\mathrm{ph}/\ln(m_0/m^*)$ as a measure of the departure from strong
coupling.

The total energy is displayed in Fig. \ref{fig:totalenergy}. The
energy decreases monotonically with increased coupling for all
lattices. The larger the coordination number, the quicker the
crossover between weak and strong coupling results. The curve lies
below the strong coupling limit for variational reasons, and also lies
below the weak coupling line. It should be noted that, for higher
coordination numbers at intermediate coupling, it is necessary to
increase the total warmup period. This is due to a meta-stable state
corresponding to a path with no kinks, with single kinks created and
then immediately destroyed. The meta-stable state has the properties
of the strong coupling (atomic) limit. Once two kinks are created
(which can take several thousand steps), the system drops out of the
metastable state into the lower energy minimum, and the correct form
for the total energy is found. Such a metastable state has been
discussed in Ref. \onlinecite{kabanov1993a}. To speed up the
computation, the path is started with several kinks inserted
randomly. In this way, the meta-stable state is avoided.

The mass isotope coefficient $\alpha_{m^*}$
is shown in Fig. \ref{fig:isotopeeffect}. Note that for
$\bar{\omega} = 1$, most lattices have a small negative isotope
coefficient for small $\lambda$. At couplings $\lambda\sim 1$, the
sign of the isotope coefficient flips. For lattices with high
coordination number, the curve overshoots the strong coupling limit,
before converging on the asymptote.

In order to determine the extent to which the cubic and triangular
lattices are similar, Fig. \ref{fig:reldiff} shows the relative
differences of physical observables on the cubic and triangular
lattices. The total energy is identical to within 2\% on the
triangular and cubic lattices, followed by the effective mass, which
is within 20\% on both lattices. All quantities agree well for
$\lambda>1$.

Finally, we demonstrate that there is no qualitative change of the
physics with phonon frequency. Fig. \ref{fig:omegavar} shows the
variation of physical observables as the phonon frequency is
reduced. This has been carried out on a linear chain, so that the low
omega regime can be reached with the QMC solver. The physics remains
qualitatively similar throughout, with no fast crossover or transition
in any property.

\subsection{Excited states}

In this section we compute the excitation spectrum and density of
states for the polaron gas. The density of states could be used to
determine thermodynamic and transport properties of the free polaron
gas in a similar way to the Sommerfeld theory of metals, although some
care should be taken in 1D, where there are no polarons at half-filling
\cite{hohenadler2005}. In most materials, it is expected that the density of polarons is sufficiently low to avoid such a breakdown of the polaron picture.

\begin{figure}
\includegraphics[width=50mm,height=70mm,angle=270]{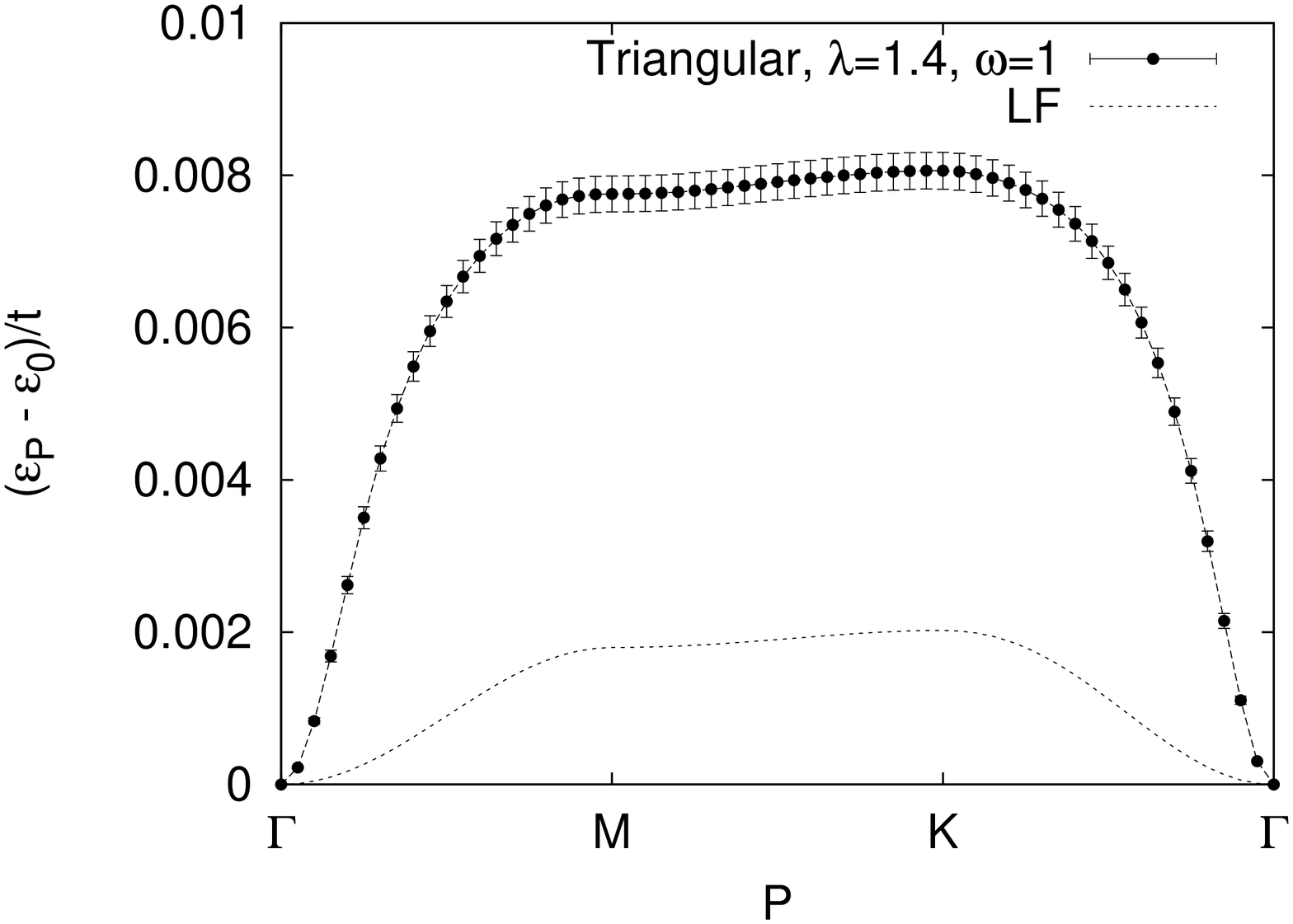}
\includegraphics[width=50mm,height=70mm,angle=270]{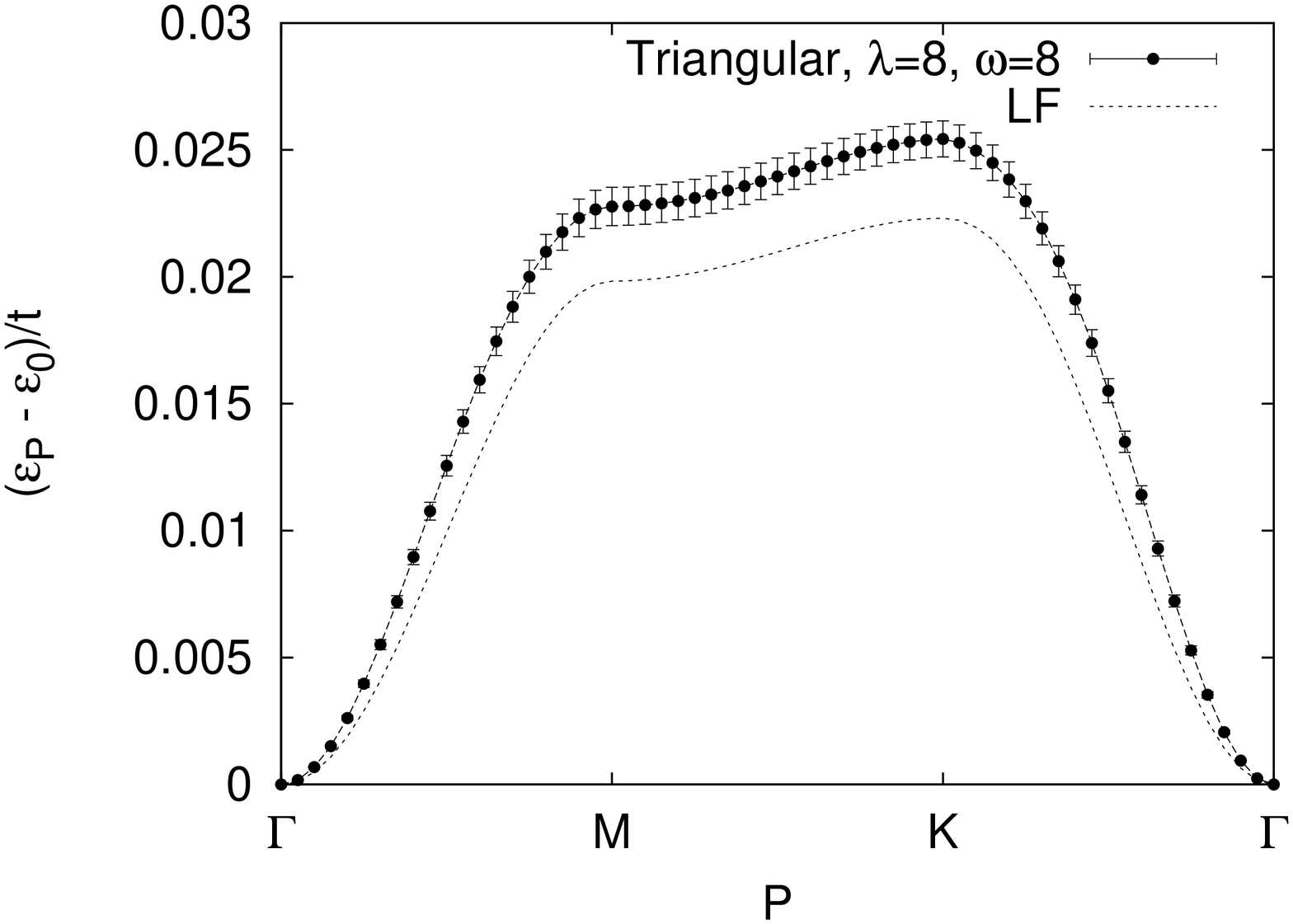}
\includegraphics[width=50mm,height=70mm,angle=270]{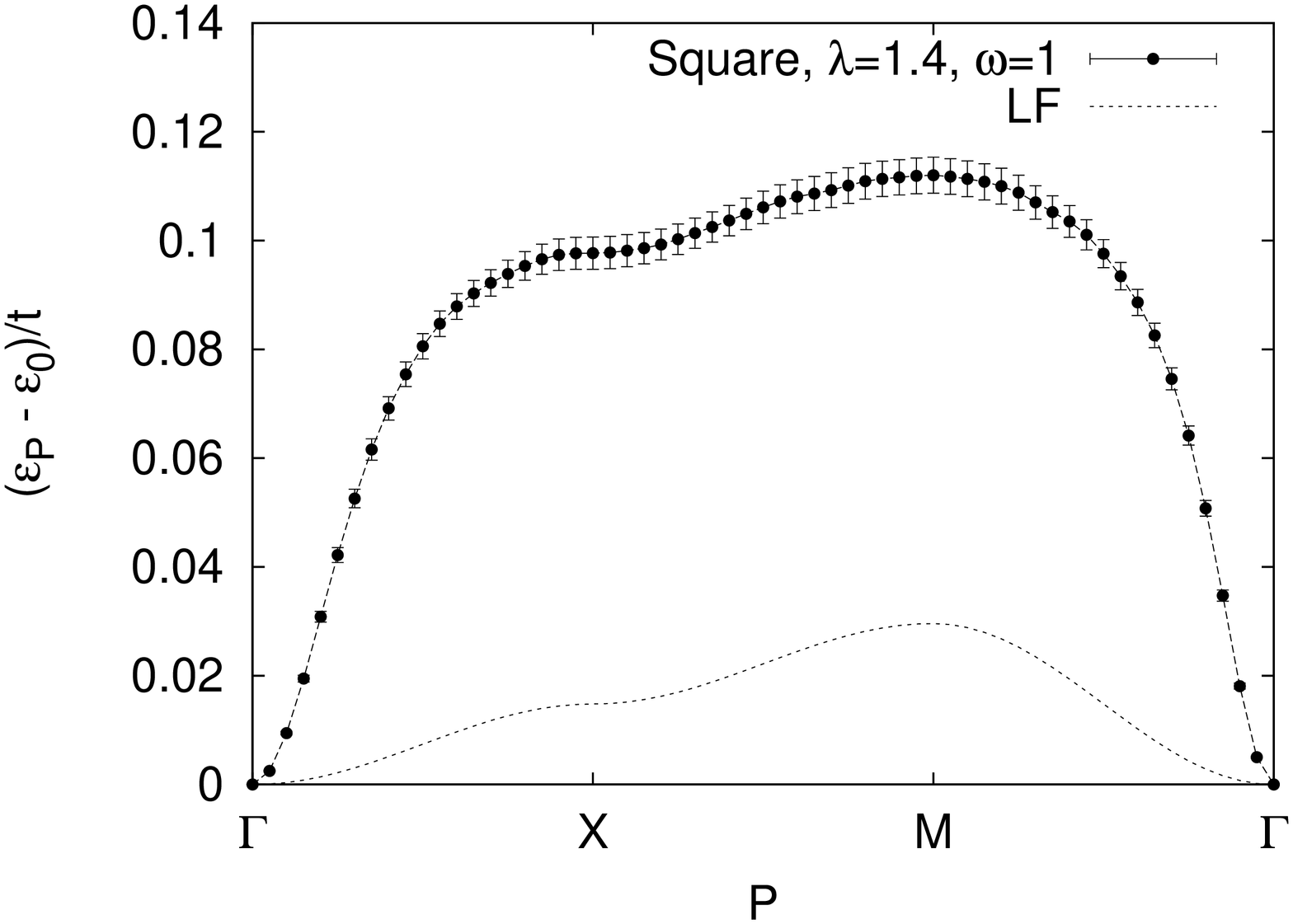}
\includegraphics[width=50mm,height=70mm,angle=270]{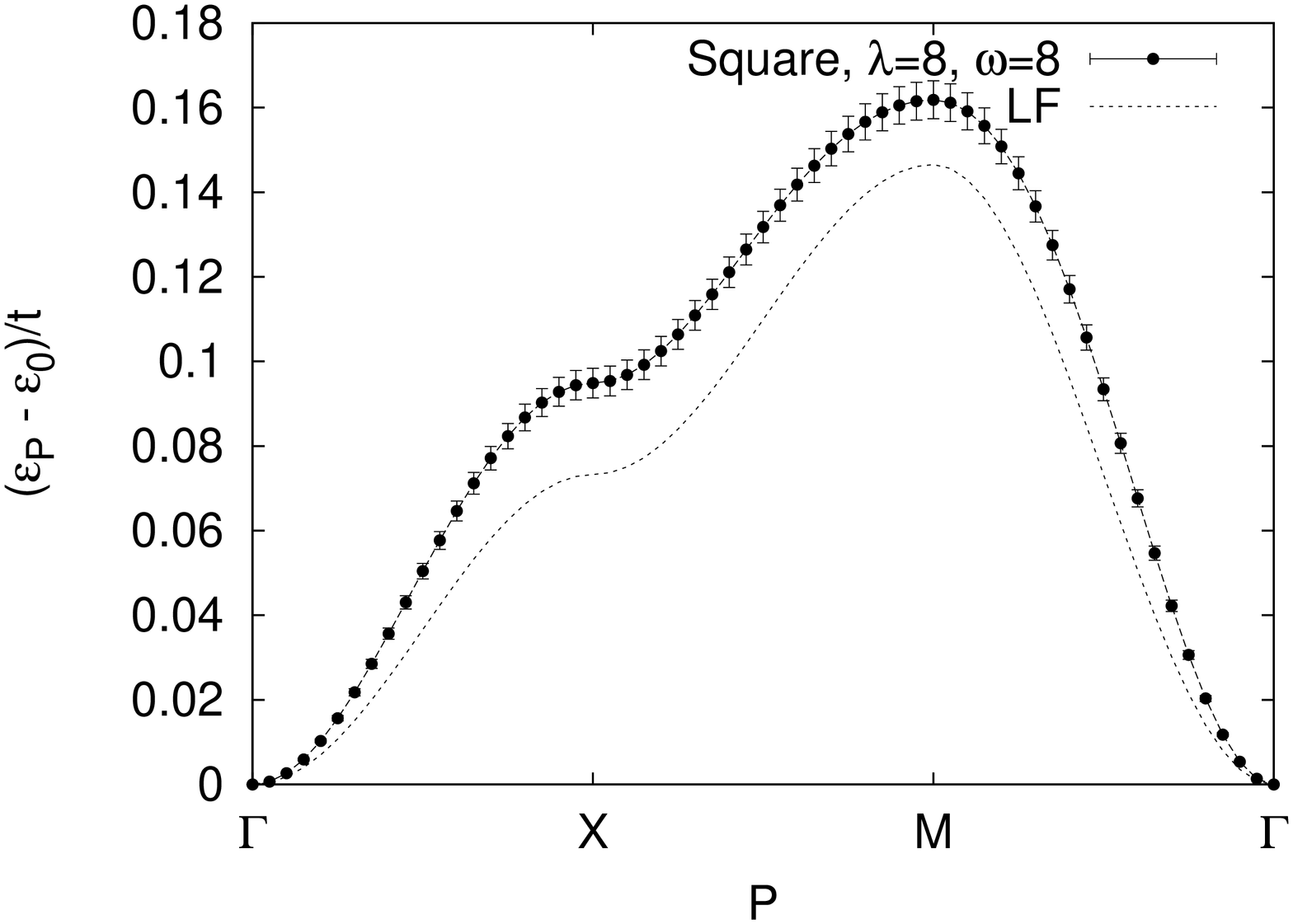}
\includegraphics[width=50mm,height=70mm,angle=270]{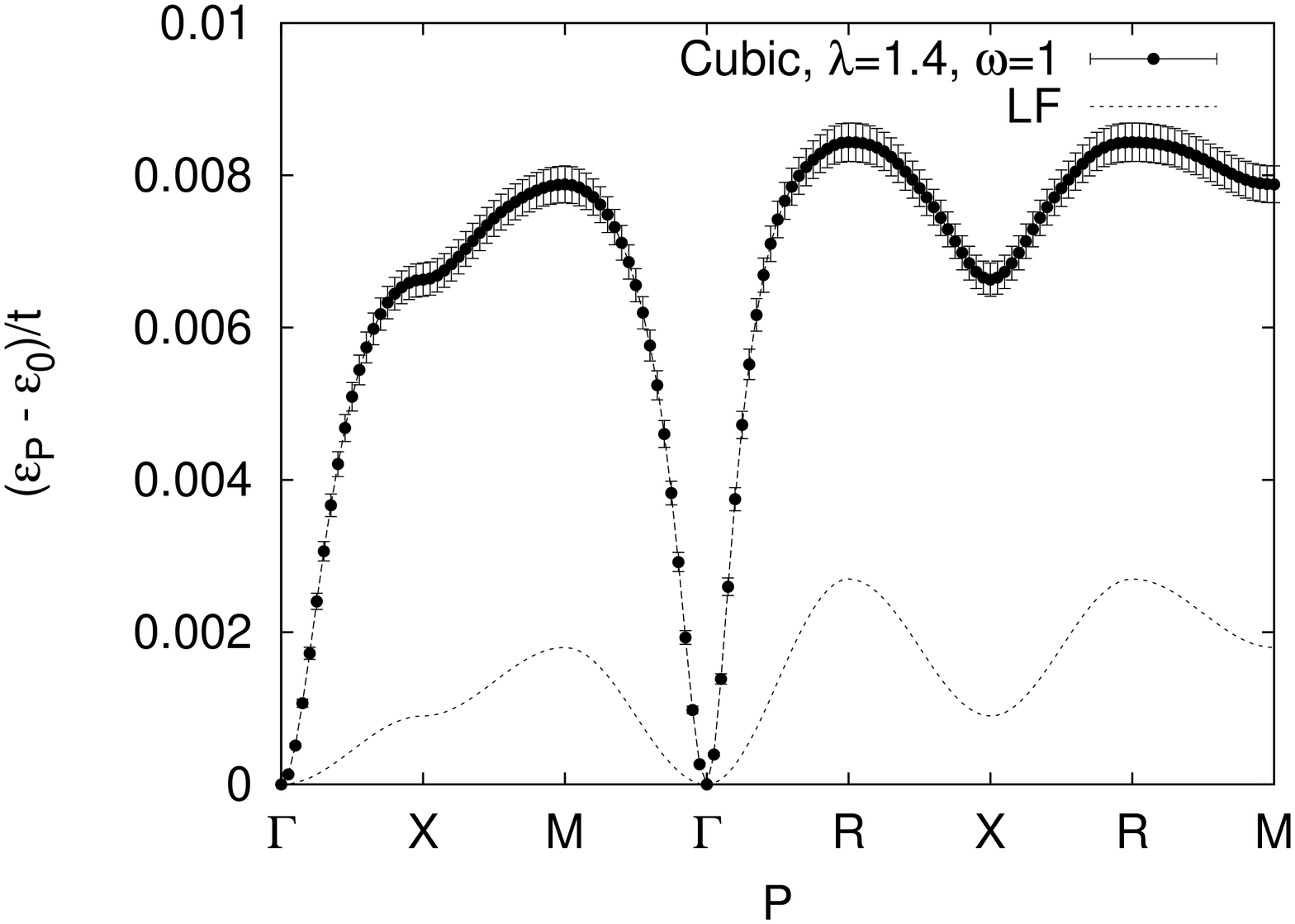}
\includegraphics[width=50mm,height=70mm,angle=270]{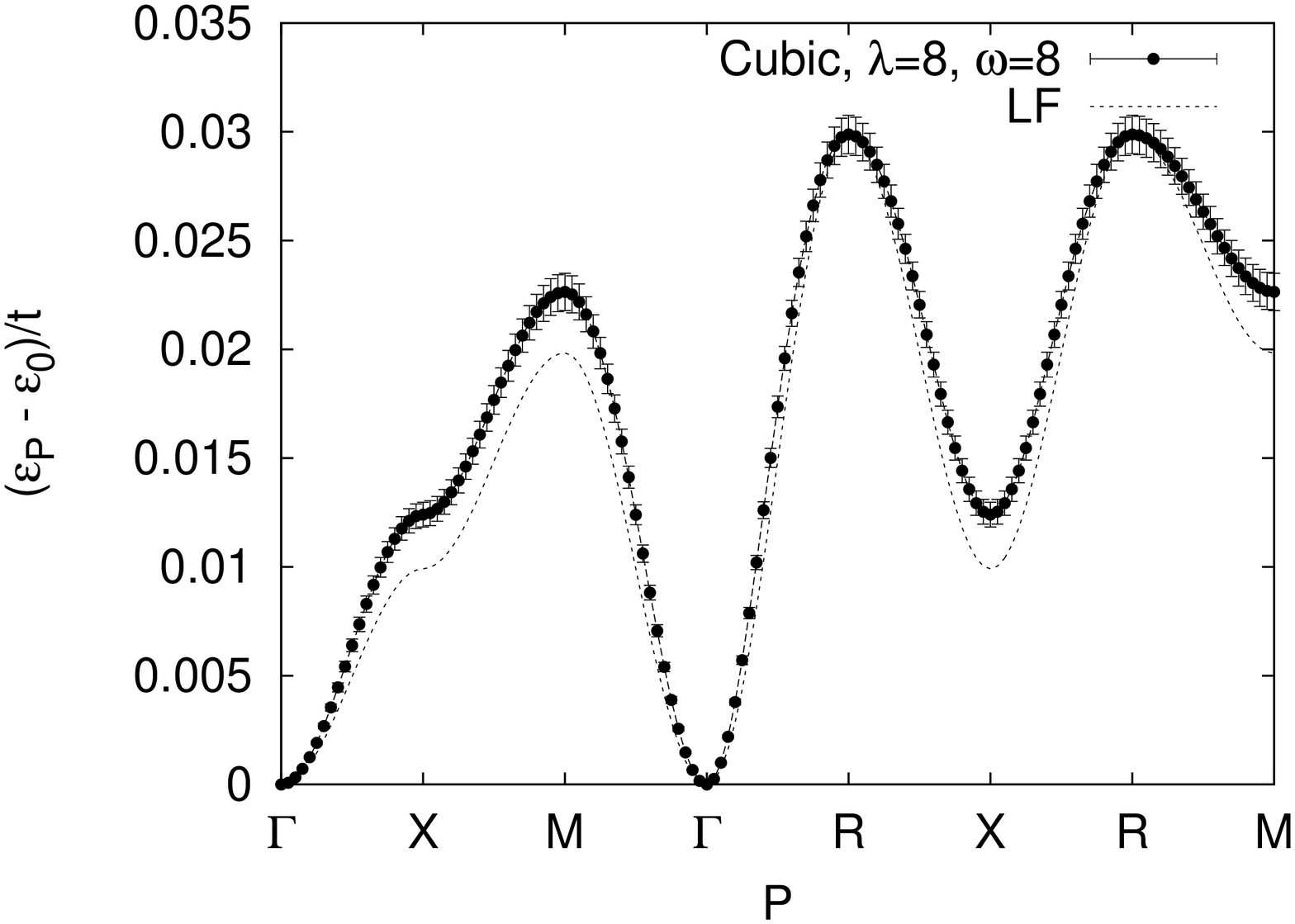}
\caption{Dispersion of the Holstein polaron in the near-adiabatic
  ($\bar{\omega}=1,\lambda=1.4$) and antiadiabatic
  ($\bar{\omega}=8,\lambda=8$) regimes. Also shown is the result from
  the Lang--Firsov approximation. The band width remains dependent on
  the coordination number, with the polaron spectra of the cubic and
  triangular lattices having similar width. Alternatively, the shape
  of the dispersion is clearly related to the dimensionality. As
  expected, the Lang--Firsov dispersion is quite accurate in the
  antiadiabatic limit and fails in the adiabatic limit.}
\label{fig:holsteinspectrum}
\end{figure}

The spectrum of the Holstein polaron in the near-adiabatic
($\bar{\omega}=1,\lambda=1.4$) and antiadiabatic
($\bar{\omega}=8,\lambda=8$) regimes is shown in Fig.
\ref{fig:holsteinspectrum} for the square, cubic and triangular
lattices. Also shown is the result from the LF transformation
(Eqn. (\ref{eqn:smallpolaron})). Polarons on triangular and cubic
lattices are not very mobile in either regime, with the coordination
number clearly playing a major role in the band-width. Alternatively,
the shape of the dispersion is strongly determined by the lattice
dimensionality.


\begin{figure}
\includegraphics[width=50mm,height=70mm,angle=270]{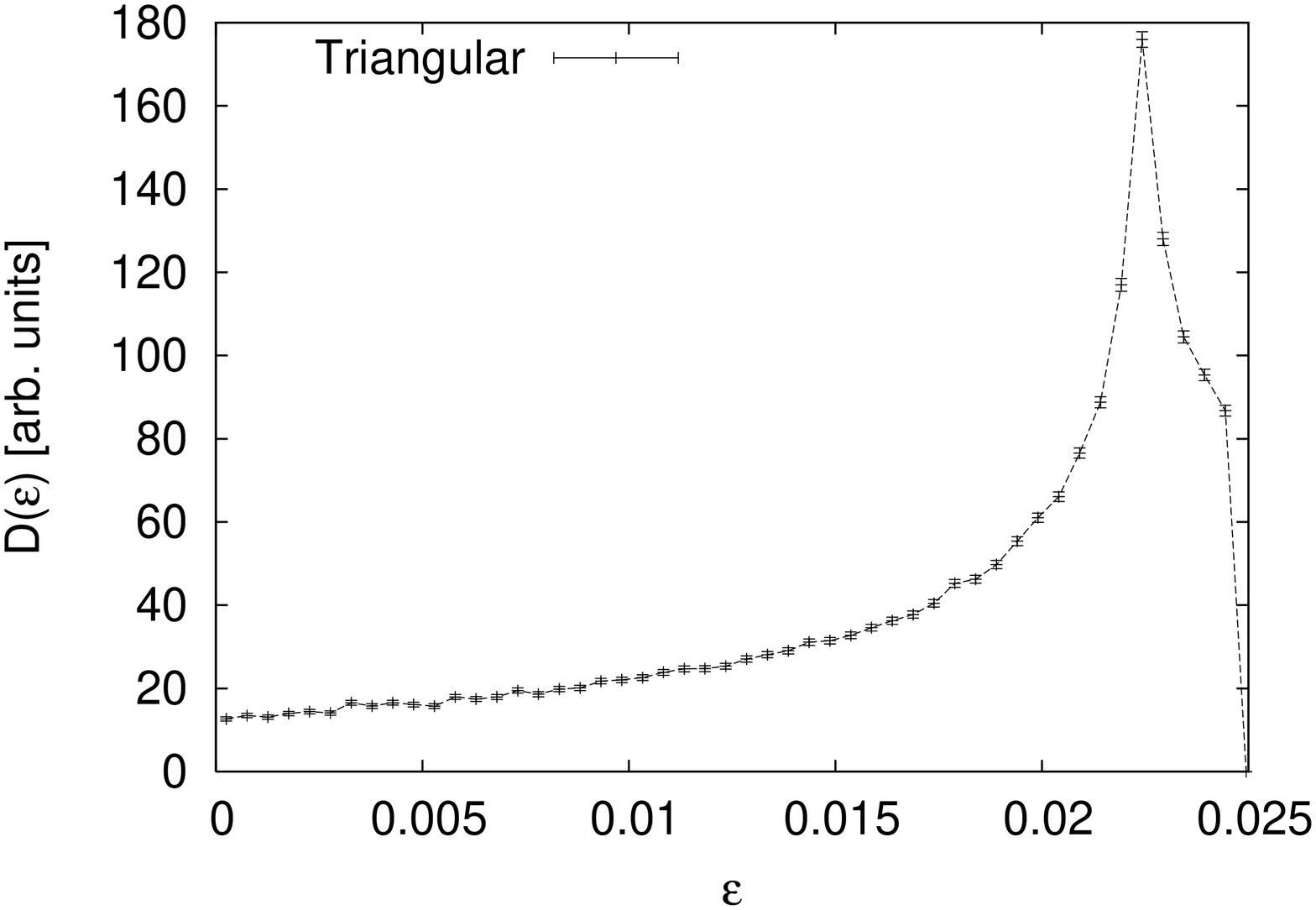}
\includegraphics[width=50mm,height=70mm,angle=270]{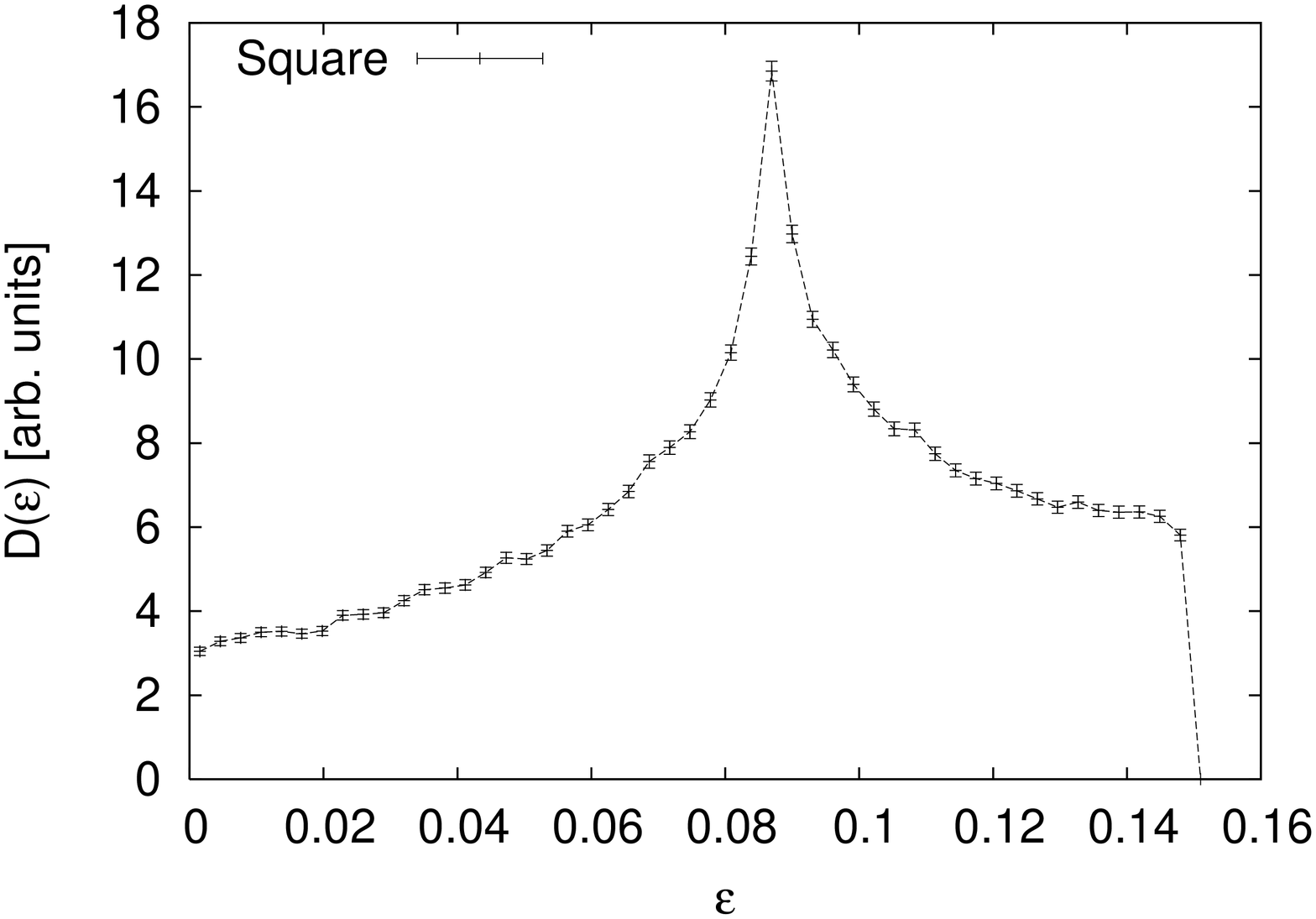}
\caption{Density of states for the Holstein interaction in the antiadiabatic limit
  ($\lambda=8$ and $\bar{\omega}=8$). Again, the effect of dimension can
  be seen in the shape of the DOS, with the logarithmic divergence
  (van Hove singularity) clearly visible for both lattices, while the
  band width is most affected by the coordination number.}
\label{fig:holsteindos}
\end{figure}

Figure \ref{fig:holsteindos} shows the density of states (DOS), $D(\epsilon)=\sum_{\bf k}\delta(\epsilon-\epsilon_{\bf k})/N$ for the
Holstein polaron in the antiadiabatic limit
($\lambda=8$ and $\bar{\omega}=8$). Again, the effect of dimension can be
seen in the shape of the DOS, with the logarithmic divergence (van
Hove singularity) clearly visible for both lattices, while the band
width is clearly determined by the coordination number (see Ref.
\onlinecite{kornilovitch1998a} for the shape of the DOS of the cubic
lattice, which has a completely different form).

\section{Effects of Screening}
\label{sec:screening}

An important question about polaron properties also involves the
effects of screening on the electron-phonon interaction. Unscreened
e-ph interactions make polarons very mobile
\cite{alebra,alexandrov1999a}, which leads to strong effects even on
the qualitative physical properties of the polaron gas. In this
section, only the triangular and square lattices are
considered.

\begin{figure}

\includegraphics[width=50mm,height=70mm,angle=270]{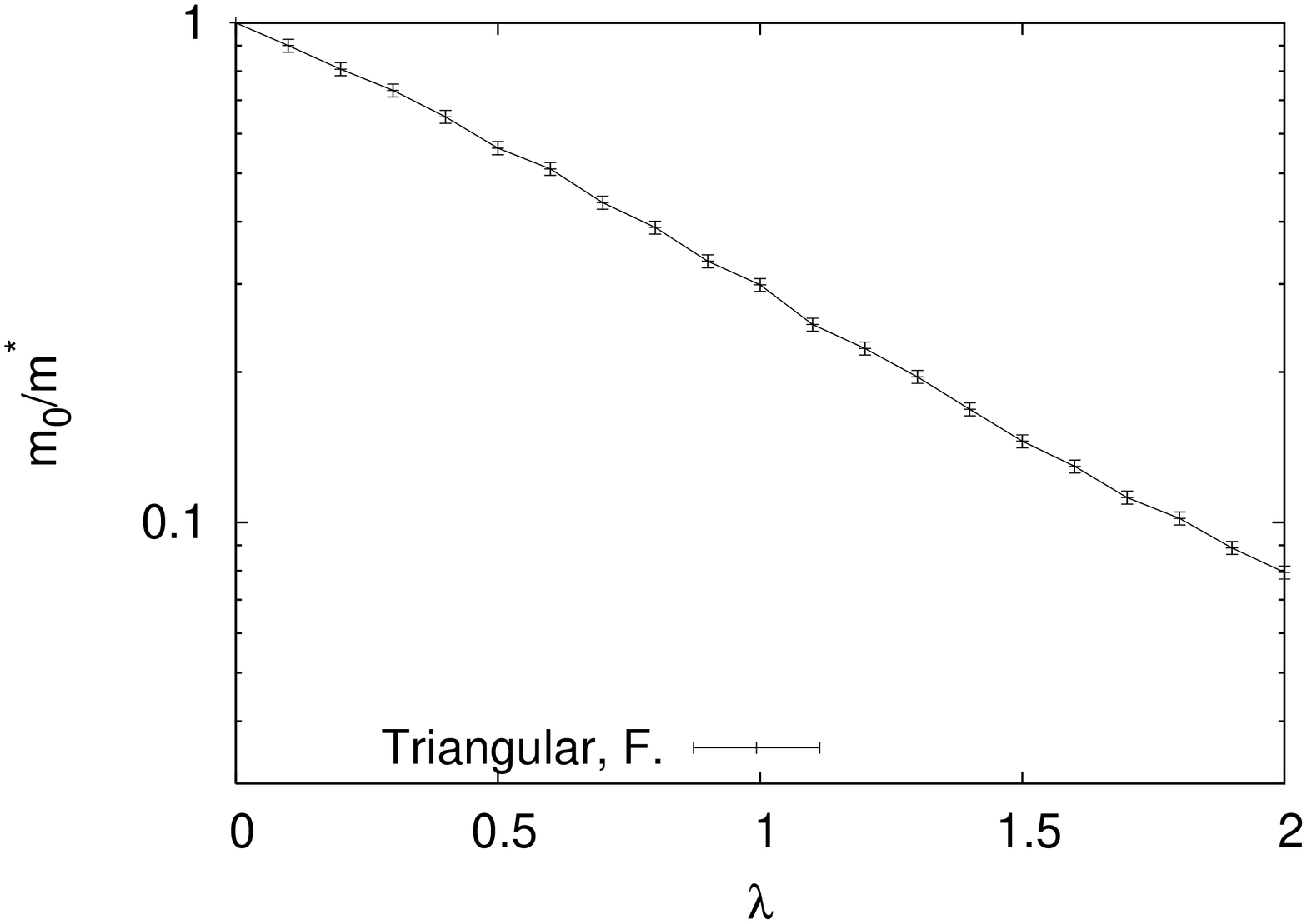}
\includegraphics[width=50mm,height=70mm,angle=270]{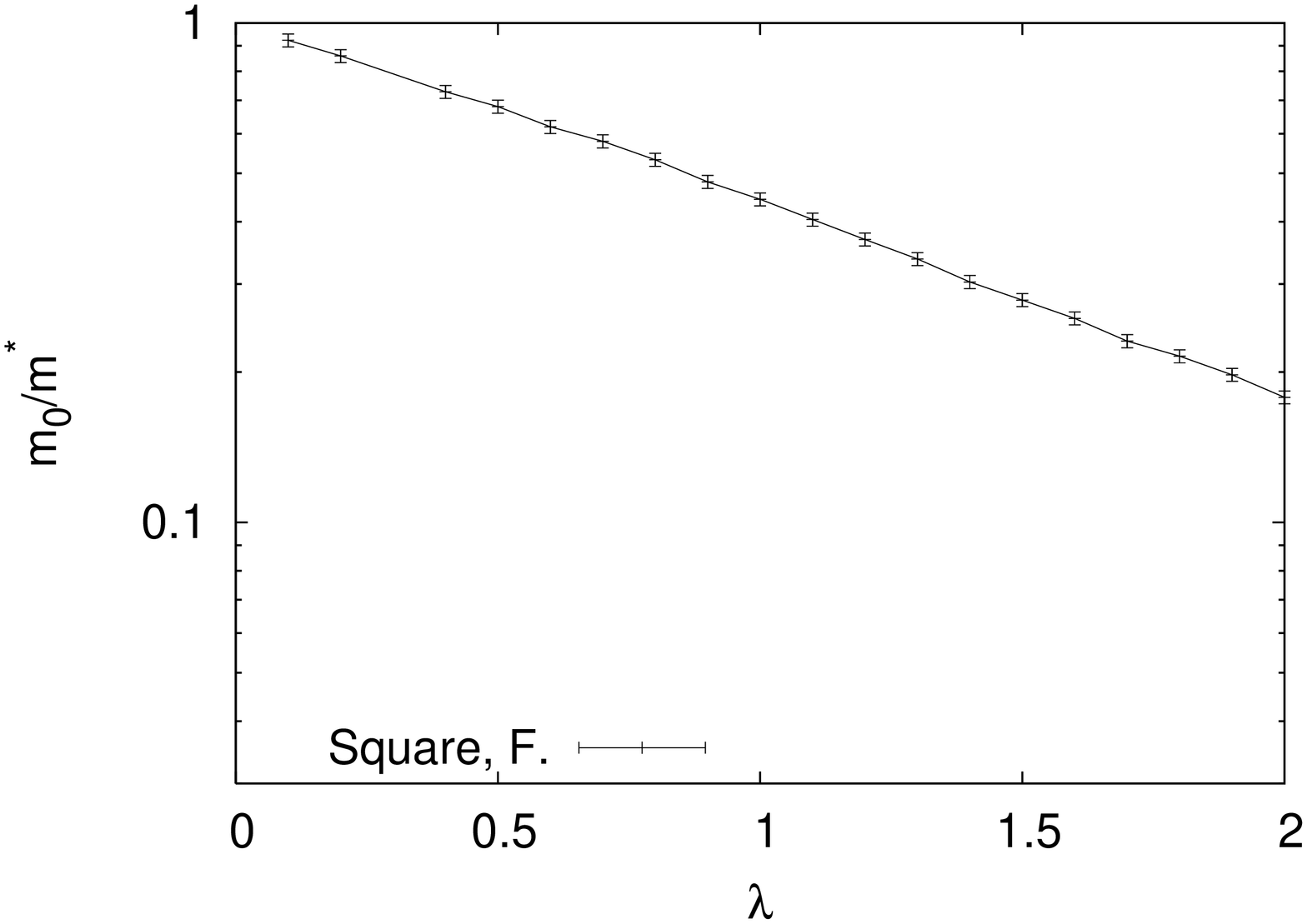}

\caption{Inverse mass for the Fr\"{o}hlich polaron with $\bar{\omega}=1$ ($R_{sc}\rightarrow\infty$). Note that the
  polaron is far more mobile than in the Holstein case. The results
  for the two lattices are very similar, indicating that in the
  case of long range interactions, the dimensionality of the lattice
  plays a much more important role in the polaron dynamics, with the
  specific form of the lattice unimportant.}
\label{fig:imfrol}
\end{figure}

In Fig. \ref{fig:imfrol}, the inverse effective mass for the Fr\"{o}hlich interaction is plotted.
The first point to note is that while the inverse mass in the strong
coupling limit depended strongly on the coordination number in the
Holstein model, this is no longer true for the Fr\"ohlich polaron.  
In terms of the way the polaron is formed, the phonon cloud is spread 
out over several lattice sites. This reduces the dependence on lattice
type.

The computation of the energy spectrum is limited by the bandwidth. For very 
large polaron energies $\epsilon_{\mathbf{k}}>k_{B}T$, there is a sign problem introduced 
by the average of a cosine term (Eqn. (\ref{eq:eleven})). This means that 
the spectrum can only be computed effectively for small bandwidths. This is 
not the case for the pure Fr\"{o}hlich polaron, which is significantly more 
mobile than the Holstein polaron. In order to circumvent this problem and 
obtain some information about long range potentials, the screened Fr\"{o}hlich
interaction with $R_{sc}=1$ is used, which has some properties of both
the Holstein and Fr\"{o}hlich polarons.

\begin{figure}
\includegraphics[width=50mm,height=70mm,angle=270]{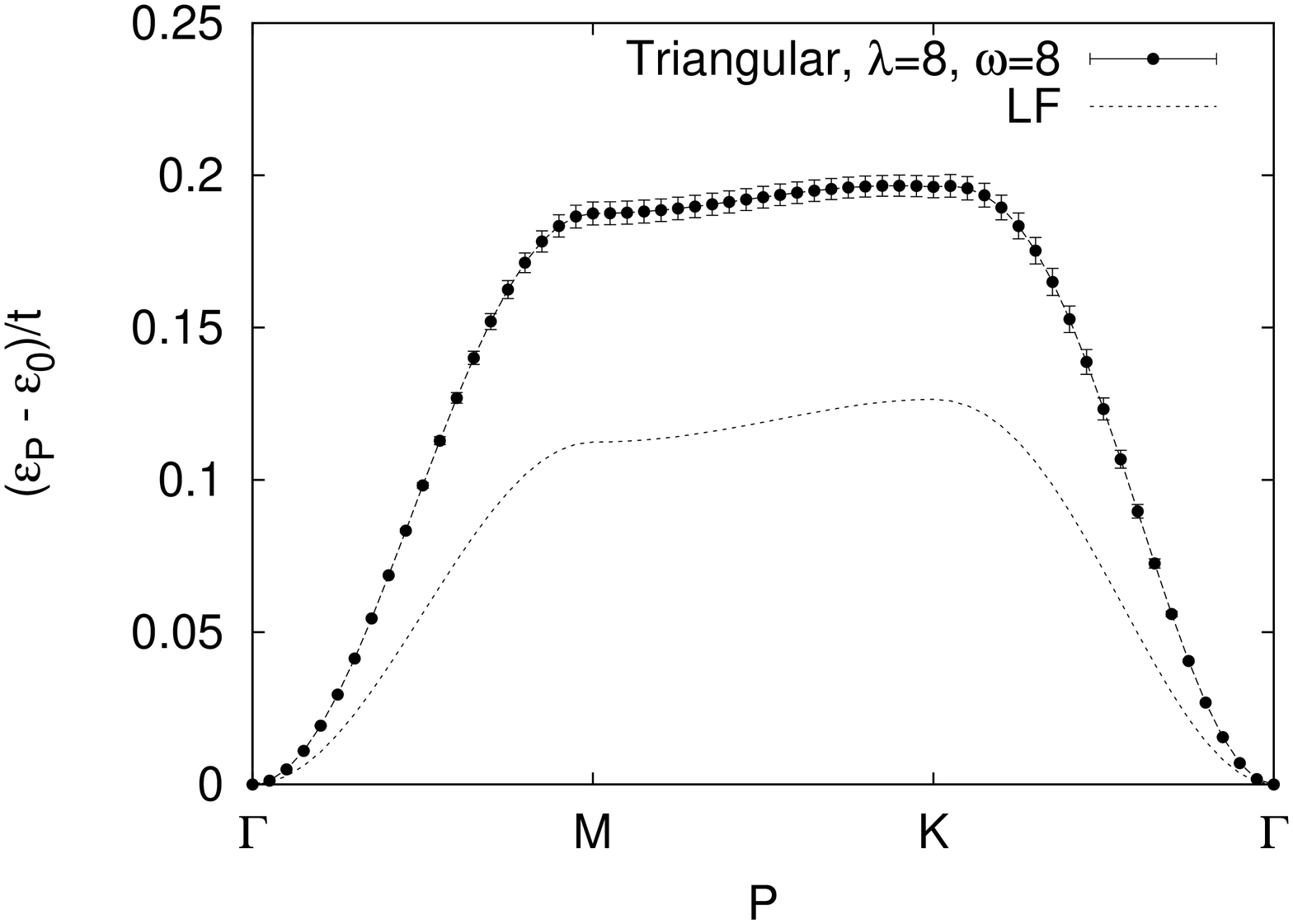}
\includegraphics[width=50mm,height=70mm,angle=270]{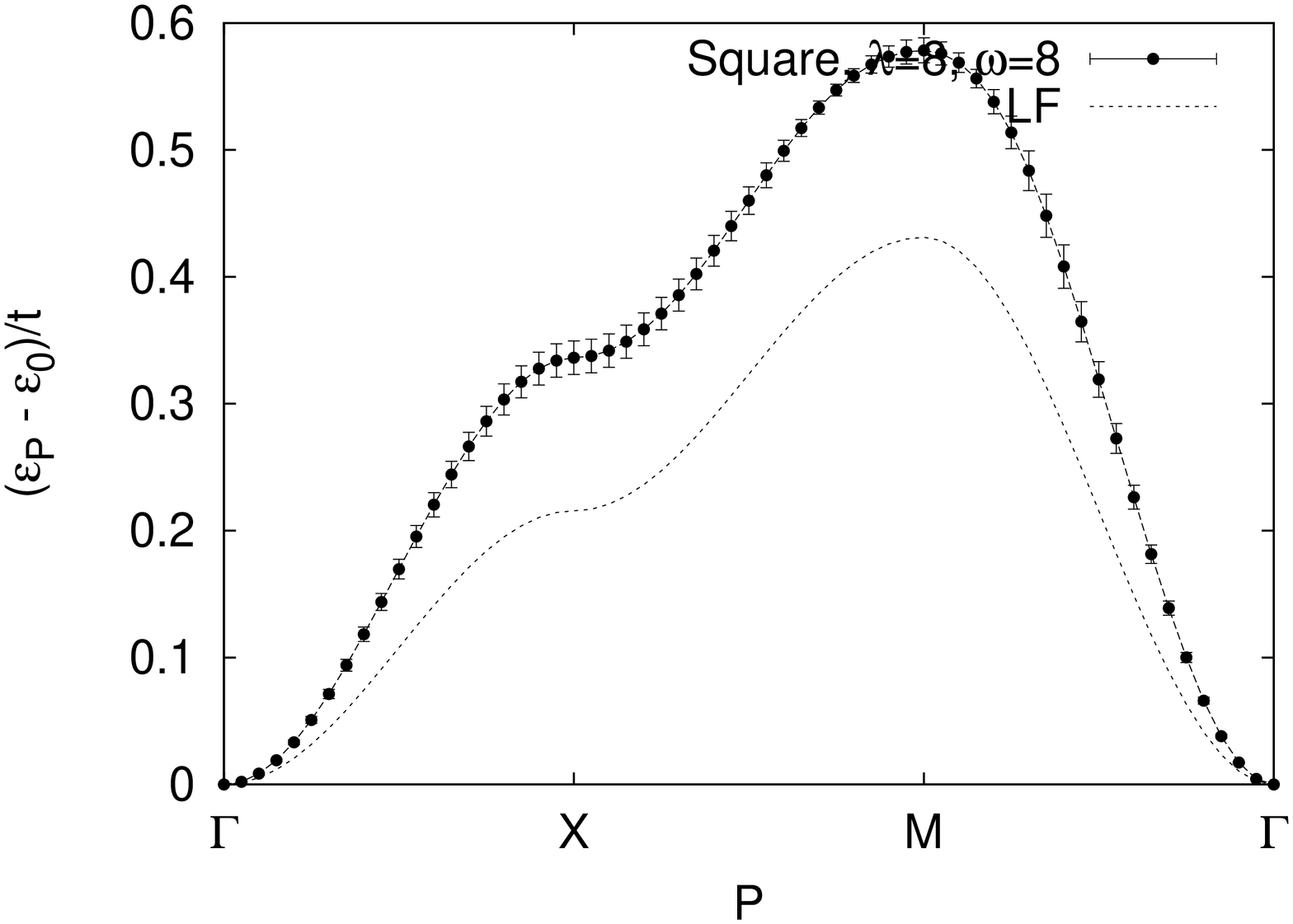}

\caption{Spectrum for a screened Fr\"{o}hlich polaron with
  $R_{\mathrm{sc}}=1$, $\lambda=8$ and $\bar{\omega}=8$. The bandwidth is much larger than in the case
  of the Holstein interaction. The Lang-Firsov result
  is also shown (light dotted line). It can be seen that the bandwidths of the triangular
  and square lattices have the same order of magnitude as the screening radius is increased.}
\label{fig:scholsteinspectrum}
\end{figure}

\begin{figure}
\includegraphics[width=50mm,height=70mm,angle=270]{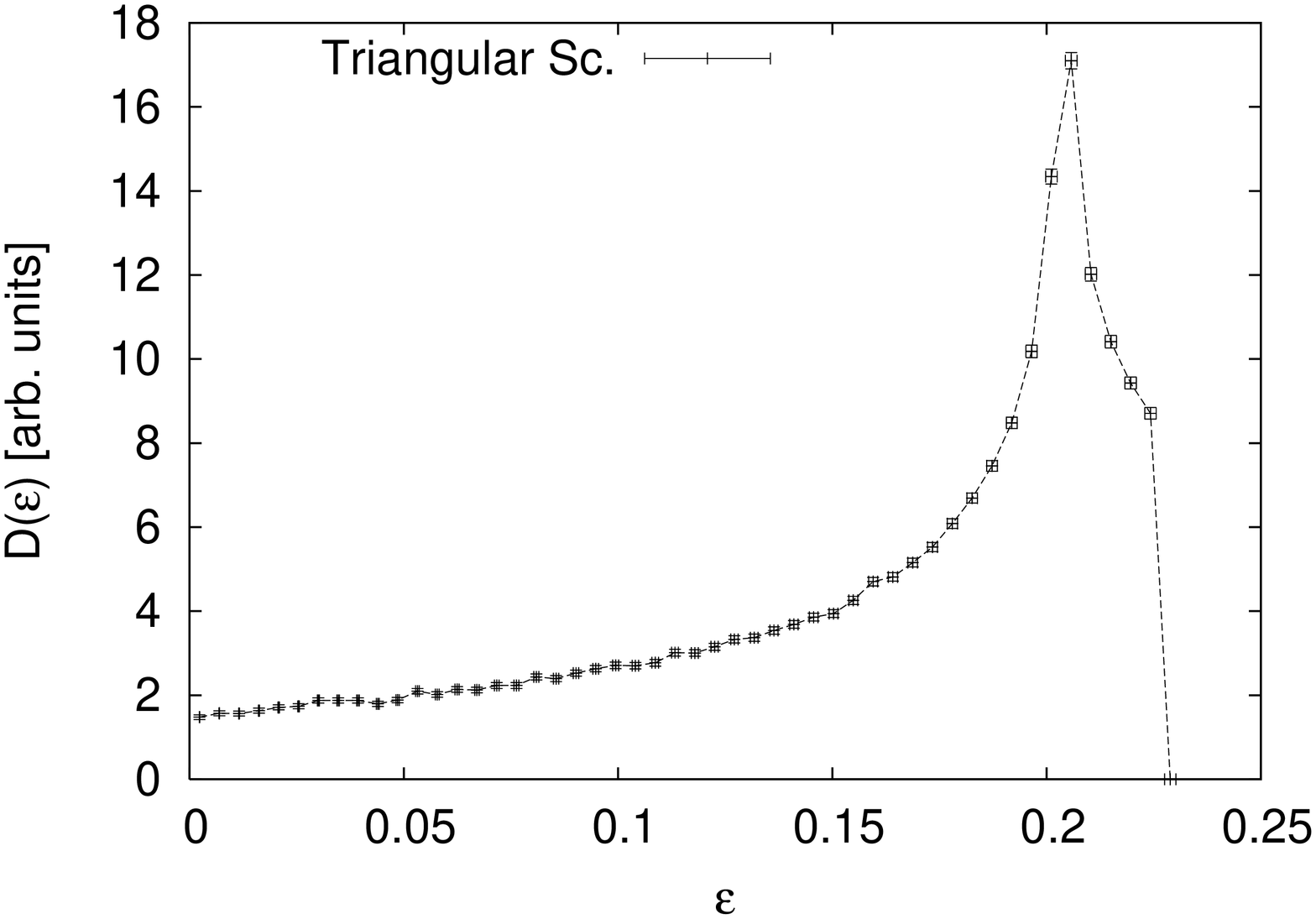}
\includegraphics[width=50mm,height=70mm,angle=270]{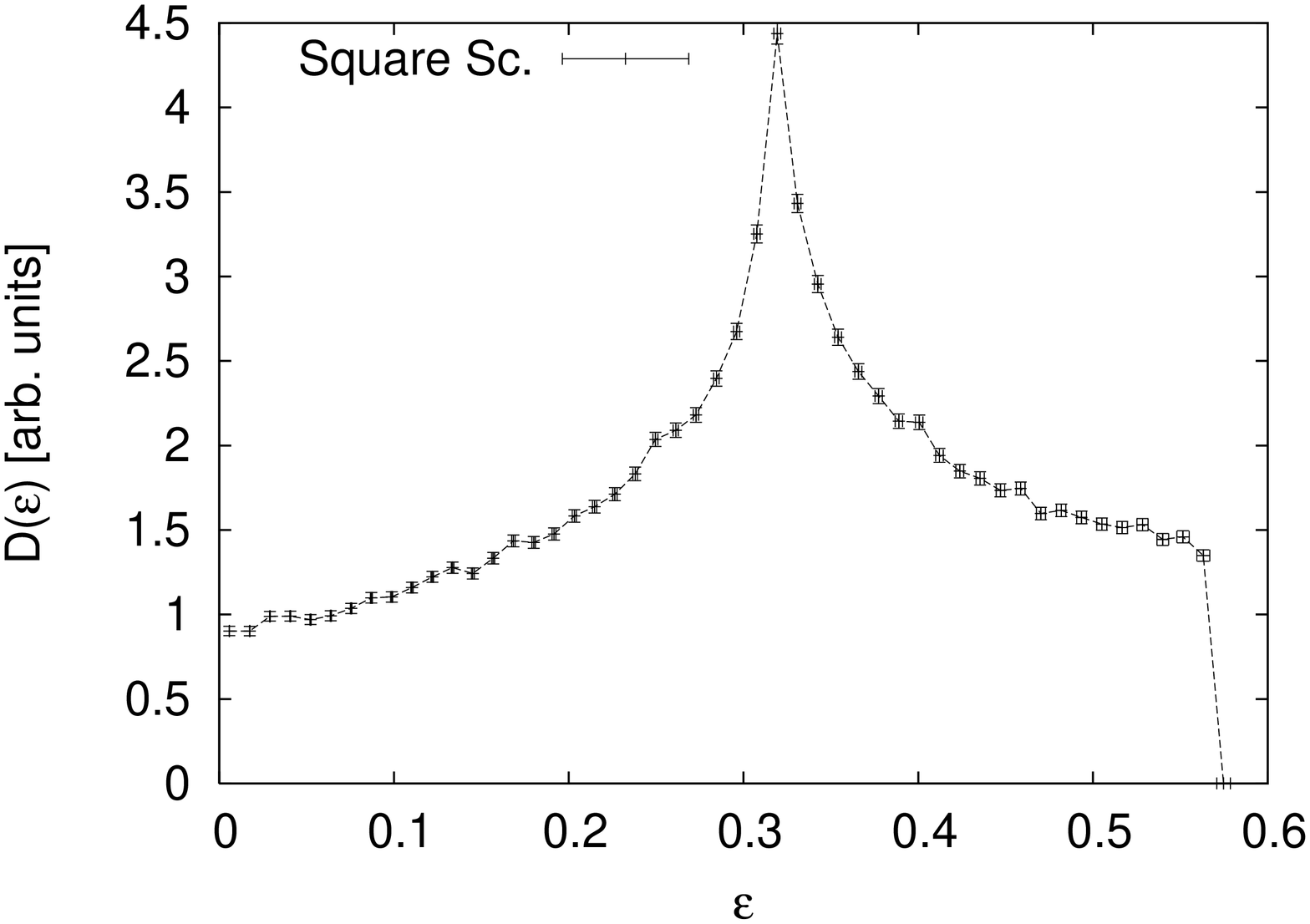}

\caption{DOS for a screened Fr\"{o}hlich polaron with 
  $R_{\mathrm{sc}}=1$, $\lambda=8$ and $\bar{\omega}=8$. Again, the lattice dimensionality determines
  the presence of van Hove singularities. The Fr\"{o}hlich polarons on
  both lattices are expected to have similar transport properties.}
\label{fig:scholsteindos}
\end{figure}

The band structure and density of states (DOS) of the screened
Fr\"{o}hlich polaron are shown in Figs. \ref{fig:scholsteinspectrum}
and \ref{fig:scholsteindos} respectively. Even for such a small
screening length, the effect on the spectrum is dramatic. Again, the
shape is given by the lattice type, but now, while the triangular
polaron is less mobile than the square polaron, the ratio of bandwidths is much closer to unity than for
the Holstein polaron. This is a little surprising, since in this case
the polaron extends over very few lattice spacings but the effects of
lattice are essentially annulled.

\begin{figure}
\includegraphics[width=50mm,height=70mm,angle=270]{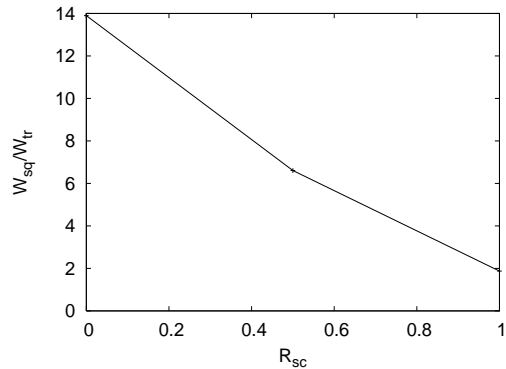}

\caption{Ratio of the effective bandwidth of the screened Fr\"{o}hlich polaron on square and triangular lattices. $\lambda=1.4$ and $\bar{\omega}=1$. Note
  how the ratio of the bandwidths converges with increasing
  $R_{\mathrm{sc}}$. This shows that the effects of lattice type are
  essentially washed out for long range potentials.}
  \label{fig:last}
\end{figure}

Finally to determine how the relative bandwidths become similar in the
Fr\"{o}hlich polarons on the triangular and square lattices, Fig.
\ref{fig:last} shows the evolution of the ratio of the effective
bandwidths as the screening length $R_{sc}$ is increased in the near adiabatic limit. The ratio of
the bandwidths converges to a constant value $\sim 1$ with increased screening
length, indicating that the effect of the lattice type is essentially
washed out for long range potentials.

\section{Summary}
\label{sec:summary}

We have studied the effects of lattice type on polaron dynamics using
a continuous time quantum Monte-Carlo technique. The effective mass,
isotope coefficient, ground state energy, phonon number, density of
states and polaron spectrum were calculated for the Holstein and
Fr\"{o}hlich polarons. Several lattices were investigated: linear, triangular,
square, cubic, face-center-cubic, hypercubic, hexagonal and body-center-cubic. The results were compared with analytic forms from weak- and
strong-coupling perturbation theory.

The overriding factor for polaron dynamics in the Holstein model is
found to be the number of nearest neighbors. For instance, the ground
state properties for the Holstein polaron on the triangular lattice
($z=6$) compare most closely with the cubic lattice ($z=6$) and not
the square lattice (which might na\"{i}vely be expected since square
and triangular lattices have the same dimensionality). Alternatively,
when spectral properties such as the DOS and spectrum are considered,
the lattice type is strongly responsible for the shape of the
spectrum/DOS, but the band width is most closely related to the number
of nearest neighbors. This is particularly obvious when the
Lang-Firsov antiadiabatic limit is considered.

When an extended Fr\"{o}hlich polaron is considered, the picture
changes significantly. In that case, it is dimensionality which most
strongly determines the form of the dynamic quantities (such as the
effective mass), and not the number of nearest neighbors, since the
lattice distortion surrounding the polaron is spread over a number of 
lattice sites. The band-widths of the triangular and cubic lattices 
converge as the screening length is increased, even for moderate length 
scales of the potential of a couple of lattice sites.

This work was supported by  EPSRC (UK) (grant EP/C518365/1).


\begin{thebibliography}{200}


\bibitem{zhao1997a}  
G.M. Zhao and D. E. Morris, 
Phys. Rev. B {\bf 51}, R16487 (1995); 
G.-M. Zhao, M. B. Hunt, H. Keller, and K. A. M\"uller, 
Nature (London) {\bf 385}, 236 (1997); 
R. Khasanov, D. G. Eshchenko, H. Luetkens, E. Morenzoni, T. Prokscha, A. Suter,
N. Garifianov, M. Mali, J. Roos, K. Conder, and H. Keller,
Phys. Rev. Lett. {\bf 92}, 057602 (2004).

\bibitem{lanzara2001a}  
A. Lanzara, P.V. Bogdanov, X.J. Zhou, S.A. Kellar, D.L. Feng, E.D. Lu, 
T. Yoshida, H. Eisaki, A. Fujimori, K. Kishio, J.I. Shimoyana, T. Noda, 
S. Uchida, Z. Hussain and Z.X. Shen, 
Nature (London) {\bf 412}, 510 (2001); 
G-H. Gweon, T. Sasagawa, S.Y. Zhou, J. Craf, H. Takagi, D.-H. Lee, and A. Lanzara,
Nature (London) {\bf 430}, 187 (2004).

\bibitem{mic} 
D. Mihailovi\'{c}, C.M. Foster, K. Voss, and A.J. Heeger, 
Phys. Rev. B{\bf 42}, 7989 (1990).

\bibitem{ita}  
P. Calvani, M. Capizzi, S. Lupi, P. Maselli, A. Paolone, P. Roy, S.W. Cheong, 
W. Sadowski, and E. Walker, 
Solid State Commun. {\bf 91}, 113 (1994).

\bibitem{ega}  
T. Egami, 
J. Low Temp. Phys. {\bf 105}, 791 (1996).

\bibitem{zhao1996a} 
G.-M. Zhao, K. Conder, H. Keller, and K. A. M\"uller,
Nature (London) {\bf 381}, 676 (1996).

\bibitem{alebra}  
A.S. Alexandrov, Phys. Rev. B{\bf 53}, 2863 (1996);  
Phys. Rev. Lett. {\bf 82}, 2620 (1999);
A.S. Alexandrov and A.M. Bratkovsky,
Phys. Rev. Lett. {\bf 84}, 2043 (2000).

\bibitem{edwards2002}
D.M.Edwards, Adv. Phys. {\bf 51}, 1259 (2002)

\bibitem{alemot}
A.S. Alexandrov and N.F. Mott, 
Rep. Prog. Phys. {\bf 57}, 1197 (1994); 
{\em Polarons and Bipolarons} (World Scientific, Singapore, 1995).

\bibitem{dev}
J.T. Devreese, 
in {\em Encyclopedia of Applied Physics}, vol. 14, p. 383, VCH Publishers (1996).

\bibitem{migdal1958a} 
A.B. Migdal, 
Zh. Eksp. Teor. Fiz. {\bf 34}, 1438 (1958) [Sov. Phys. JETP {\bf 7}, 996 (1958)].

\bibitem{alemaz} 
A.S. Alexandrov, V.N. Grebenev, and E.A. Mazur,
Pis'ma Zh. Eksp. Teor. Fiz. {\bf 45}, 357 (1987)[JETP Lett. {\bf 45}, 455 (1987)].

\bibitem{mar} 
F. Dogan and F. Marsiglio, 
Phys. Rev. B {\bf 68}, 165102 (2003).

\bibitem{hag} 
J.P. Hague, 
J. Phys.: Condens. Matter {\bf 15}, 2535 (2003); J.P.Hague, J. Phys.: Condens. Matter {\bf 17}, 5663 (2005) and references therein.

\bibitem{alebook} 
A.S. Alexandrov, \emph{Theory of Superconductivity: From Weak to Strong Coupling} 
(IoP Publishing, Bristol, 2003).

\bibitem{lang1962a}
I.G. Lang and Yu.A. Firsov, 
Zh. Eksp. Teor. Fiz. {\bf 43}, 1843 (1962) 
[Sov. Phys. JETP {\bf 16}, 1301 (1963)].

\bibitem{ciuchi1997a} 
S.Ciuchi, F. de Pasquale, S. Fratini, and D. Feinberg, 
Phys. Rev. B {\bf 56}, 4494 (1997).

\bibitem{alekab} 
A.S. Alexandrov, V.V. Kabanov, and D.K. Ray,
Phys. Rev. B {\bf 49}, 9915 (1994); 
H. Fehske, H. R\"oder, G. Wellein, and A. Mistriotis, 
Phys. Rev. B {\bf 51}, 16582 (1995); 
G. Wellein, H. R\"oder, and H. Fehske, 
Phys. Rev. B {\bf 53}, 9666 (1996); 
F. Marsiglio, Physica C {\bf 244}, 21 (1995); 
W. Stephan, Phys. Rev. B {\bf 54}, 8981 (1996); 
M. Capone, W. Stephan, and M. Grilli, 
Phys. Rev. B {\bf 56}, 4484 (1997).

\bibitem{tru} S.A. Trugman, J. Bon\v{c}a, and L.-C. Ku, 
Int. J. Mod. Phys. B {\bf 15}, 2707 (2001); L.-C.Ku, S.A.Trugman and J.Bon\v{c}a, Phys. Rev. B {\bf 65}, 174306 (2002) and references therein.

\bibitem{qmc} 
J.E. Hirsch and E. Fradkin,
Phys. Rev. Lett. {\bf 49}, 402 (1982); 
Phys. Rev. B {\bf 27}, 4302 (1983); 
E. Fradkin and J.E. Hirsch, Phys. Rev. B {\bf 27}, 1680 (1983); 

\bibitem{DeRaedt}
H. De Raedt and A. Lagendijk, 
Phys. Rev. Lett. {\bf 49}, 1522 (1982); 
Phys. Rev. B {\bf 27}, 6097 (1983); 
Phys. Rev. B {\bf 30}, 1671 (1984); 
Phys. Rep. {\bf 127}, 234 (1985), and references therein.

\bibitem{White}
E. Jeckelmann and S.R. White, 
Phys. Rev. B {\bf 57}, 6376 (1998);
E. Jeckelmann, C. Zhang, and S.R. White, 
{\em ibid.} {\bf 60}, 7950 (1999);
C. Zhang, E. Jeckelmann, and S.R. White, 
{\em ibid.} {\bf 60}, 14092 (1999);

\bibitem{kornilovitch1998a} 
P.E. Kornilovitch, Phys. Rev. Lett. {\bf 81}, 5382 (1998);
Phys. Rev. B {\bf 60}, 3237 (1999).

\bibitem{hohenadler2005}
M.Hohenadler, D.Neuber, W. von der Linden, G.Wellein, J.Loos, H.Fehske, Phys. Rev. B {\bf 71}, 245111 (2005).

\bibitem{spencer2005a} 
P.E. Spencer, J.H. Samson, P.E. Kornilovitch, and A.S. Alexandrov, 
Phys. Rev. B {\bf 71}, 184310 (2005).

\bibitem{friedman}
L.Friedman in Polarons and Bipolarons in High-$T_C$ superconductors and related materials (Cambridge University Press, Cambridge, 1995), Chapter 11.

\bibitem{hohen}
M.Hohenadler, H.G. Evertz, and W. von der Linden,
Phys. Rev. B {\bf 69}, 024301 (2004).

\bibitem{Prokofev}
N.V. Prokof'ev and B.V. Svistunov,
Phys. Rev. Lett. {\bf 81}, 2514 (1998);
A.S. Mishchenko, N.V. Prokof'ev, A. Sakamoto, and B.V. Svistunov,
Phys. Rev. B {\bf 62}, 6317 (2000).

\bibitem{Macridin}
A. Macridin, G.A. Sawatzky, and M. Jarrell,
Phys. Rev. B {\bf 69}, 245111 (2004).

\bibitem{Feynman}
R.P. Feynman,
Phys. Rev. {\bf 97}, 660 (1955);
{\em Statistical Mechanics} (Benjamin, Reading, MA, 1972), Chapter 8.

\bibitem{KornilovitchPike}
P.E. Kornilovitch and E.R. Pike,
Phys. Rev. B {\bf 55}, R8634 (1997).

\bibitem{alexandrov1999a}  
A.S. Alexandrov and P.E. Kornilovitch, 
Phys. Rev. Lett. {\bf 82}, 807 (1999).

\bibitem{Kornilovitch1999b}
P.E. Kornilovitch,
Phys. Rev. B {\bf 59}, 13531 (1999).

\bibitem{KornilovitchAlexandrov2004}
P.E. Kornilovitch and A.S. Alexandrov,
Phys. Rev. B {\bf 70}, 224511 (2004).

\bibitem{Holstein}
T. Holstein, 
Ann. Phys. {\bf 8}, 325 (1959); 
Ann. Phys. {\bf 8}, 343 (1959).

\bibitem{Fehske2000}
H. Fehske, J. Loos, and G. Wellein,
Phys. Rev. B {\bf 61}, 8016 (2000).

\bibitem{Bonca2001}
J. Bon\v{c}a and S.A. Trugman,
Phys. Rev. B {\bf 64}, 094507 (2001).

\bibitem{aleiso} 
A.S. Alexandrov, 
Phys. Rev. B {\bf 46}, R14932 (1992).


\bibitem{fro}
H. Fr\"ohlich, 
Adv. Phys. {\bf 3}, 325 (1954).



\bibitem{kabanov1993a} 
V.V. Kabanov and O.Yu. Mashtakov, 
Phys. Rev. B {\bf 47}, 6060 (1993).


\end{thebibliography}
\end{document}